\documentclass{aa}

\usepackage{graphicx}
\usepackage{txfonts}
\usepackage{natbib}
\usepackage{hyperref}
\bibpunct{(}{)}{;}{a}{}{,} 

\begin{document}

\title{Modeling protoplanetary disk SEDs with artificial neural networks}
\subtitle{Revisiting the viscous disk model and updated disk masses}

\author{\'A. Ribas\inst{1}$^,$\inst{2} \and
  C. C. Espaillat\inst{2},
  E. Mac\'ias \inst{1}$^,$\inst{2}$^,$\inst{3} \and
  L. M. Sarro\inst{4}
}

\institute{European Southern Observatory (ESO), Alonso de C\'ordova 3107, Vitacura, Casilla 19001, Santiago de
  Chile, Chile\\
  \email{aribas@eso.org}\label{inst1} \and
  Department of Astronomy, Boston University, 725 Commonwealth Avenue, Boston, MA 02215, USA\label{inst2} \and
  Joint ALMA Observatory, Alonso de C\'ordova 3107, Vitacura, Casilla 19001, Santiago, Chile\label{inst3} \and
  Dpto. de Inteligencia Artificial, UNED, Juan del Rosal, 16, 28040 Madrid, Spain\label{inst4}
}

   \date{Accepted: 6 August 2020}

   \abstract{We model the spectral energy distributions (SEDs) of 23 protoplanetary disks in the
     Taurus-Auriga star-forming region using detailed disk models and a Bayesian approach. This is made
     possible by combining these models with artificial neural networks to drastically speed up their
     performance. Such a setup allows us to confront $\alpha$-disk models with observations while accounting
     for several uncertainties and degeneracies. Our results yield high viscosities and accretion rates for
     many sources, which is not consistent with recent measurements of low turbulence levels in disks. This
     inconsistency could imply that viscosity is not the main mechanism for angular momentum transport in
     disks, and that alternatives such as disk winds play an important role in this process. We also find that
     our SED-derived disk masses are systematically higher than those obtained solely from (sub)mm fluxes,
     suggesting that part of the disk emission could still be optically thick at (sub)mm wavelengths. This
     effect is particularly relevant for disk population studies and alleviates previous observational
     tensions between the masses of protoplanetary disks and exoplanetary systems.
   }

\titlerunning{SED modeling of protoplanetary disks in Taurus-Auriga}
\authorrunning{\'A. Ribas et al.}
\maketitle

\section{Introduction}\label{sec:introduction}

The field of protoplanetary disks has undergone a revolution during in recent years thanks to the improvement
in sensitivity and spatial resolution of new facilities. In particular, the Atacama Large Millimeter/submillimeter
Array (ALMA) has produced censuses of disk dust
and gas masses in different star-forming regions \citep[e.g.,][]{Ansdell2016, Miotello2016}, revealed the
ubiquity of substructures in disks \citep[e.g.,][]{Andrews2018_DSHARP, Long2018}, and found evidence for low
viscosities in these systems \citep[e.g.,][]{Pinte2016, Flaherty2017, Dong2018}, among other results. These new
data have clearly proven that planet formation theories are still incomplete, and that some of the assumptions
that are typically made require further consideration.

One important concept that has been increasingly questioned lately is the $\alpha$-disk prescription
\citep[e.g.,][]{Shakura1973}, which is at the base of many models of protoplanetary disks. In this
prescription, the transport of angular momentum is due to viscosity, $\nu$, which is proportional to the local
sound speed $c_s$ and the scale height $H$ \citep[$\nu = \alpha c_s H$, with $\alpha$ being constant
throughout the disk, e.g.,][]{Lynden-Bell1974, Pringle1981}. The origin of this turbulent viscosity is usually
attributed to the magneto-rotational instability \citep[MRI,][]{Balbus1991}. However, theoretical studies have
shown that this effect may not be sufficiently strong in protoplanetary disks: The midplane is shielded from
ionizing radiation and its ionization fraction is low \citep{Stone1998}, and the effects of non-ideal
magnetohydrodynamics could be enough to suppress MRI everywhere in the disk
\citep{BaiStone2013_winds}. Supported by recent observational evidence of low turbulence levels, disk winds
have gained increasing relevance as the potential main mechanism of angular momentum transport in disks
\citep[e.g.,][]{BaiStone2013_winds, Simon2015, Bai2016_winds, Bai2016}.

A second paradigm shift involves disk masses and the time of planet formation. ALMA has yielded complete
censuses of disk dust masses for several star-forming regions, allowing for comparative studies as a function
of different factors \citep[age, stellar mass, environment, see e.g.,][]{Ansdell2016, Miotello2016, Cieza2019,
  Williams2019}. Among the many results derived from this wealth of data, one emerging trend is that the
masses of disks between 1-3\,Myr appear to be consistently lower than the masses of confirmed exoplanetary
systems. This suggests that dust and gas masses could be systematically underestimated and/or that
planetesimals form rapidly during the first Myr of the disk lifetime \citep[e.g.,][]{Najita2014, Manara2018},
both results having important implications for our understanding of planet formation.

Exploring scenarios like the ones mentioned above requires the use of detailed physical
models. Due to their complexity, protoplanetary disk models have a large number of free parameters, which
makes them both very flexible and computationally demanding (models frequently take several minutes or
hours to run). For these reasons, disk modeling usually requires either fixing several parameters (some of
which could be highly uncertain) or adopting more simplistic models, which limits the information that can be
obtained from them. Two recent examples of this dichotomy are found in \citet{Woitke2019} and
\citet{Ballering2019}: In the first case, the authors modeled the complete spectral energy distribution (SED)
of 27 protoplanetary disks and further expanded the models to high angular and spectral resolution data for a
subset of 14 sources. For this purpose, they used a number of state-of-the-art thermo-chemical modeling codes,
producing the most detailed disk modeling effort to date. The downside of using such a complex scheme,
however, is that they are forced to fix several free parameters in the process, and a statistical treatment is
not within computational reach. On the other hand, \citet{Ballering2019} fitted 132 SEDs using
simpler models, but this allowed them to analyze their results in a statistical (Bayesian) manner. Because of
the high computational demand of combining both approaches, studies using both complex models and a Bayesian
approach have been limited to individual cases \citep[e.g.,][]{Ribas2016, Wolff2017}. Extending these efforts
to more sources while maintaining sufficiently detailed models and a statistical treatment could reveal
possible caveats in our current theories, but such an analysis has remained unfeasible so far.

Here we present a novel approach to fit samples of protoplanetary disks, combining physically motivated models
with neural networks to speed up the process. We train an artificial neural network (ANN) to mimic the
D'Alessio Irradiated Accretion Disk models \citep{DAlessio1998, DAlessio1999, DAlessio2001, Dalessio2005,
  DAlessio2006}, which we then use to fit 23 SEDs of disks in the Taurus-Auriga star-forming region with a
Bayesian framework. The results allow us to explore the performance of the $\alpha$-disk prescription when
confronted with observations from a statistical perspective. Section~\ref{sec:ann} describes the models and
the training of the artificial neural network. We present the sample of protoplanetary disks in
Sect.~\ref{sec:sample}, and the fitting process is then described in Sect.~\ref{sec:fitting}. The results are
presented in Sect.~\ref{sec:results}, and we discuss their implications in Sect.~\ref{sec:discussion}. A
summary of our main findings can be found in Sect.~\ref{sec:summary}.

\section{Speeding up disk models with artificial neural networks} \label{sec:ann}

Before we describe our use of an ANN, we first clarify its exact purpose in our study. The application of
ANNs is growing rapidly both in the astronomical and the machine learning communities due to their enormous
flexibility, and are regularly used for classification purposes or to derive parameter values based on
observations. A good example of such usage would be to consider an observed SED and to employ an ANN
(previously trained with disk models) to predict its possible disk parameters. This method, however, does not
provide probability distributions for the parameters.

Our goal is to infer the probability distributions (posterior distributions) for the different parameters in
our disk models, which requires the use of Bayesian statistics. The high dimensionality of this problem makes
it unfeasible to approach this task using regular grids of models and, instead, Markov chain Monte Carlo
(MCMC) methods are needed to properly sample posterior distributions. However, this technique involves the
evaluation of a large number (usually over hundreds of thousands) of models arbitrarily determined by the
evolution of the chains and, therefore, models need to be computed quickly enough for the process to be
feasible. Given the timescales of radiative transfer models, they are typically not combined with MCMC
methods. To solve this problem, in this study we adopted an ANN for a different purpose than those described
above: Its aim is to mimic physically motivated models that are computationally demanding, that is, given the
same input parameters, the ANN will output the same SED as the disk models (but much faster). With this tool,
the likelihood can be estimated as usual with the observed data and the model SED. The ANN in this paper can
be thought of as an interpolator function that quickly provides the SED at any point in the parameter space,
but it does not perform any parameter prediction itself (in contrast with the most standard use of ANNs).

\subsection{The DIAD models}\label{sec:DIAD_models}

In this paper, we used the D'Alessio Irradiated Accretion Disk models \citep[DIAD, ][]{DAlessio1998,
  DAlessio1999, DAlessio2001, Dalessio2005, DAlessio2006}. DIAD employs an $\alpha$-disk prescription and
solves the hydrostatic equilibrium and energy transport in the disk self-consistently. The models also include
a distribution of dust grain sizes and a simple parametrization of the settling of large particles toward the
disk midplane. Based on the input physical parameters, DIAD estimates the SED at the requested
wavelengths. The relevant parameters considered in this study are:

\begin{itemize}

\item Stellar parameters, $M_*$, $T_*$, and $R_*$.

\item The mass accretion rate in the disk, $\dot{M}$.

\item The disk viscosity, characterized by the adimensional $\alpha$ parameter \citep{Shakura1973}.

\item Dust settling, $\epsilon$ (adimensional). DIAD includes two different dust populations, one of small
  grains in the upper layers and a second one that also includes larger grains in the disk
  midplane. $\epsilon$ is defined as the ratio between the gas-to-dust mass ratio in the upper layers and the
  standard value of this ratio \citep{DAlessio2006}. Lower $\epsilon$ values imply a higher dust depletion in
  the disk atmosphere, and thus more settled disks. Changing $\epsilon$ alters the total mass of grains in the
  disk atmosphere, and the mass of large grains in the disk midplane is modified accordingly so that the total
  dust mass in the disk remains unchanged.

\item Maximum grain sizes for the dust populations of the upper layer and in the disk midplane, $a_{\rm max, upper}$
  and $a_{\rm max, midplane}$. For each of these populations, a power-law grain size distribution is assumed
  following $dn/da \propto a^{-p}$ from $a_{\rm min}$ to $a_{\rm max}$, where $a$ is the grain size and $p=3.5$. In
  both cases, a minimum grain size of $a_{\rm min}=0.005 \mu$m is assumed.

\item The disk radius, $R_{\rm disk}$.

\item The inclination of the disk, $i$. 

\item The dust sublimation temperature, $T_{\rm wall}$, and scaling of the inner edge, or wall, of the
  disk, $z_{\rm wall}$ (in units of the pressure scale height). These two parameters control the location and
  surface area of the inner wall. In particular, $T_{\rm wall}$ is the temperature at which dust
  sublimates. The inner walls of disks are probably puffed-up and/or curved \citep[e.g.,][]{Natta2001,
    Dullemond2001, Isella2005, Dullemond2010}, but the DIAD models assume a flat, vertical wall and so, to
  account for the possible increase in surface area due to curvature, the emission from the wall is scaled in
  this work by a factor of $z_{\rm wall}$. We refer the reader to \citet{McClure2013} and
  \citet{Manzo-Martinez2020} for examples of how the DIAD models can be used to approximate curved walls,
  which is outside the scope of this work.

\end{itemize}

Therefore, there are 12 free parameters in DIAD that we consider in this study. Examples of the effect of
different disk parameters on the SED can be found in Appendix~\ref{appendix:parameters}. We adopted a standard
gas-to-dust mass ratio of 100, and a dust composition of 40\,\% graphites, 30\,\% olivines, and 30\,\%
pyroxenes.  Graphite and silicate opacities were calculated using Mie theory and optical constants from
\citet{Draine1984} and \citet{Dorschner1995}, respectively. The emission from the stellar photosphere was
  computed using the empirical colors of young stars provided in \citet{Pecaut2013}, which are interpolated
  based on the input stellar temperature and then scaled to yield the corresponding luminosity (determined by
  the input stellar temperature and radius). The accretion luminosity permeating the disk was calculated
assuming that the material is accreted from a magnetospheric radius corresponding to 5 stellar radii, and with
a shock temperature of 8000\,K. Furthermore, a blackbody component with this temperature was added to the SED
and scaled to have half of the total accretion luminosity \citep[since the accretion shock occurs at the
stellar surface and only half of it is visible at any moment, e.g.,][]{Hartmann2016}. All models were calculated at a
distance of 100\,pc. In each case, DIAD computed the fluxes corresponding to 100 wavelengths, distributed in
logarithmic space from 0.3\,$\mu$m to 3\,cm. This resulted in a good SED coverage across the entire wavelength
range of interest. We note that the disk mass ($M_{\rm disk}$) is not a free parameter in DIAD, since it is
set by the input parameters: $\alpha$ and $\dot{M}$ determine the surface density profile $\Sigma(r)$
\citep[$\Sigma \propto \alpha/\dot{M}$, see eq. 37 in][]{DAlessio1998}, and the total disk mass can then be
computed by integrating $\Sigma(r)$ from the inner radius \citep[$R_{\rm wall}$, determined by the dust
sublimation temperature and opacity following][]{Dalessio2005} to the outer radius ($R_{\rm disk}$). Our final
models include additional parameters such as distance and extinction, but these do not need to be considered in
DIAD since they can be accounted for later in the process. Section~\ref{sec:fitting} provides a description of
the complete model used.

The main advantage of the DIAD models is that the inputs are physical parameters, and the resulting disk
structure is physically consistent. On the other hand, the models include a significant number of free
parameters, some of which cannot be directly measured or are correlated with others. Observationally, many
disks have estimates for some parameters but lack measurements for others. In other cases, the same parameter
has been determined using different methods for different sources, which could introduce systematic
biases. These issues complicate the study of general trends using pre-existing detailed modeling of individual
objects, and a better approach for this problem would be to combine a Bayesian framework and homogeneous
modeling of a large sample of disks, which is the scope of this work. However, a typical SED calculation with
DIAD requires 1-2\,hours, and such a statistical study is not feasible with DIAD alone.

\subsection{The artificial neural network}\label{subsec:ann}

In order to solve the aforementioned problem, a massive improvement in the calculation time of the
models is needed. For this purpose, we have trained an ANN that, given the same input parameters, yields the same output
SED as DIAD but in significantly less time. Here we describe the ANN and the training process, while more
details are provided in Appendix~\ref{appendix:ANN}.

In its most common form, an ANN is a collection of nodes arranged in layers, each of them connected to the nodes
in the next layer. These connections have associated weights which, when set to appropriate values, can make
the ANN reproduce the expected behavior \citep[see][for an introduction to ANNs]{Bailer-Jones2002}. The input
parameters propagate through the layers (weighted with the corresponding values) to produce an output. In
our case, the ANN has 12 input nodes, corresponding to the 12 input parameters of the DIAD models. Likewise,
there are 100 output nodes, each of them corresponding to the flux at one of the 100 wavelengths.

The process of finding appropriate values for the weights is called training, and requires a training set: a
dataset for which both the input values (in our case, the 12 input parameters for DIAD) and the correct
outputs (the SED from DIAD, i.e., the fluxes at the requested wavelengths) are known. The training set is used
to determine weight values by feeding the input parameters into the ANN and comparing the resulting outputs
with the correct ones. The weights are then updated to minimize the difference with respect to the expected
output, and the process is repeated iteratively until the desired precision is reached. We used three
different datasets for this purpose: a training set on which the training is performed, a validation set that
is used to evaluate when convergence has been reached, and a blind set that we used to estimate the accuracy
of the ANN (how far its predictions are from the true DIAD models). The training set comprises 70000 models,
17000 models were used for validation purposes, and the blind set contains 5000 models. To maximize the amount
of information in the training set, these models were randomly distributed across the parameter space instead
of using a regular grid (see Appendix~\ref{appendix:training_sample} for details).

We used the feedforward multilayer perceptron regressor implemented in the scikit-learn Python package
\citep{scikit-learn} to create and train the ANN. There is no standard procedure for finding an appropriate
structure for an ANN, and it is usually selected through trial and error by testing different architectures
until the results reach the desired precision.  After different tests, we chose an architecture with two
hidden layers, each containing 250 nodes. A scheme of the architecture of the ANN is shown in the appendix
(Fig.~\ref{fig:ANN_architecture}). Details about the training process are provided in
Appendix~\ref{appendix:ANN_details}.

\begin{figure}
  \centering
  \includegraphics[width=\hsize]{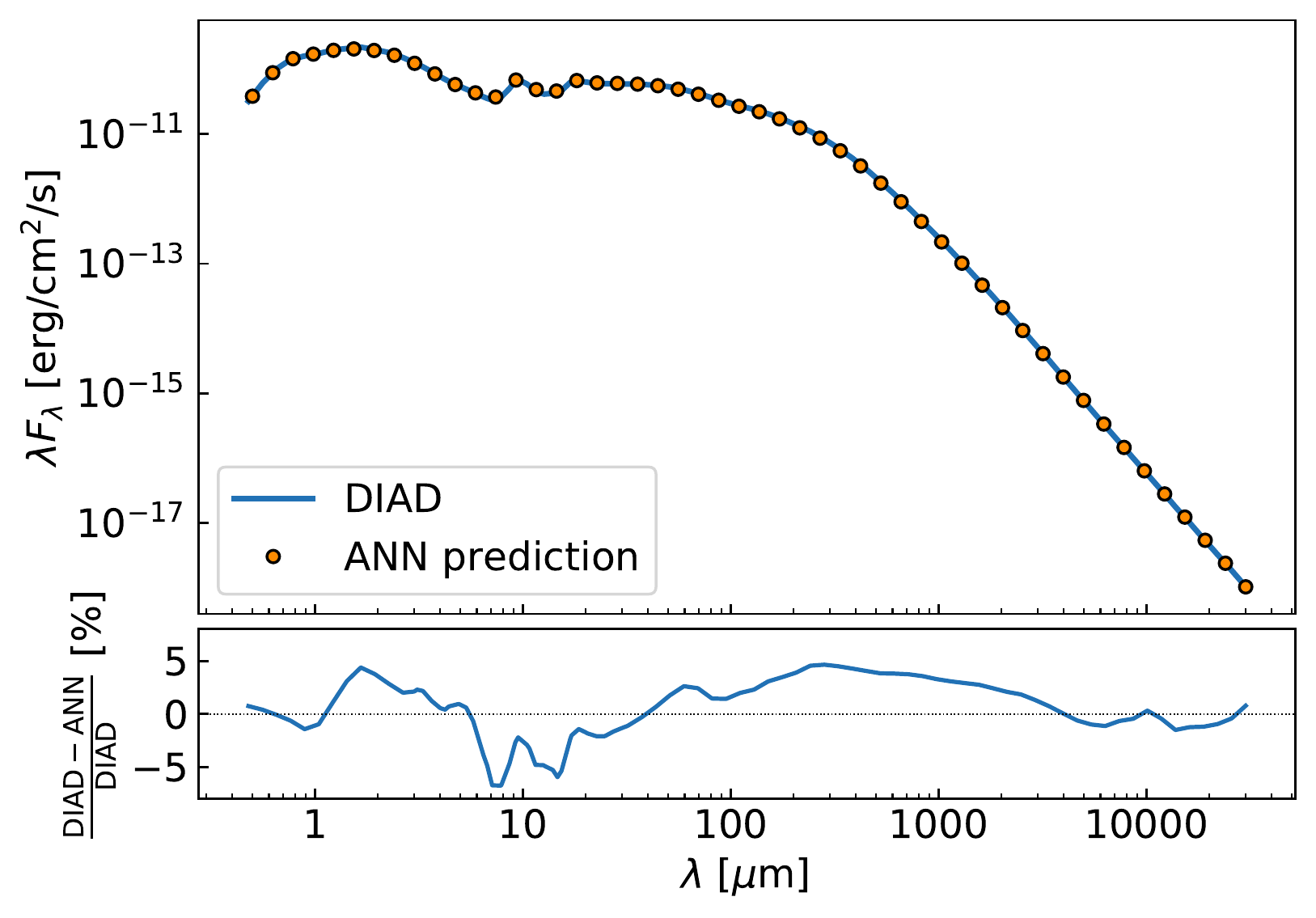}
  \caption{\textit{Top}: example of an SED calculated using the DIAD models (solid blue line), and the
    prediction from the artificial neural network (orange dots) for the same model
    parameters. \textit{Bottom}: difference (residual) between the DIAD model and the corresponding
    prediction. This SED belongs to the blind sample and was not used during the training of the ANN.}\label{fig:SED_ANN_example}
\end{figure}

The accuracy of the ANN (i.e., how far the predicted SEDs are from the DIAD ones) was estimated using the blind
sample of 5000 additional models, and we adopted a 10\,\% uncertainty for it (see Appendix~\ref{appendix:ANN}
for details). An example of an SED computed with DIAD and the corresponding prediction from the ANN is shown
in Fig.~\ref{fig:SED_ANN_example}.

The advantage of the ANN with respect to the standard DIAD models is the time required to evaluate SEDs, which
is a few milliseconds. Compared to the $\sim$\,1-2\,hours that DIAD requires, this represents an
improvement of $\sim10^6$ in computation time, allowing for detailed statistical analysis of large samples
of protoplanetary disks.

\subsection{A second ANN for disk masses}\label{sec:ann_diskmass}

The disk mass can be considered as an additional output in the DIAD models, since it is set by the input
parameters. The ANN described in Sect.~\ref{subsec:ann} cannot determine this quantity because it bypasses all
physical calculations and yields the SED directly. However, the models used for the training set in the
previous section have disk masses calculated by DIAD, and can thus be used to train a second ANN: one that outputs
disk masses as a function of input parameters. We will refer to this second ANN as ANN$_{\rm disk mass}$. The
value of the disk mass depends on all the DIAD parameters described in Sect.~\ref{sec:DIAD_models} except for
the inclination of the disk $i$ and the scaling of the wall $z_{\rm wall}$. Thus, it has 10 inputs and one
single output, the disk mass. Because ANN$_{\rm disk mass}$ only predicts one output (which also has a more
linear behavior than the SED), a precision of 5\,\% was achieved using a single layer with 250 nodes (see
Appendix~\ref{appendix:ANN} for details).

We note that ANN$_{\rm disk mass}$ is not used for fitting observations: it simply calculates the disk mass
corresponding to a given set of parameters. Once the posterior distributions for the model parameters are
derived from the observed SED, they are provided to ANN$_{\rm disk mass}$ to obtain the
posterior distribution of the disk mass as well.

\section{Sample}\label{sec:sample}

Once the ANN has been trained, it can be used to fit SEDs of protoplanetary disks. In this first
study, we focused on the population of protoplanetary disks in the Taurus-Auriga region \citep[see][for a
review on the region]{Kenyon2008_Taurus}. This association has been a common benchmark for star-formation
studies given its proximity \citep[140-160\,pc, e.g.,][]{GaiaDR2} and its young age \citep[$\sim$2\,Myr,
e.g.,][]{Kenyon1995,Andrews2013}. Numerous studies have identified a population of over 400 young stellar
objects (YSOs) in this region, many of them displaying infrared excess characteristic of disk-harboring
systems \citep[e.g.,][]{Luhman2004Taurus, Monin2010, Rebull2010, Rebull2011, Luhman2017}.

In this work, we modeled a sample of 23 objects selected from the 161 T~Tauri stars and brown dwarfs with
\emph{Spitzer}/IRS spectra studied in \citet{Furlan2011}. We used the dataset presented in
\citet{Ribas2017}, which merged the photometric compilation for sources in Taurus-Auriga by
\citet{Andrews2013} with ancillary photometry, far-IR photometry from the \emph{Herschel} Space observatory
\citep{Herschel}, and \emph{Spitzer}/IRS (mid-IR) and \emph{Herschel}/SPIRE (far-IR/submm) spectra when
available. After a number of validation steps in \citet{Ribas2017}, the sample was reduced to 154
sources. Additionally, here we included ALMA 1.3\,mm photometry from \citet{Long2019}.

Recently, \citet{Luhman2017} updated the census of Taurus members, including previous and new spectroscopic
measurements of their spectral types (SpTs). We used these SpTs to derive effective temperatures for all the
sources in our sample using the relation in \citet{Pecaut2013}. We considered sources with $T_*$ between
3000\,K and 6000\,K (corresponding to $\sim$M5 and F9-G0, respectively) to select T~Tauri stars, and adopted
an uncertainty of 100\,K. This temperature range decreases the sample size to 118
objects. We also used a spectral type for Haro 6-13 of K5.5 as reported in \citet{Herczeg2014}, since the
corresponding stellar mass is in better agreement with the dynamical mass estimate in \citet{Simon2019}.

Parallax measurements are available from the \emph{Gaia} DR2 catalog \citep{GaiaDR2} for most sources in the
sample, and were retrieved by cross-matching the 2MASS coordinates of these objects with the \emph{Gaia} catalog
using a 1\,\arcsec radius. Five objects had parallaxes compatible (within uncertainties) with distances smaller
than 100\,pc or larger than 200\,pc, and were discarded.

Several sources were also removed from the sample due to the following reasons:

\begin{enumerate}

\item Substructures are ubiquitous in protoplanetary disks \citep[e.g.,][]{Andrews2018_DSHARP, Long2018}, and
  their presence requires a more detailed treatment of their radial structure that was not included in our
  ANN. In particular, inner cavities have an important effect in the SED, and we only considered sources that
  are not known to have large inner cavities in them. To our knowledge, the list of such sources in our sample
  includes AA Tau \citep{Loomis2017}, CIDA~1 \citep{Pinilla2018b}, CIDA~9 \citep{Long2018}, CoKu~Tau/4
  \citep{Forrest2004, Dalessio2005}, DM~Tau \citep{Calvet2005, Andrews2011}, GG~Tau \citep{Guilloteau1999,
    Dutrey2014}, GM~Aur \citep{Macias2018}, IRAS~04125+2902 \citep{Furlan2011}, IP~Tau \citep{Espaillat2011},
  LkCa~15 \citep{Pietu2006, Espaillat2007b, Espaillat2008}, RY~Tau \citep{Isella2010}, UX~Tau~A
  \citep{Espaillat2007b}, and V410~X-ray~6 \citep{Furlan2011}. In addition, \citet{Furlan2011} also identified
  other (pre)transitional disk candidates based on their IRS spectra, such as MHO~3, IRAS~04370+2559, GK~Tau,
  JH~112~A, 2MASS~J04202606+2804089, and 2MASS~J04390525+2337450, and they were also excluded. The presence of
  rings in the outer regions of the disk does not significantly affect the SED, so we did not remove sources
  with this morphology \citep[e.g., CI~Tau, DS~Tau;][]{Clarke2018, Long2019}.

\item Stellar light from edge-on disks is highly extincted by the disk itself, and their central objects are
  usually poorly characterized. Models of edge-on disks are also very sensitive to small changes in
  inclination values, settling, and dust properties, which poses a significant challenge for accurately training
  the ANN. Therefore, we excluded disks with inclinations $i \geq 70$\,\degr. The edge-on disks in Taurus, as
  compiled in \citet{Furlan2011}, include seven sources: 2MASS J04202144+2813491, 2MASS J04333905+2227207,
  2MASS J04381486+2611399, Haro~6-5B, HH~30, IRAS~04260+2642, and ZZ~Tau~IRS. Additionally, HK~Tau~B is 
  known to be an edge-on disk \citep{McCabe2011}, \citet{Tripathi2017} found MHO~6 and 2MASS~J04334465-2615005
  to have inclinations above 70\,\degr, and \citet{Long2019} measured i$\sim$70\,\degr for HN~Tau~A. These sources
  were also excluded from our analysis.

\item As mentioned in Sect.~\ref{sec:DIAD_models}, DIAD computes the emission from the inner wall of the disk
  based on its temperature and solid angle, which is then re-scaled to account for the unknown true shape of
  the wall. For almost face-on disks, the solid angle subtended by the wall becomes very small, and a very
  large scaling factor is required. To avoid training the ANN with a very large range of wall heights, we
  discarded DR Tau, which has an inclination close to face-on \citep[$i \leq 5$\,\degr,][]{Long2019}.

  \item RW Aur, DO Tau, and SU Aur were also removed from the sample because they display long tails of dust and
  gas. This suggests past flyby encounters, which could have strongly altered the disk structures
  \citep[e.g.,][]{Rodriguez2018, Winter2018, Akiyama2019}. Their photometric measurements may also be affected by
  the surrounding material.
  
\item Binary stars and multiple systems can have a major impact in the properties, structure, and lifetime of
  protoplanetary disks \citep[e.g.,][]{Artymowicz1994, Cieza2009, Kraus2012}.  Hydrodynamical simulations and
  detailed information about the system components and orbits are required to take these effects into
  consideration, and thus we excluded them from our analysis. We cross-matched the list of 133 Taurus sources
  with companions closer than $<$\,300\,au in \citet{Kraus2012} and discarded those that overlapped with our
  sample.  Additionally, we removed the known close binaries DQ~Tau \citep{Mathieu1997}, UZ~Tau~E
  \citep{Prato2002}, DK~Tau and IT~Tau \citep[since both components host disks and are closer than 3.5'',
  e.g.,][]{Akeson2014}, and the triple system T~Tau \citep{Dyck1982, Koresko2000}.

\item Given the important degeneracies in SED modeling and the number of free parameters involved, we required
  that the objects have at least one photometric measurement at wavelengths $\geq$\,500\,$\mu$m. This
  requirement ensures that all these objects have coverage up to the submm regime.

\end{enumerate}

We note that some sources meet more than one of these criteria (e.g., binaries can carve gaps in their disks). Thirty-two objects
remained after this process. After attempting to fit their SEDs, another 9 objects are discarded (see Sect.~\ref{sec:fits}),
resulting in a final sample of 23 sources with well-sampled SEDs from the optical to (sub)mm wavelengths and
successful fits.

\section{Fitting process} \label{sec:fitting}

\subsection{The complete model}\label{sec:model}

\begin{figure*}
  \centering
  \includegraphics[width=0.95\hsize]{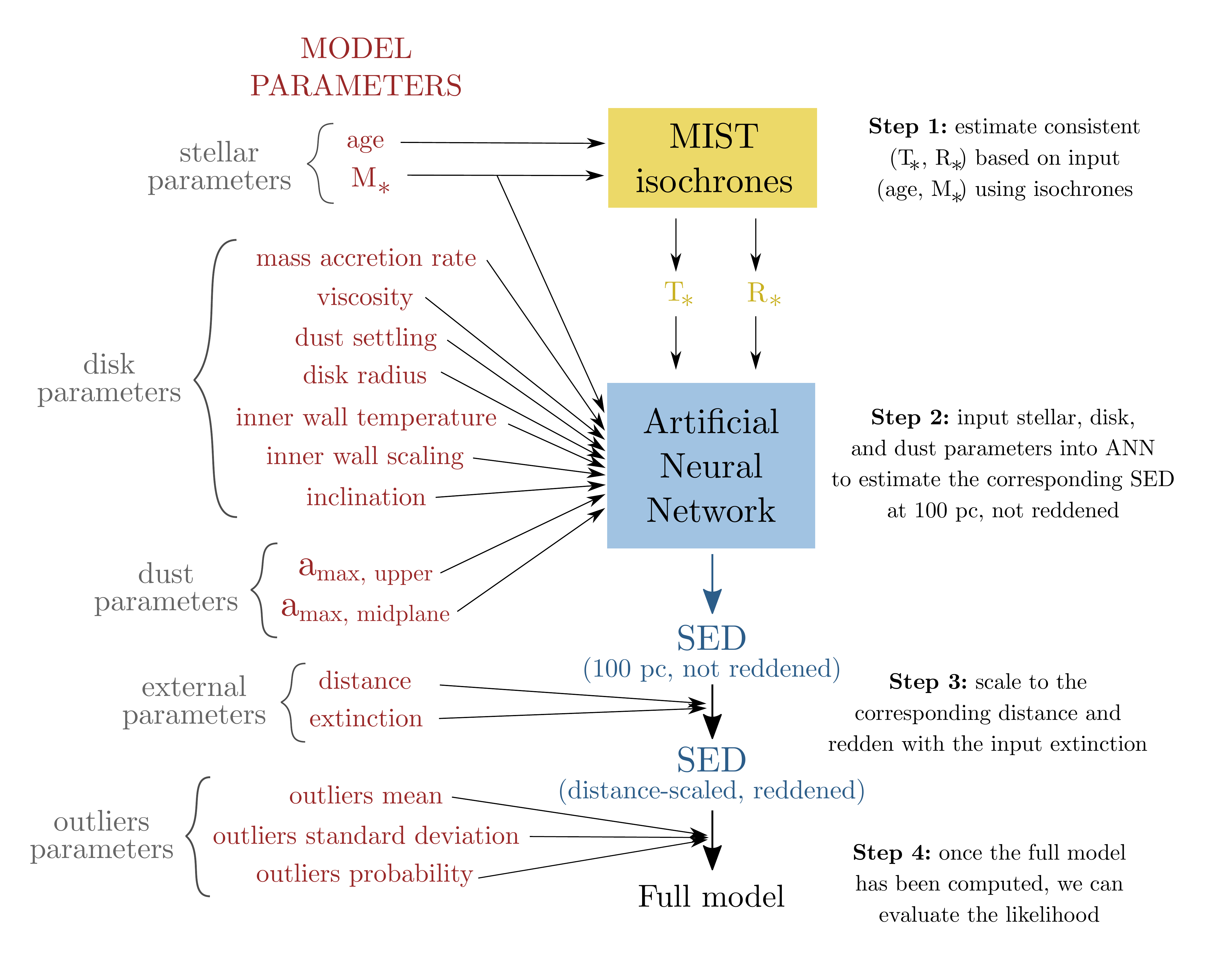}
  \caption{Scheme of the complete model used in this study. The input parameters (16 in total) are shown to
    the left (red). The middle column shows the flow of the model calculation, and each step in the
    calculation is explained on the right. We note that the stellar mass $M_*$ is used both in the MIST
    isochrones and in the ANN calculations. The final SED obtained through this process, together with tree
    additional parameters that account for potential outlier photometry, is used to estimate the model
    likelihood (see Sect.~\ref{sec:bayesian}).}\label{fig:model_scheme}
\end{figure*}

DIAD (and thus the ANN) provides SED estimates based on input parameters, but some additional parameters such
as interstellar extinction ($A_V$) and distance ($d$) are also needed to model observations. Additionally,
although in principle one could use different combinations of the stellar parameters ($M_*$, $T_*$, and $R_*$) in
DIAD, most of these combinations are not consistent with stellar evolution. Thus, every model computed in
  our fitting procedure explored different values of stellar and disk parameters, $A_V$, and distance following
  these steps:

\begin{enumerate}

\item The two fundamental parameters that determine stellar properties in our modeling procedure are the
  age and mass of the star ($M_*$). Therefore, we adopted these two as input free parameters in our
  model. For any given age-$M_*$ pair, the first step was to use the MESA Isochrones and Stellar Tracks
  \citep[MIST,][]{Paxton2011, Paxton2013, Paxton2015, Dotter2016, Choi2016} to obtain the corresponding $T_*$
  and $R_*$ values. We note that $T_*$ and $R_*$ are determined by the age and $M_*$ and thus are not free
  parameters in the model. Although age is not a free parameter in DIAD, we used this approach to ensure that
  the combinations of $M_*$, $T_*$, and $R_*$ explored during the fitting process are consistent with
  evolutionary tracks and that its effect in the uncertainties of other parameters is accounted for in our
  results.

\item This set of consistent stellar parameters is fed to the ANN together with the remaining disk and
  dust parameters: the mass accretion rate ($\dot{M}$), the disk viscosity ($\alpha$), dust settling
  ($\epsilon$), the maximum grain sizes in the disk atmosphere and midplane ($a_{\rm max, upper}$ and
  $a_{\rm max,midplane}$), the disk size ($R_{\rm disk}$), the inclination ($i$), the dust sublimation temperature
  ($T_{\rm wall}$) and the scaling factor of the inner wall ($z_{\rm wall}$). The ANN then yields the
  corresponding SED at 100\,pc, and without any interstellar extinction.

\item The SED output from the ANN is then scaled to the corresponding distance ($d$), and reddened by $A_V$
  using the extinction law by \citet{McClure2009} following \citet{Ribas2017}. $A_V$ is the interstellar
  extinction, and does not include the self-extinction produced by the disk itself (which is already accounted
  for in DIAD when needed).

\end{enumerate}

With this setup, the model has 13 free parameters: the age and mass of the central source (which determine the
remaining stellar parameters through the MIST isochrones), the remaining DIAD parameters, and the distance and
interstellar extinction to the source. Additionally, three more free parameters were included to account for
possible outlier fluxes (see Sect.~\ref{sec:likelihoods}). Therefore, there are a total of 16 free
  parameters involved in the modeling process, all of which are varied during the fitting to compute posterior
  probability distributions. A scheme of the complete model is shown in
Fig.~\ref{fig:model_scheme}.

\subsection{The Bayesian framework}\label{sec:bayesian}

Our model has a significant number of free parameters, some of which are unconstrained by observations,
degenerate, or may have multimodal posteriors. For these reasons, we adopted a Bayesian framework
in our analysis, which allows us to incorporate pre-existing information about some of the model parameters,
as well as to naturally account for unconstrained/uncertain parameters in the resulting posterior
distributions.

\subsubsection{Photometric and spectroscopic uncertainties}\label{sec:uncertainties}

Uncertainties of the observed data are crucial in model fitting, since underestimated uncertainties can give
excessive weight to some data points and bias inference results. In our case, the photometric data were mostly
gathered from the comprehensive compilation by \citet{Andrews2013}, which used tens of studies to create SEDs
for YSOs in the Taurus-Auriga region. While this yielded detailed SEDs for most objects, the heterogeneous
origins of the data imply that their uncertainties were also estimated in different ways and, in some cases,
could be missing systematic uncertainties. Therefore, we set the minimum relative uncertainty to 10\,\% for
all the photometry to account for the heterogeneity of the data.

We also face a problem when fitting both photometry and spectra together: Because spectra have a much larger
number of data points, the fitting process is dominated by these, even though they may only cover a fraction
of the whole wavelength range probed by the SEDs. While there is no standard procedure to take this into
account, we tried to mitigate this effect by weighting the spectra so that they correspond to a certain
(smaller) effective number of points $N$. This is achieved by increasing their uncertainties by a factor of
$\sqrt{N_{\rm spectrum}/N}$, where $N_{\rm spectrum}$ is the number of points in the corresponding
spectrum. After various tests, we found $N$=10 and $N$=5 to be good compromises for the weights of IRS and
SPIRE spectra, respectively (although the SPIRE spectra are featureless, the IRS ones contain the silicate
features at 10 and 20\,$\mu$m, informative of the properties of small grains in the disk atmosphere). We note
that these numbers do not have a physical justification and were chosen heuristically.

\subsubsection{Priors}\label{sec:priors}

Bayesian analysis requires the use of priors, which encompass pre-existing knowledge of the parameters (both
from physical arguments or previous measurements). For most parameters, we chose either flat or Jeffreys priors
(which we used for parameters that could change by orders of magnitude to give equal probabilities to each
decade interval) within reasonable ranges. Our tests also showed that using uniform priors instead of Jeffreys
did not significantly alter the results. For the inclination, we used a Gaussian distribution when an
observational measurement exists, otherwise the observational prior was used (i.e., uniform in $\cos(i)$; see
Appendix~\ref{appendix:priors_incl}). For the distance, an observational prior (i.e., proportional to $d^2$)
between 100 and 200\,pc was used (an additional term is included in the likelihood for sources with available
parallax information from the \emph{Gaia} DR2, see next section and Appendix.~\ref{appendix:likelihoods}). We
did not include prior information about disk sizes from resolved observations given their strong dependence on
the wavelength due to dust radial migration \citep[e.g.,][]{Tazzari2016}, which is not included in our
models. A complete list of the priors used as well as the references for inclination measurements (when
available) are provided in Appendix~\ref{appendix:priors}.

\begin{figure*}[h]
\centering
  \includegraphics[width=0.33\hsize]{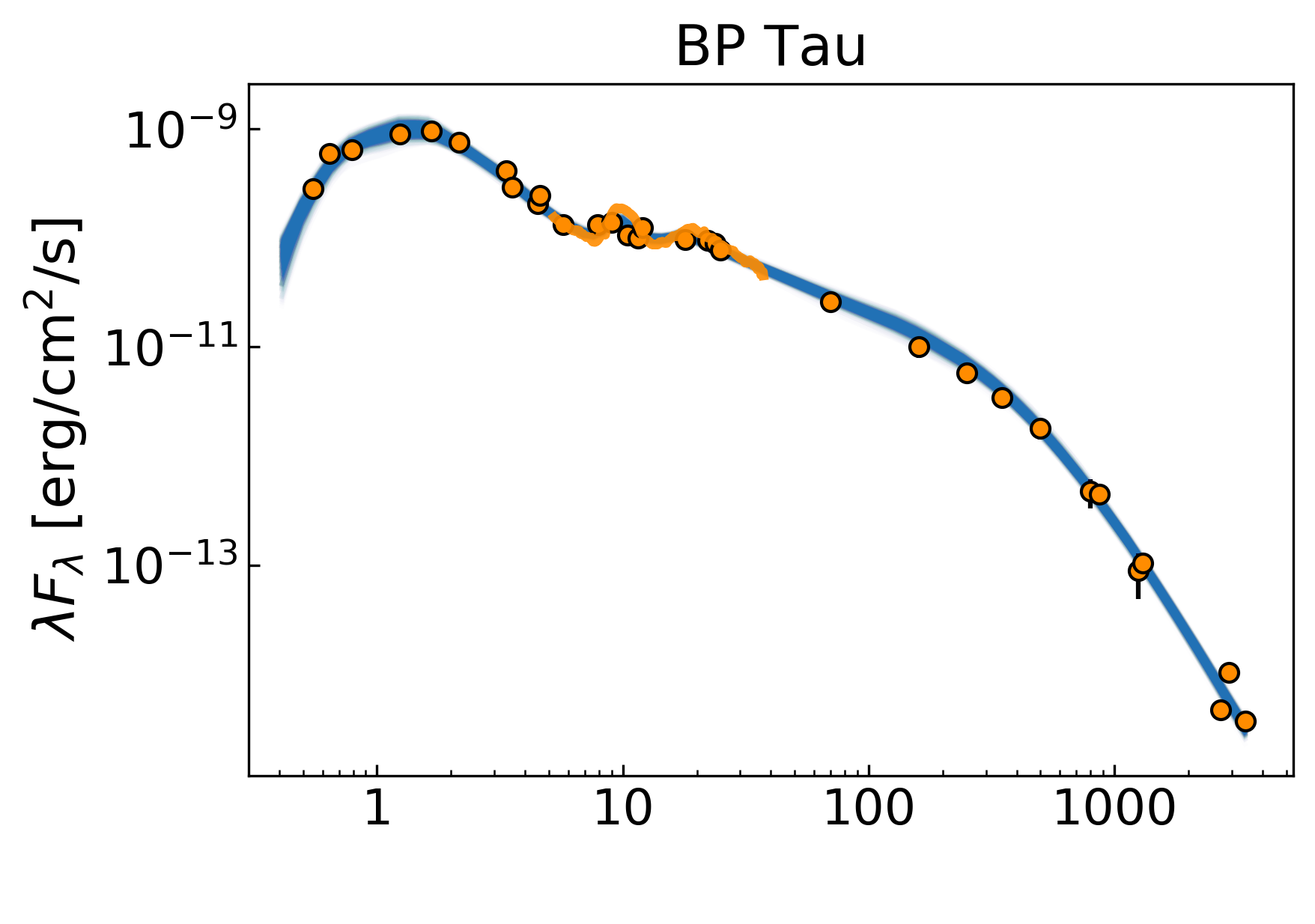}\hfill
  \includegraphics[width=0.33\hsize]{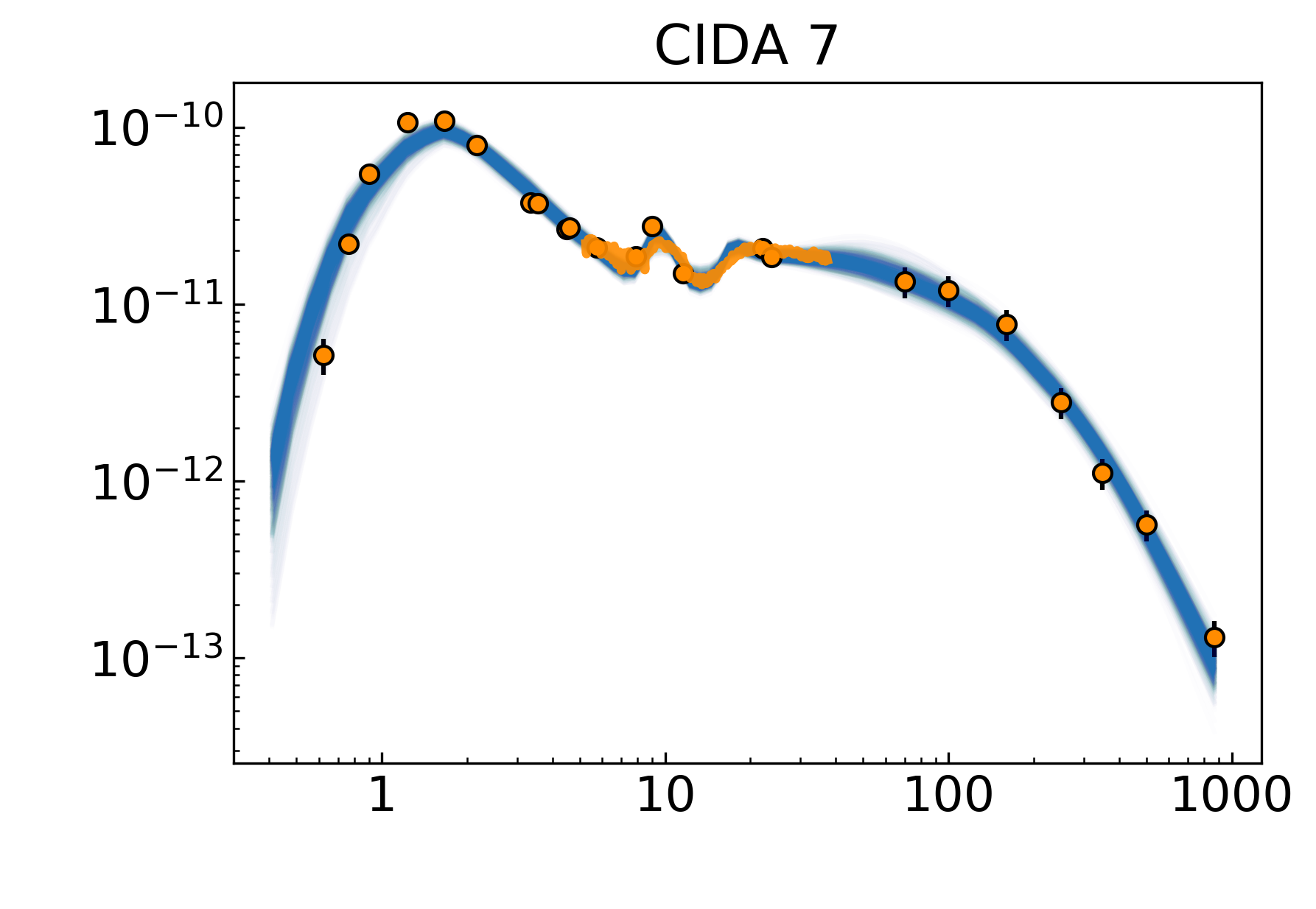}\hfill
  \includegraphics[width=0.33\hsize]{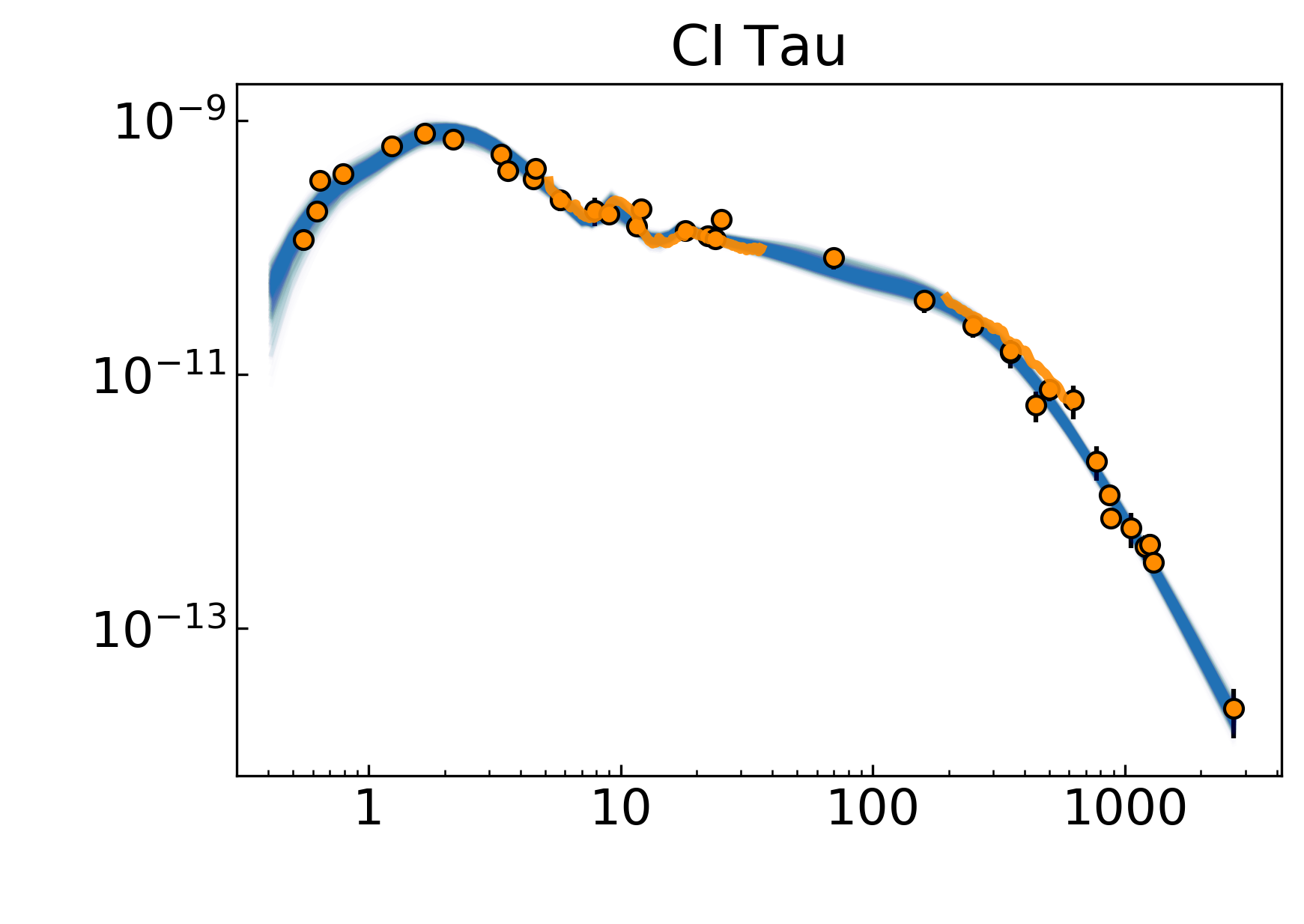}\\
  \includegraphics[width=0.33\hsize]{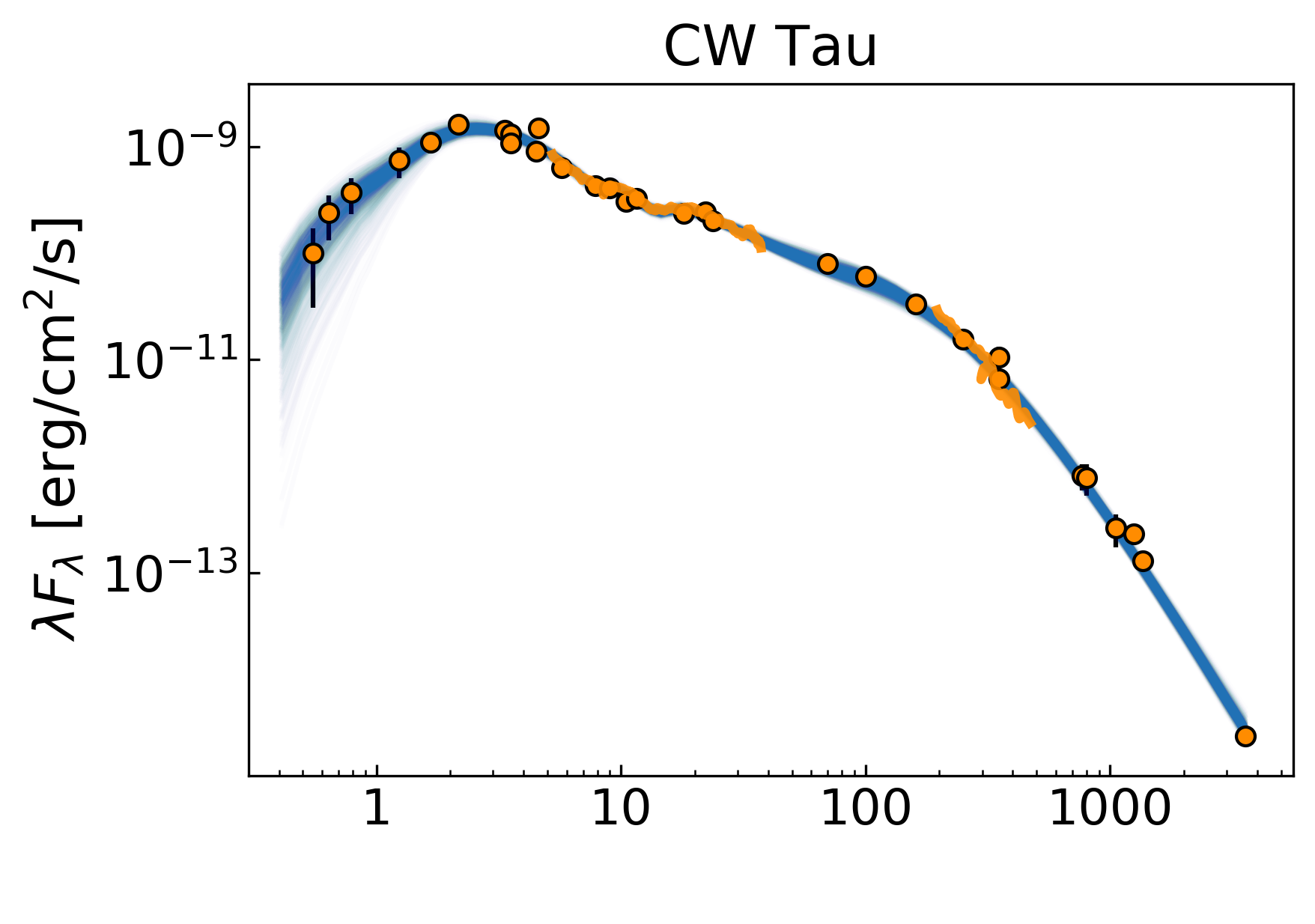}\hfill
  \includegraphics[width=0.33\hsize]{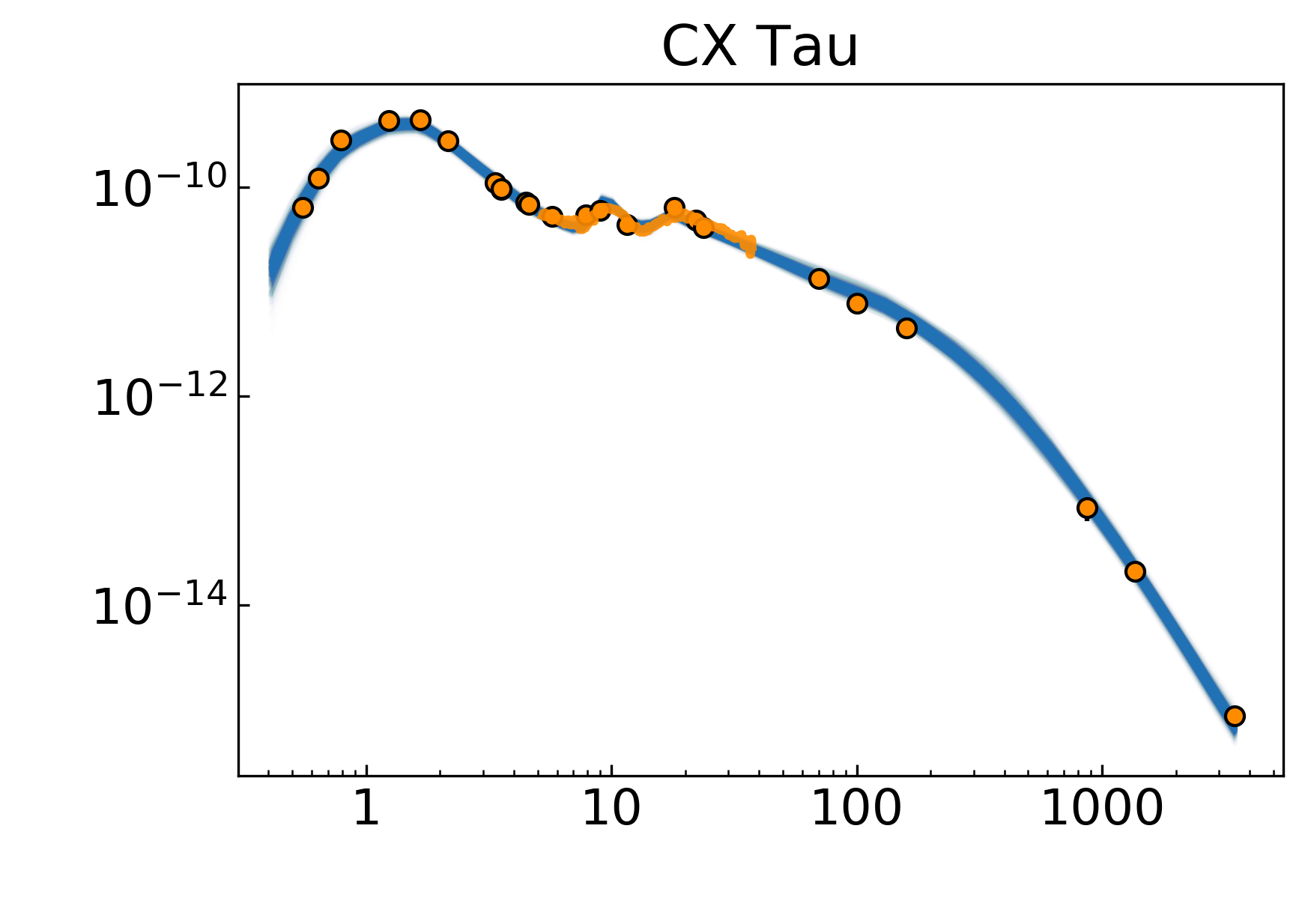}\hfill
  \includegraphics[width=0.33\hsize]{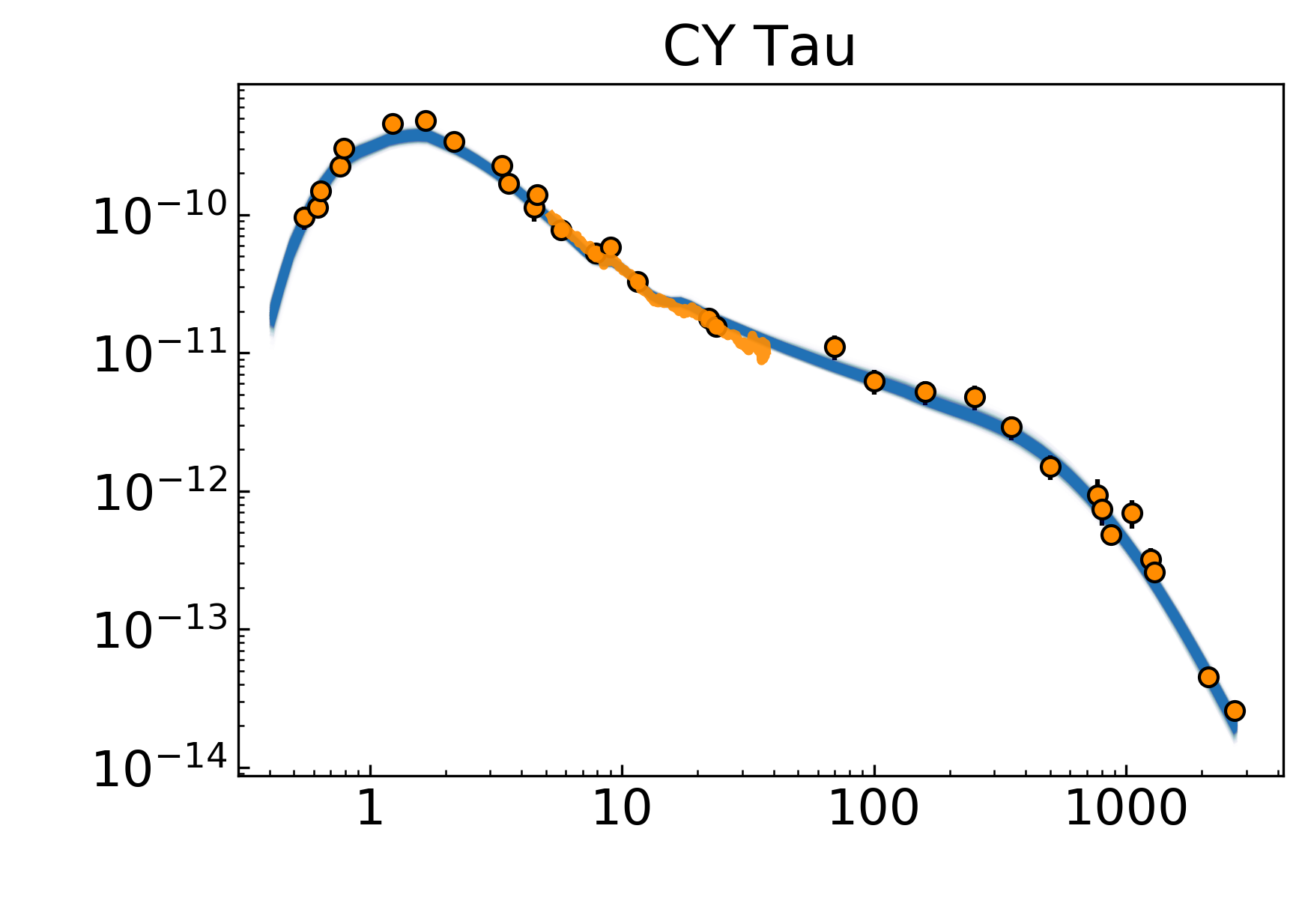}\\
  \includegraphics[width=0.33\hsize]{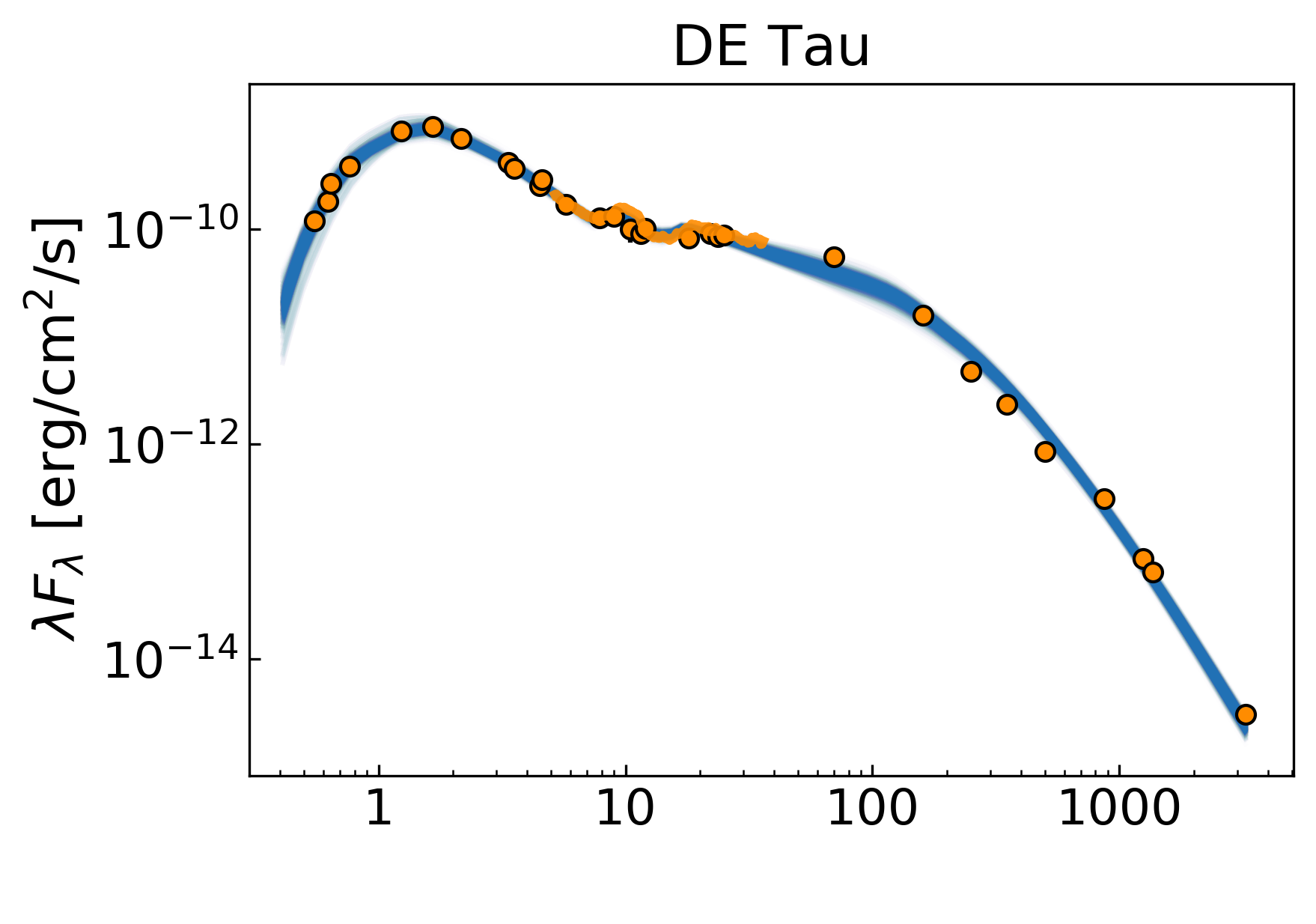}\hfill
  \includegraphics[width=0.33\hsize]{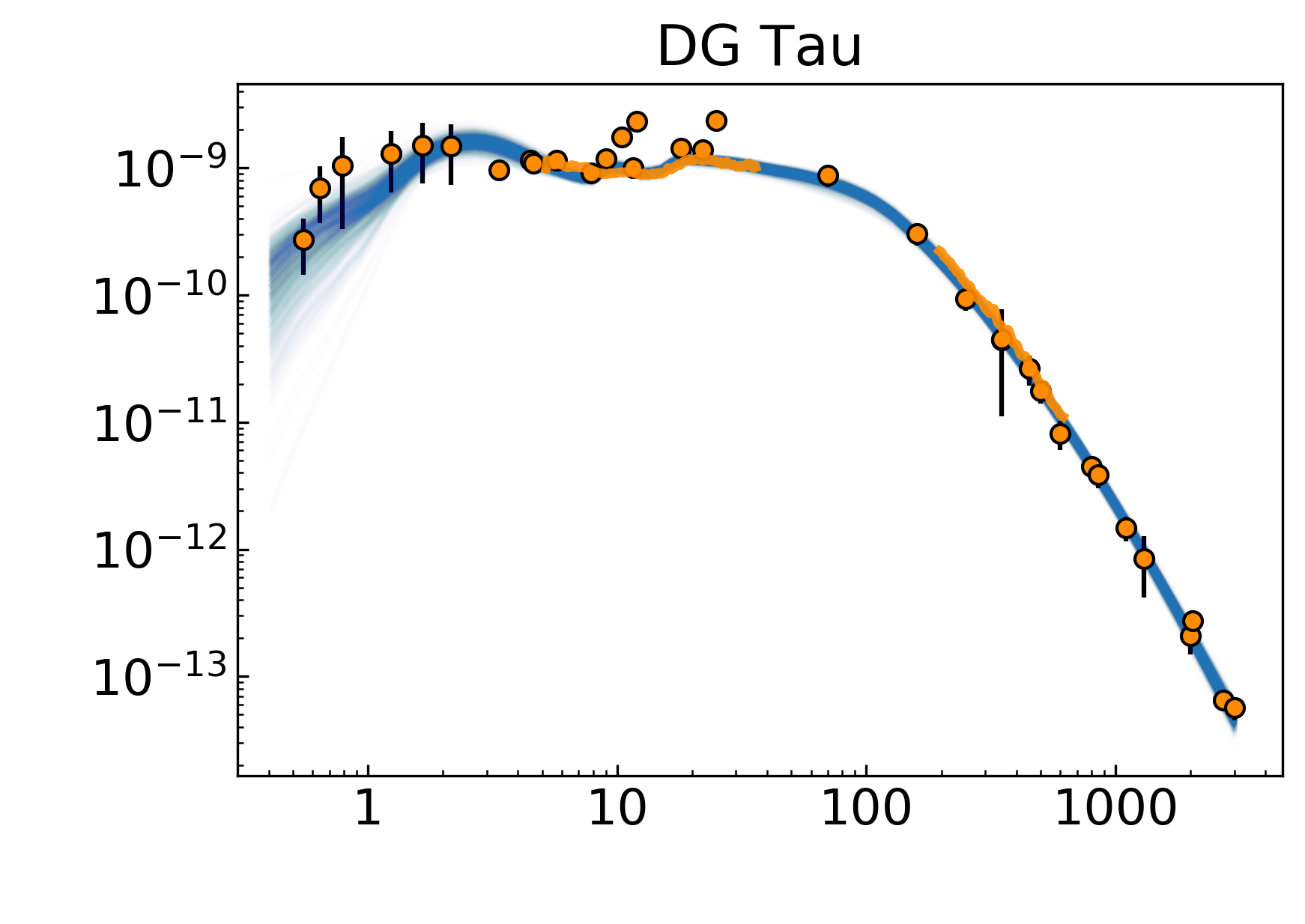}\hfill
  \includegraphics[width=0.33\hsize]{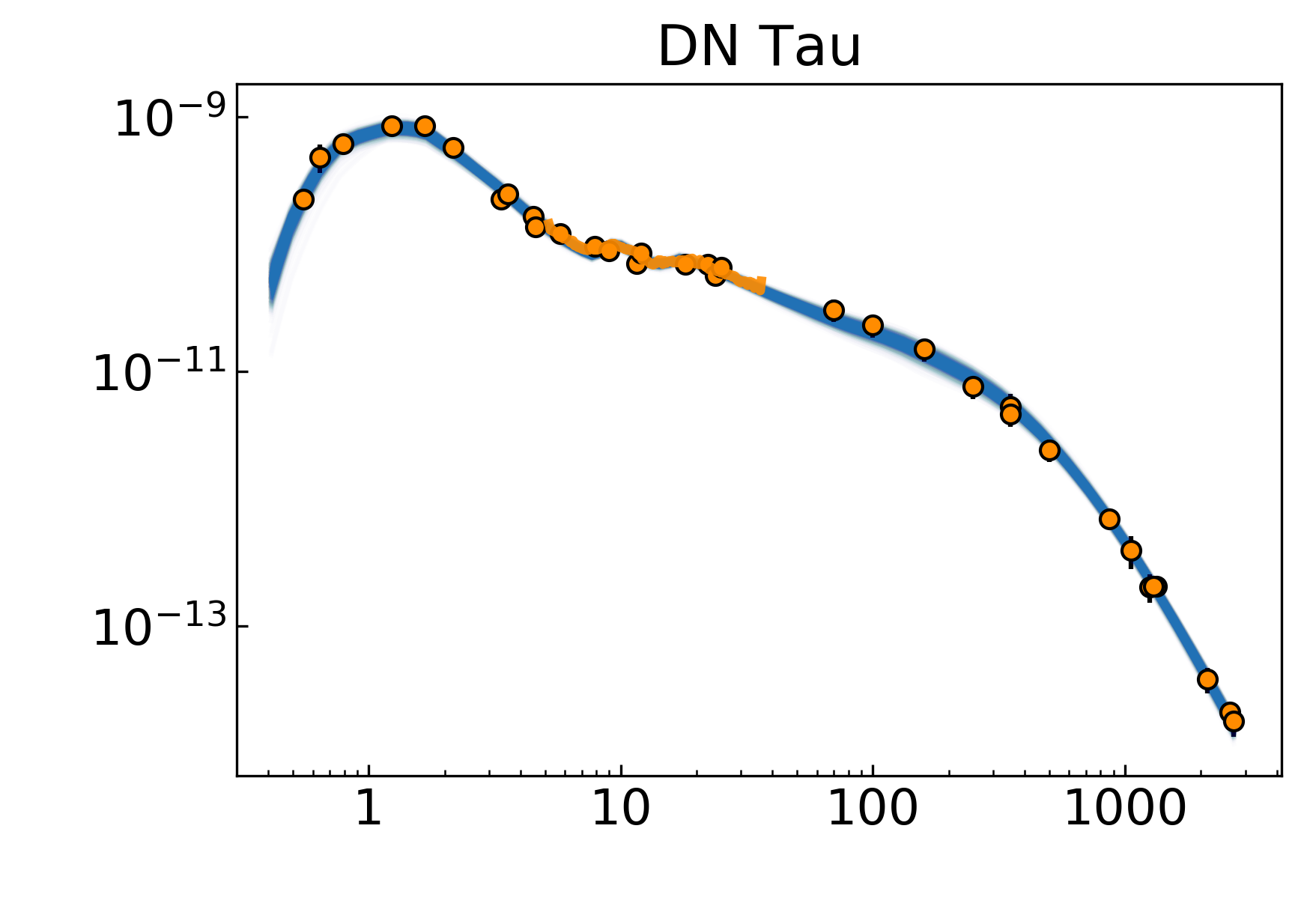}\\
  \includegraphics[width=0.33\hsize]{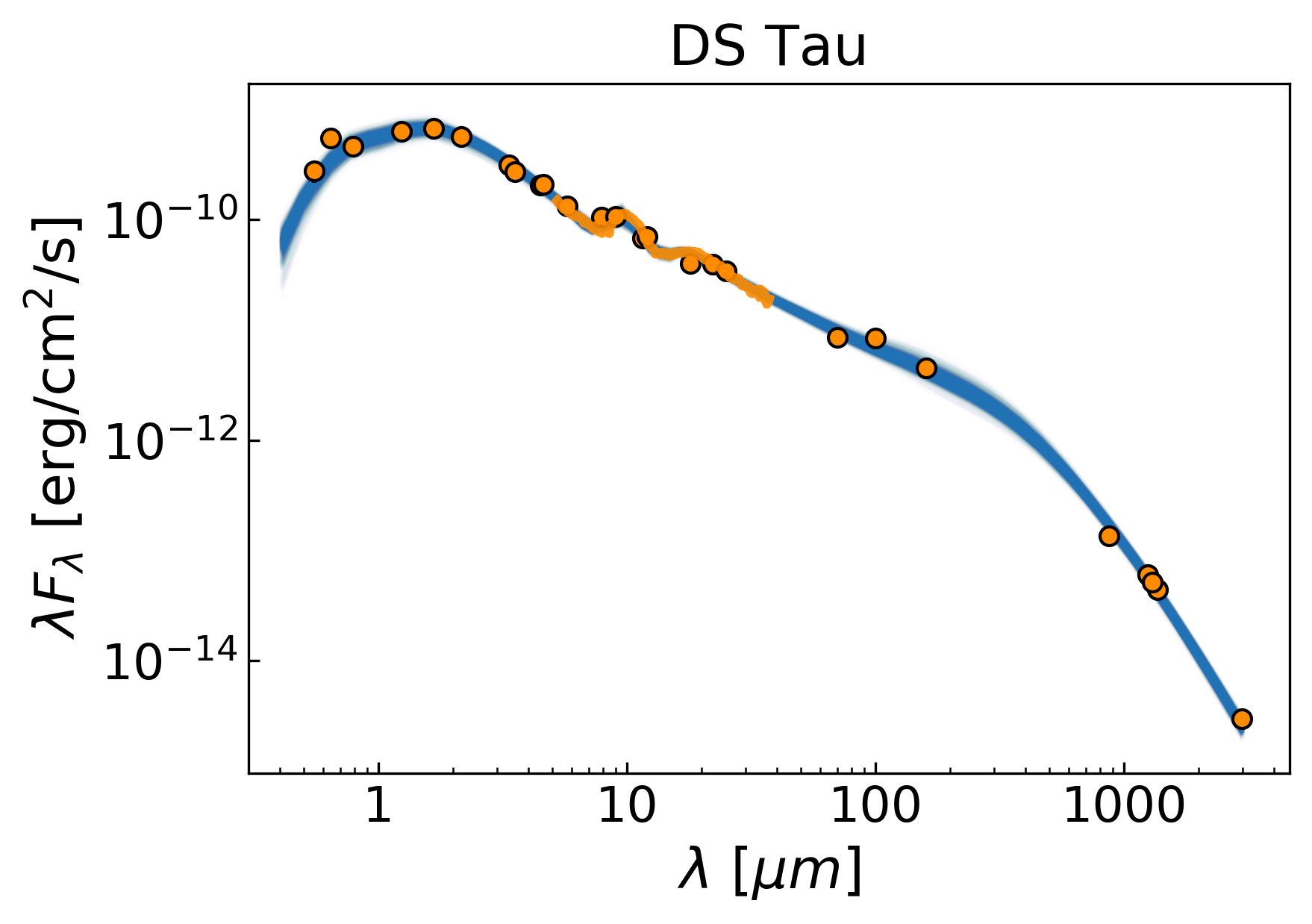}\hfill
  \includegraphics[width=0.33\hsize]{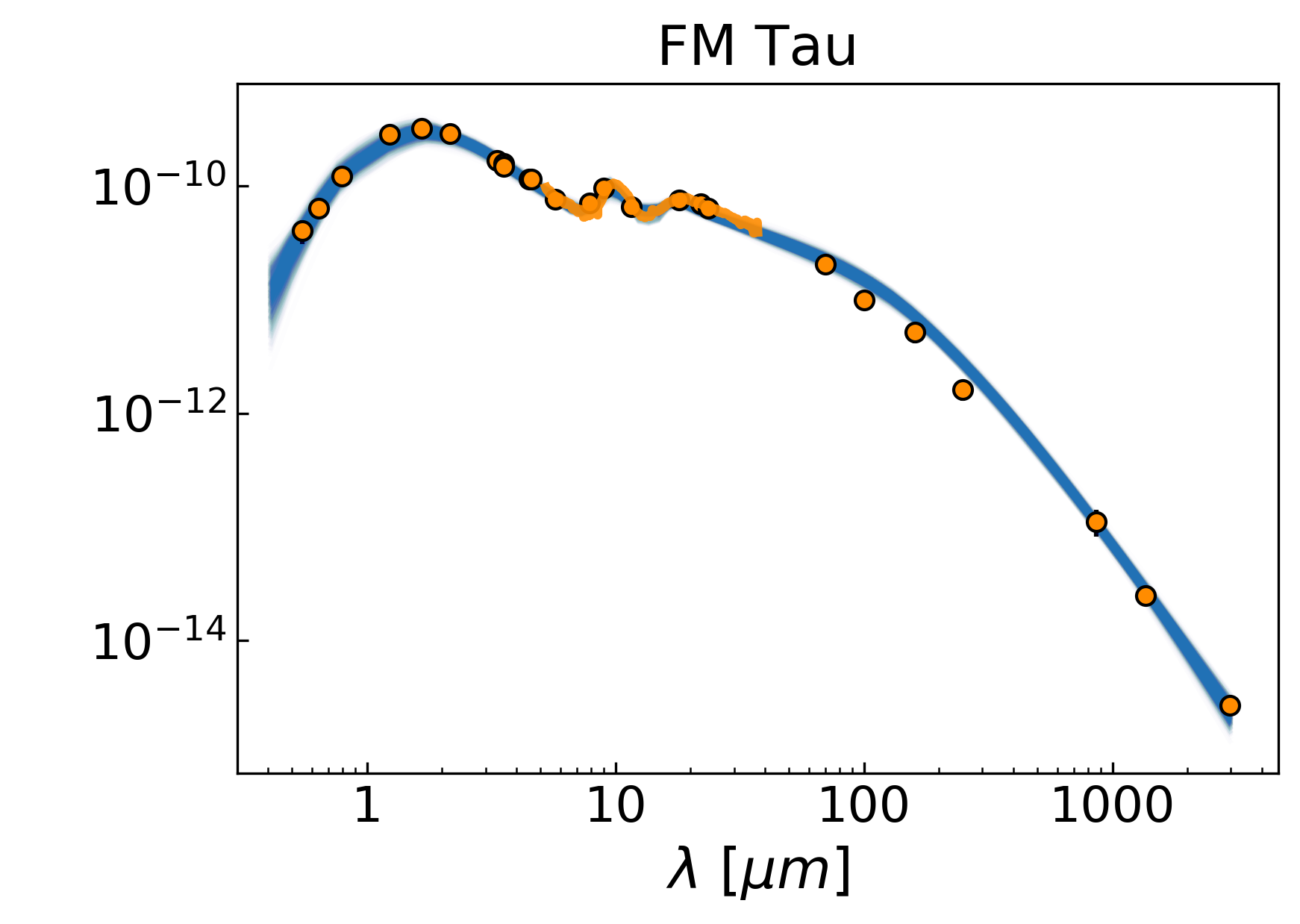}\hfill
  \includegraphics[width=0.33\hsize]{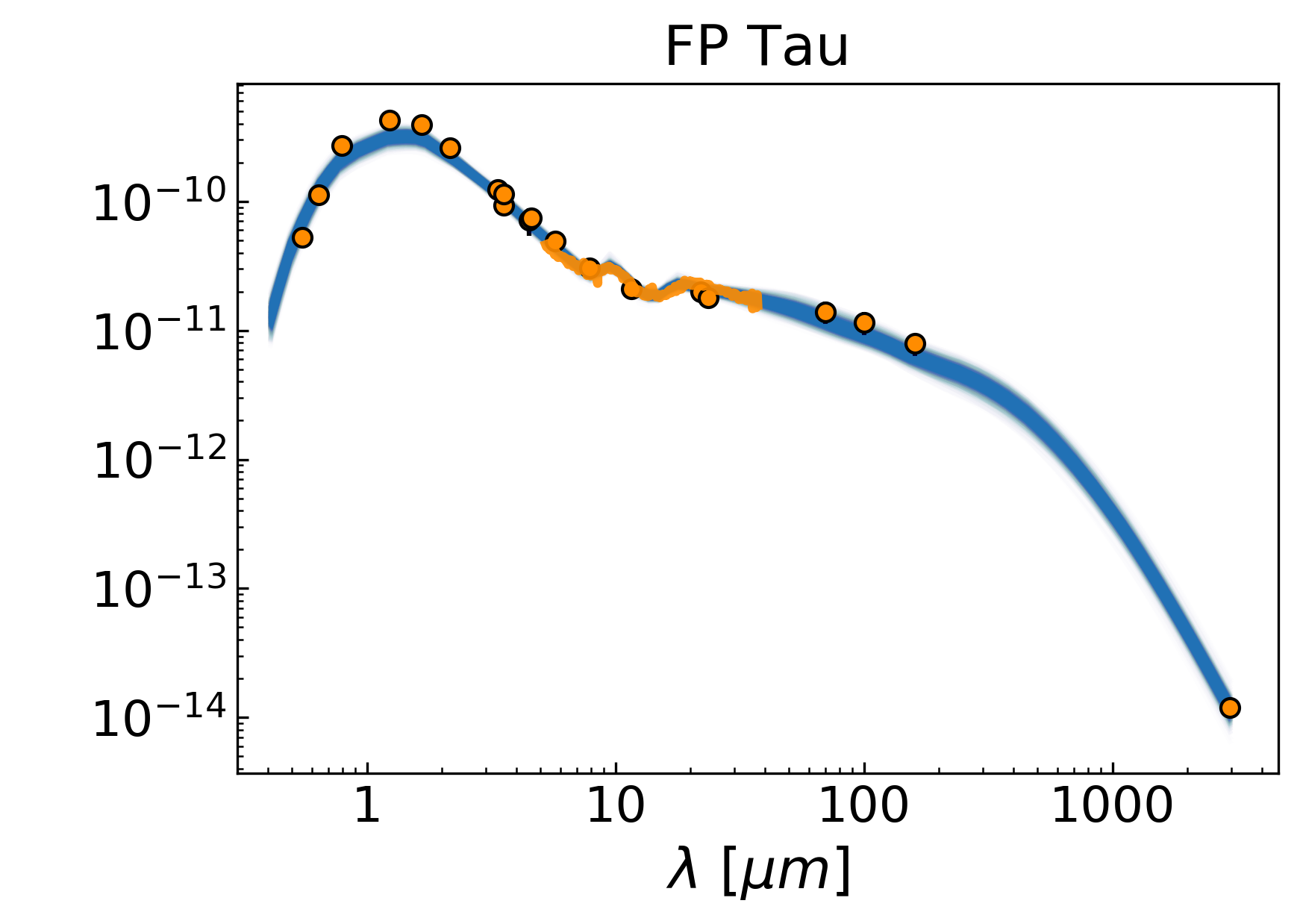}\\  
  \caption{SEDs of the 23 sources successfully fit in this study. The observed photometric data (orange dots)
    and the IRS/\emph{Spitzer} and SPIRE/\emph{Herschel} spectra (orange line, when available) are shown. We
    also plot the results of the modeling process by randomly selecting 1000 models (blue lines) from the
    posterior distributions.}\label{fig:successful_fits}
\end{figure*}

\subsubsection{Likelihood functions}\label{sec:likelihoods}

For each object, up to four different datasets can exist in our sample, namely: photometry, spectroscopy (from
IRS/\emph{Spitzer}, SPIRE/\emph{Herschel}, or both), a SpT measurement, and a parallax value from the
\emph{Gaia} DR2. Therefore, our likelihood function contains up to four different terms, and can be expressed
as:

\begin{equation}
\mathcal{L} = \mathcal{L}_{\rm phot} \, \mathcal{L}_{\rm spect}\, \mathcal{L}_{\rm T_*} \mathcal{L}_{\rm parallax}.
\end{equation}

These likelihood functions follow the standard case for Gaussian uncertainties. Additionally, the
photometric likelihood incorporates a mixture model to account for potential outlier photometric points.  The
mathematical description for each of these is provided in Appendix~\ref{appendix:likelihoods}. For objects
with no parallax measurement available, the parallax term $\mathcal{L}_{\rm parallax}$ is not included in the
likelihood.

Only photometric data between 0.5\,$\mu$m and 5\,mm were used for the fitting process: shorter wavelengths may
include significant emission from accretion shocks, which is also highly variable
\citep[e.g.,][]{Ingleby2013, Robinson2019}, while longer ones could be affected by free-free emission from
photoevaporative winds or gyrosyncroton emission from stellar activity \citep[][]{Pascucci2012, MacGregor2015,
  Macias2016}.

\subsection{Estimation of posterior distributions}\label{sec:posteriors}

Once the prior and likelihood functions were defined, we evaluated the posterior distribution of each object
using the \texttt{emcee} \citep{emcee} Python implementation of the Goodman \& Weare's Affine Invariant Markov
Chain Monte Carlo ensemble sampler \citep{Goodman2010}. We used 50\,walkers in each case, and their initial
positions were randomly distributed within the prior ranges. To avoid individual walkers getting stuck in
local minima, we first ran 25000 steps. We then used the median and standard deviation of the walker positions
to define a second set of locations from where we ran 75000 more steps, adding up to a total of 100000
steps. The autocorrelation times of the chains were then computed, and we checked that they were smaller than
a tenth of the chains length. If that was not the case, the chain was evolved for another 10000 steps, and the
process was repeated until the criterion was met. Finally, another 5000 steps were run to estimate the
posterior distributions. The masses corresponding to these final 5000 steps were also computed using
ANN$_{\rm disk mass}$ as described in Sect.~\ref{sec:ann_diskmass} to create the probability distributions for
this quantity.

\addtocounter{figure}{-1}

\begin{figure*}[h]
  \centering
  \includegraphics[width=0.33\hsize]{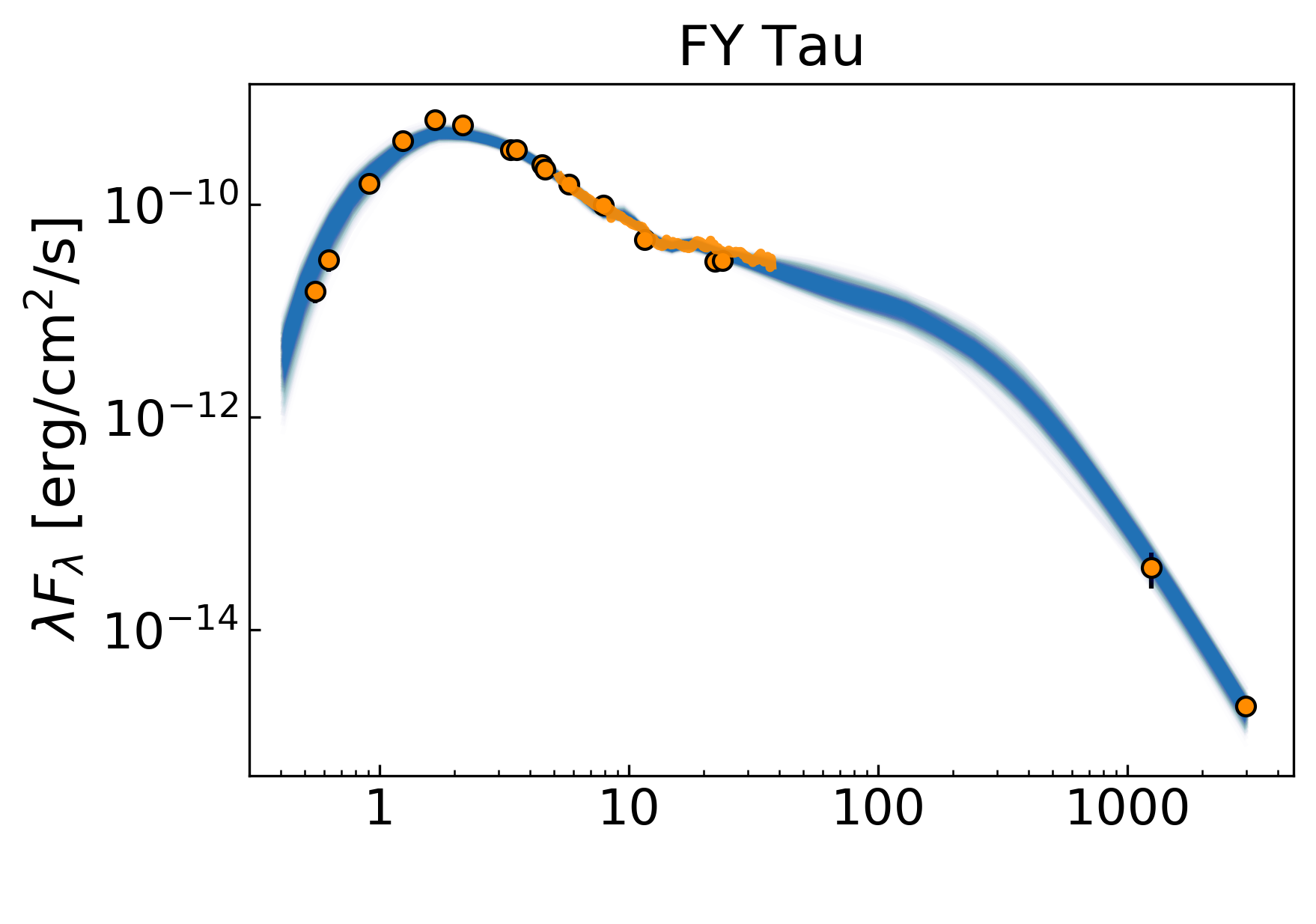}\hfill
  \includegraphics[width=0.33\hsize]{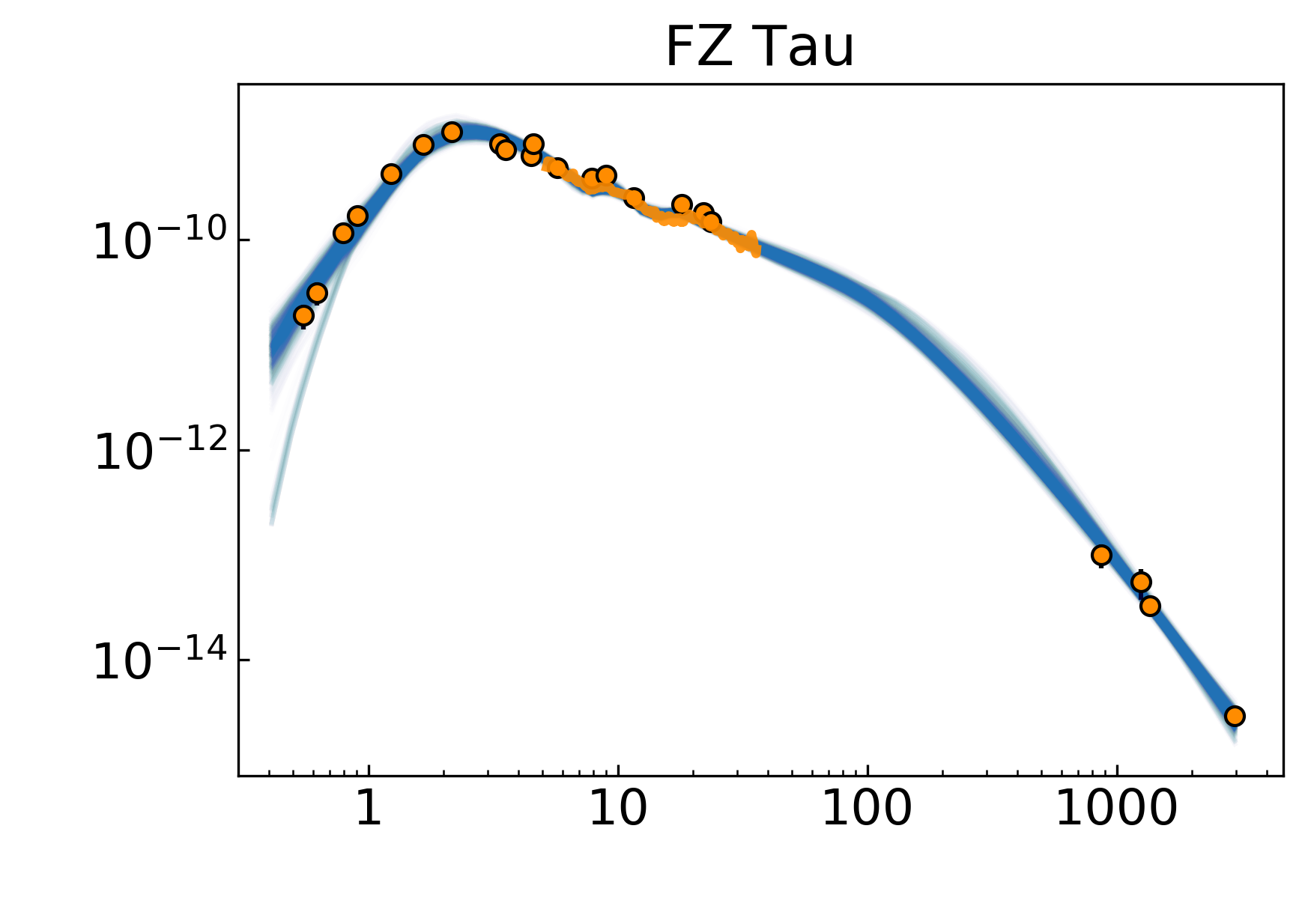}\hfill
  \includegraphics[width=0.33\hsize]{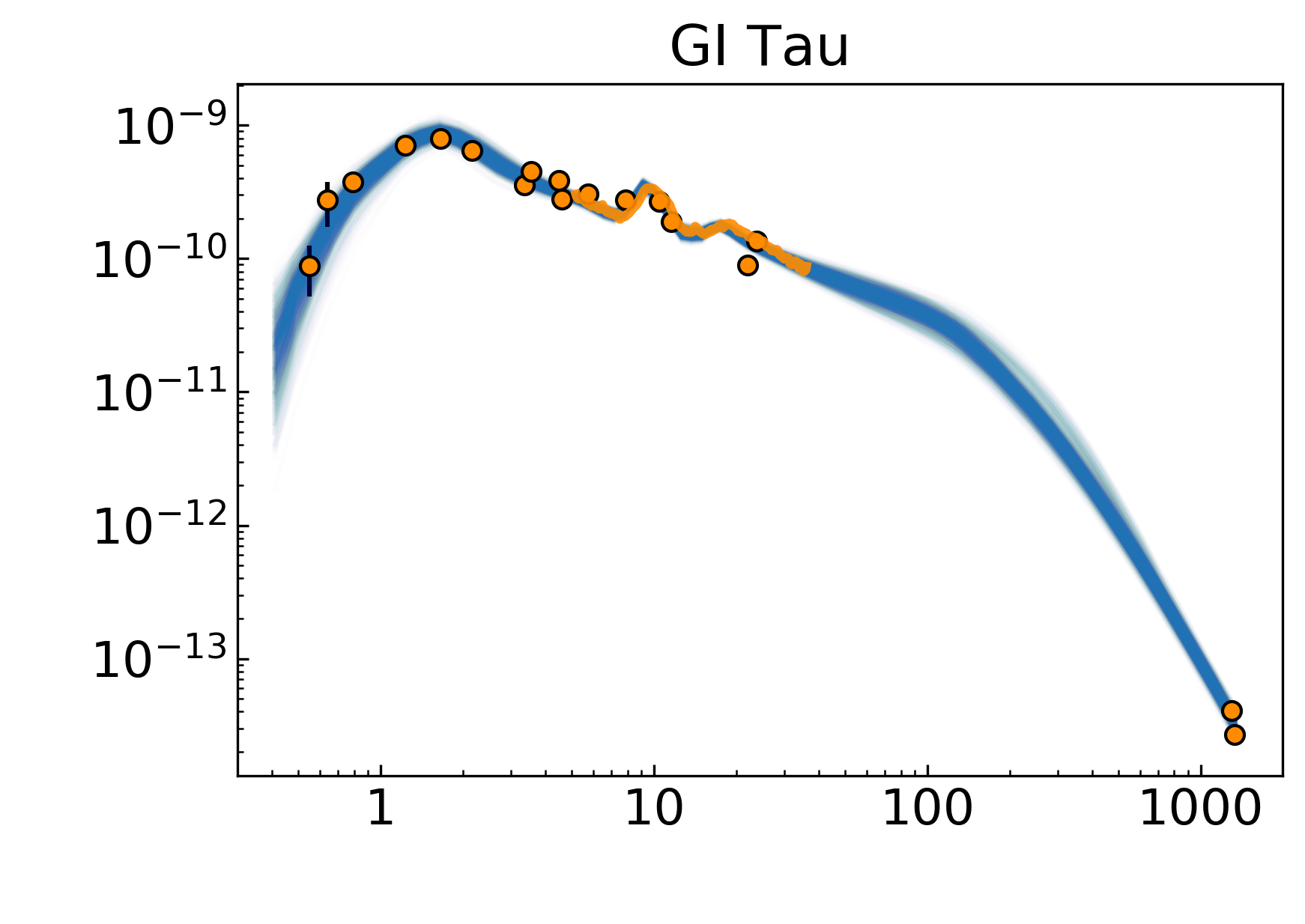}\\
  \includegraphics[width=0.33\hsize]{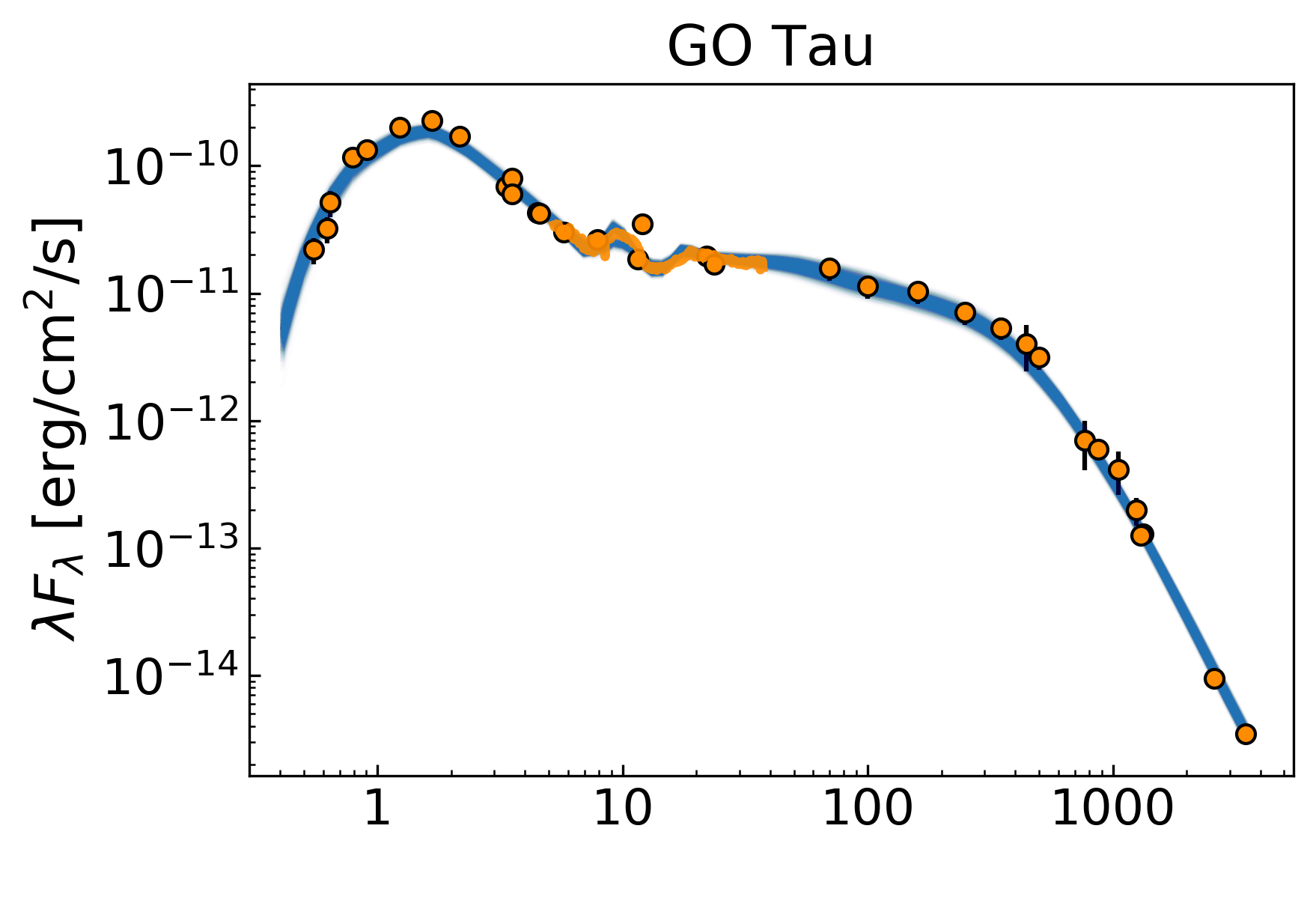}\hfill
  \includegraphics[width=0.33\hsize]{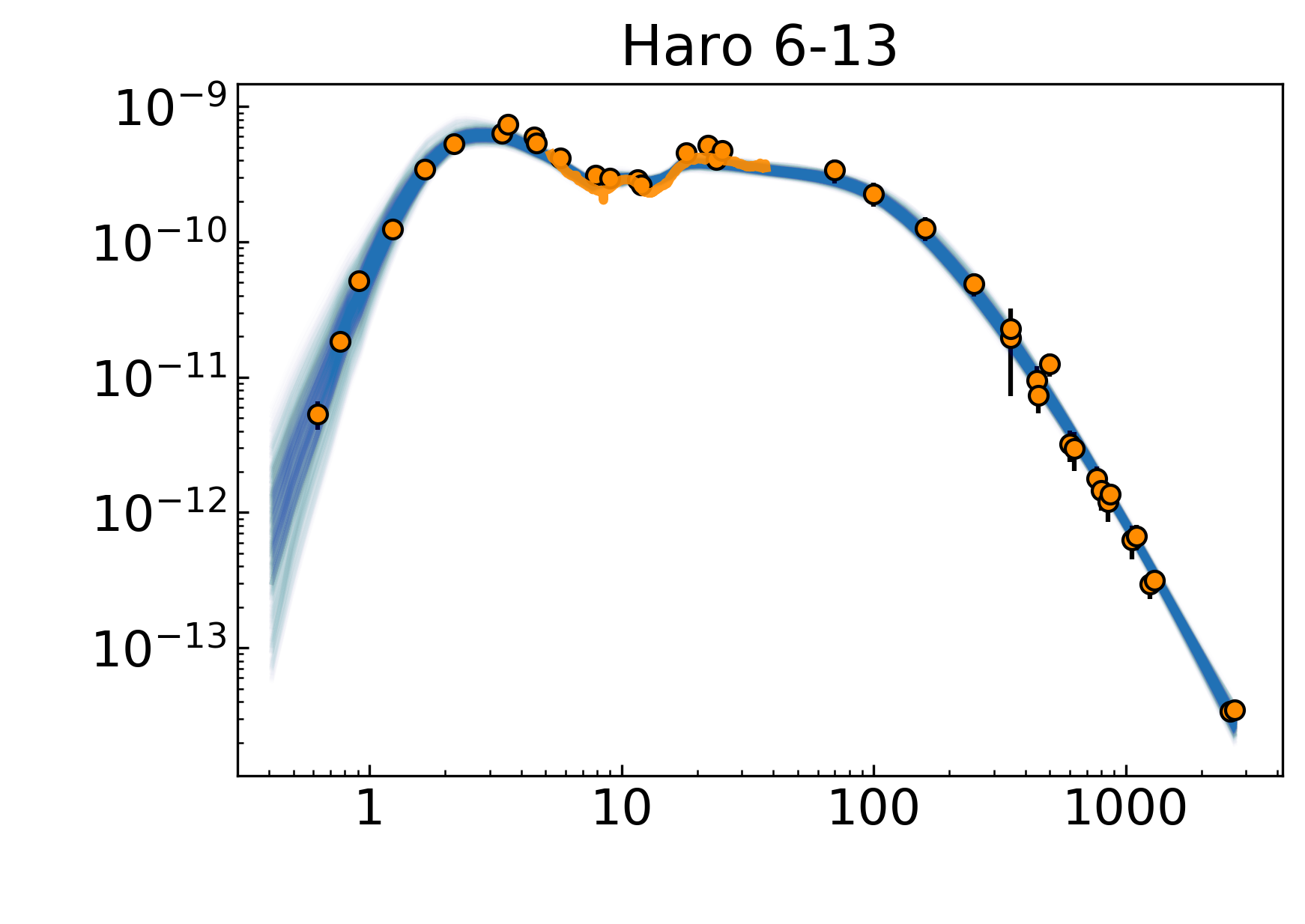}\hfill
  \includegraphics[width=0.33\hsize]{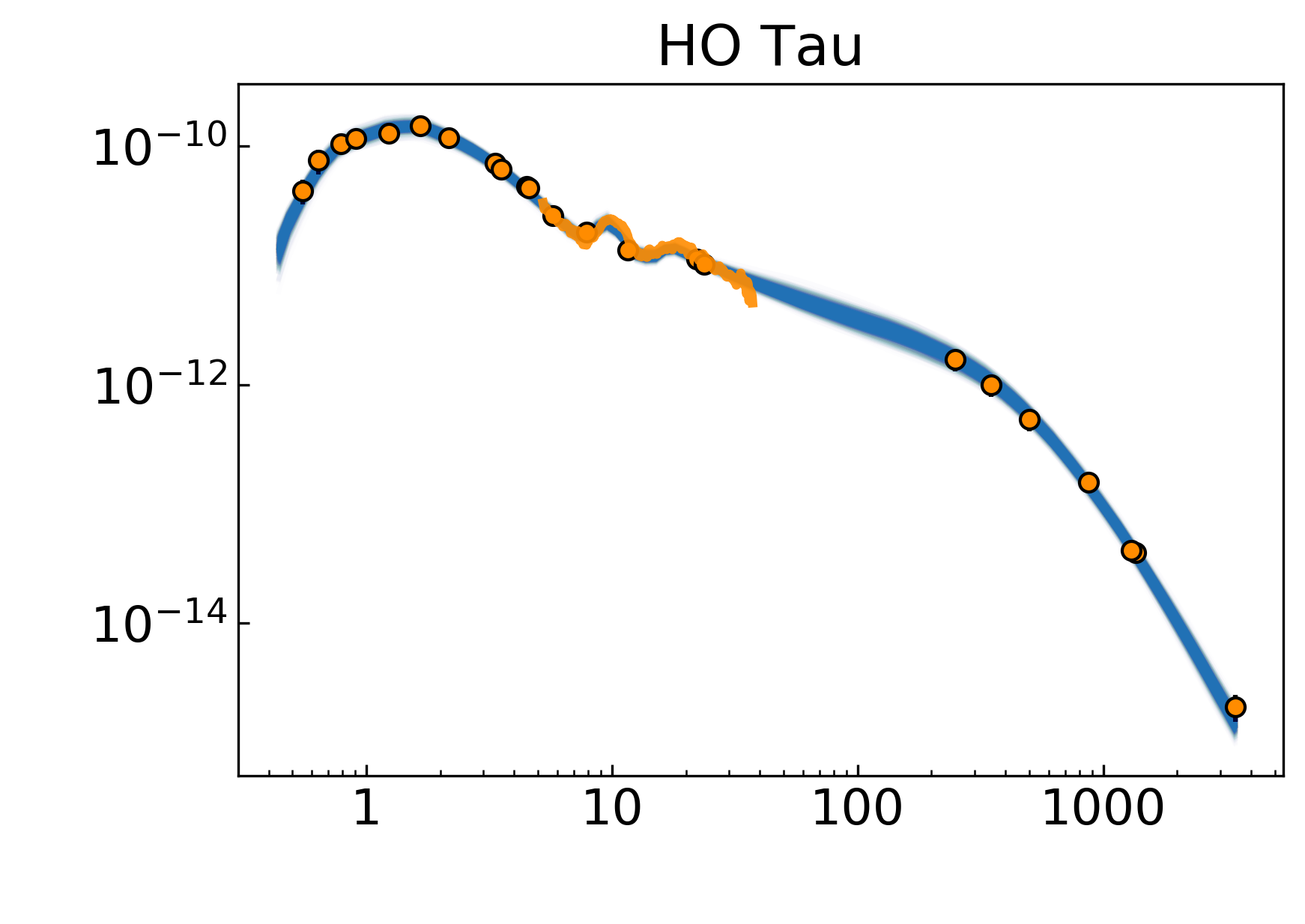}\\
  \includegraphics[width=0.33\hsize]{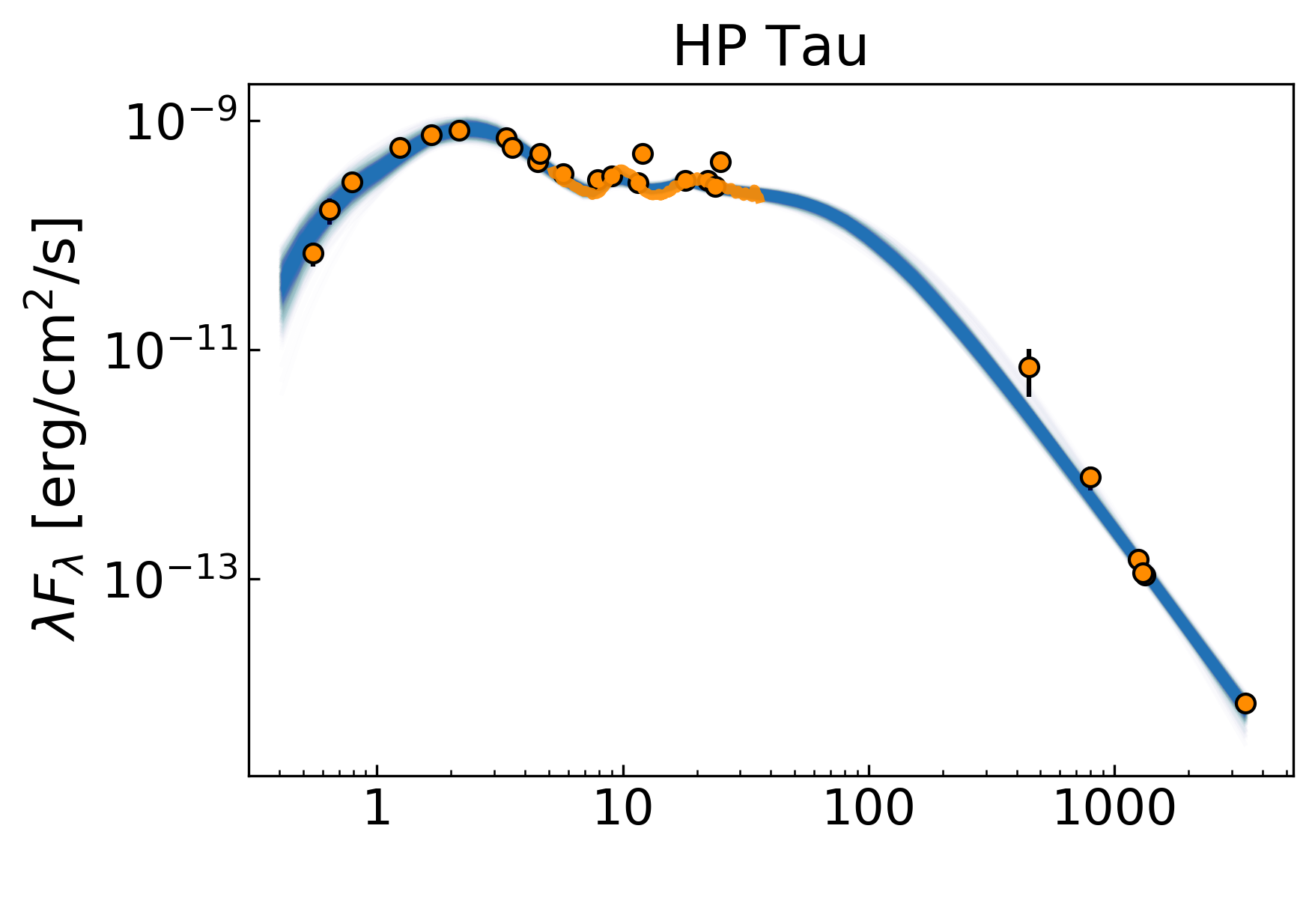}\hfill
  \includegraphics[width=0.33\hsize]{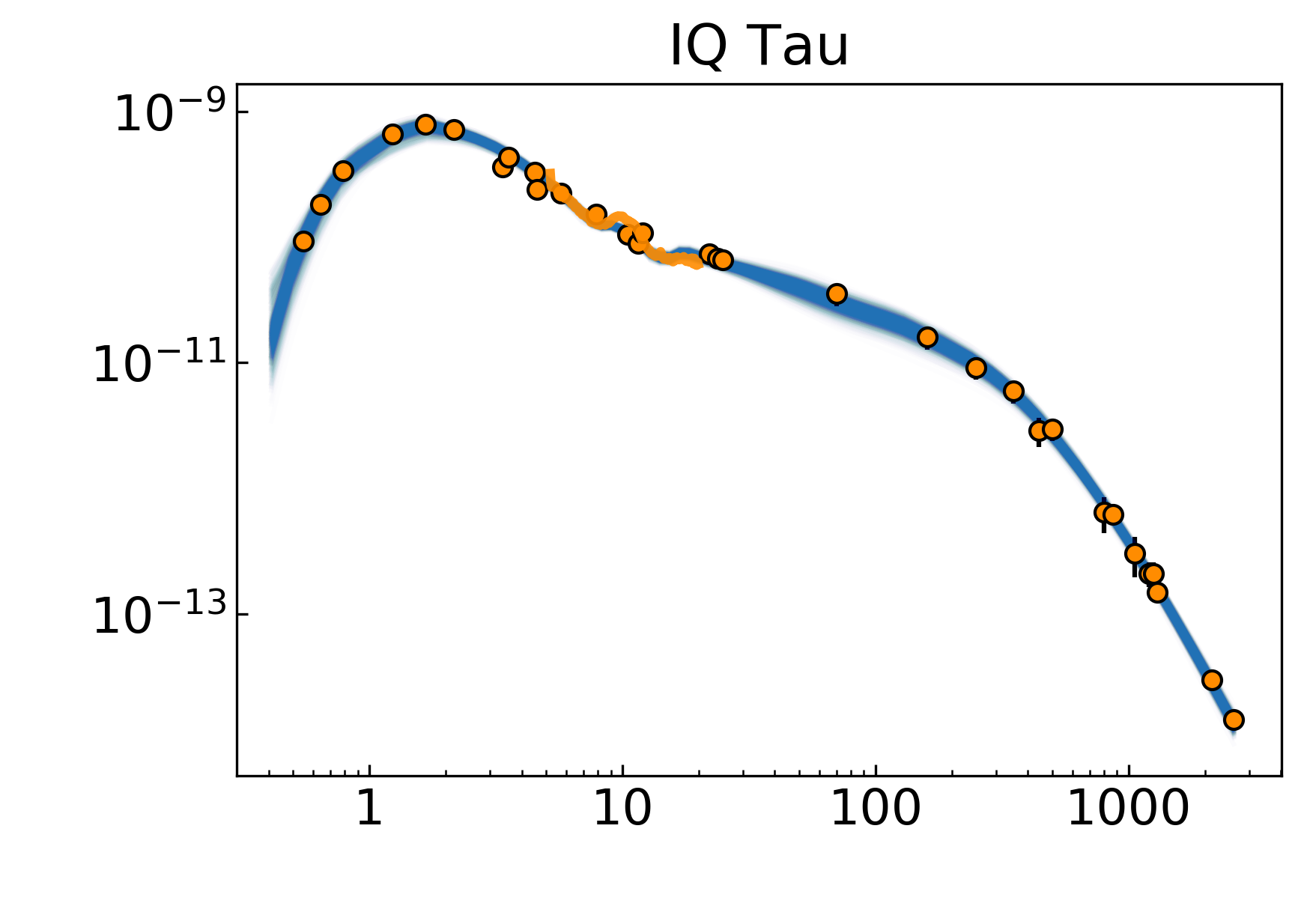}\hfill
  \includegraphics[width=0.33\hsize]{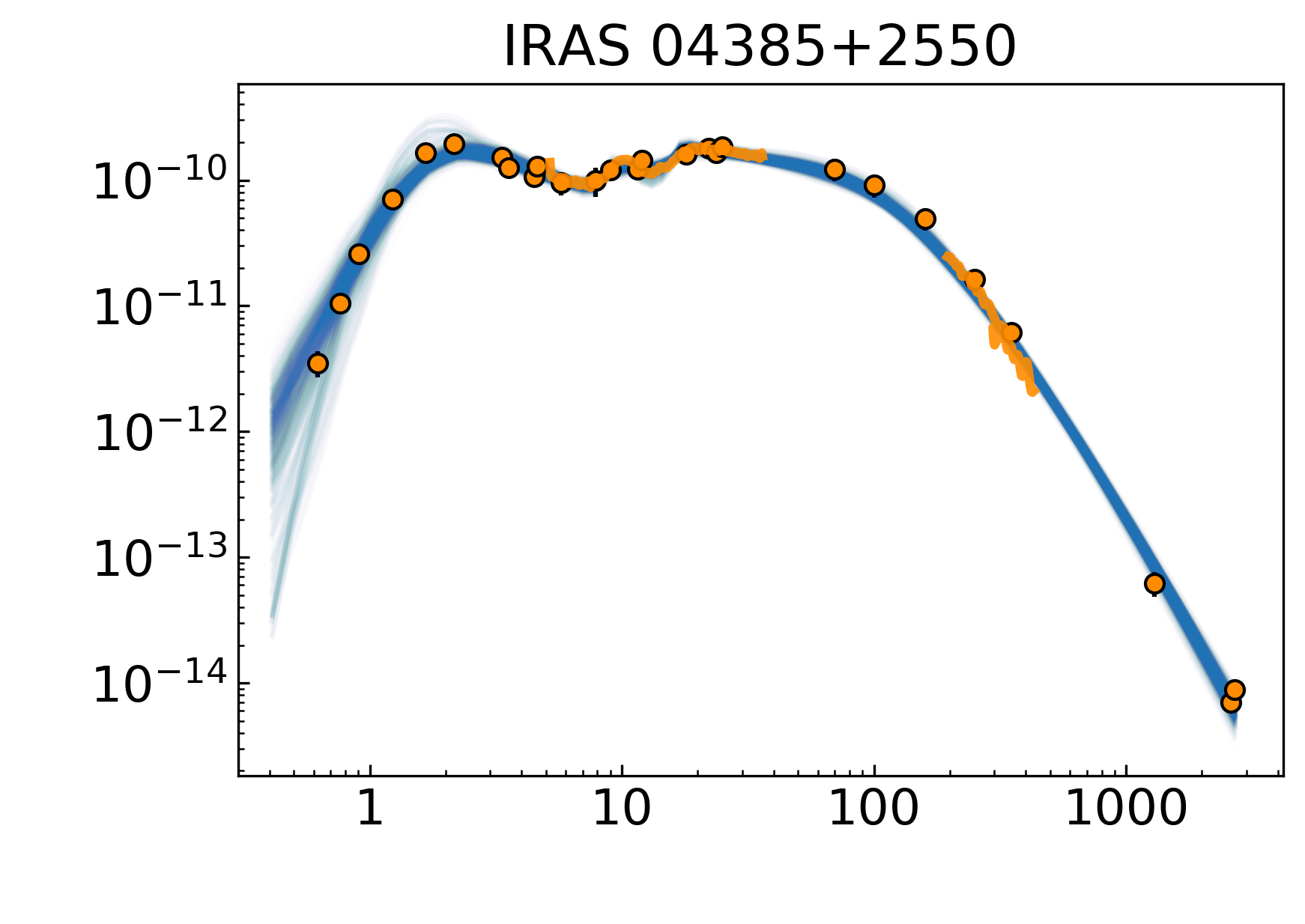}\\  
  \includegraphics[width=0.33\hsize]{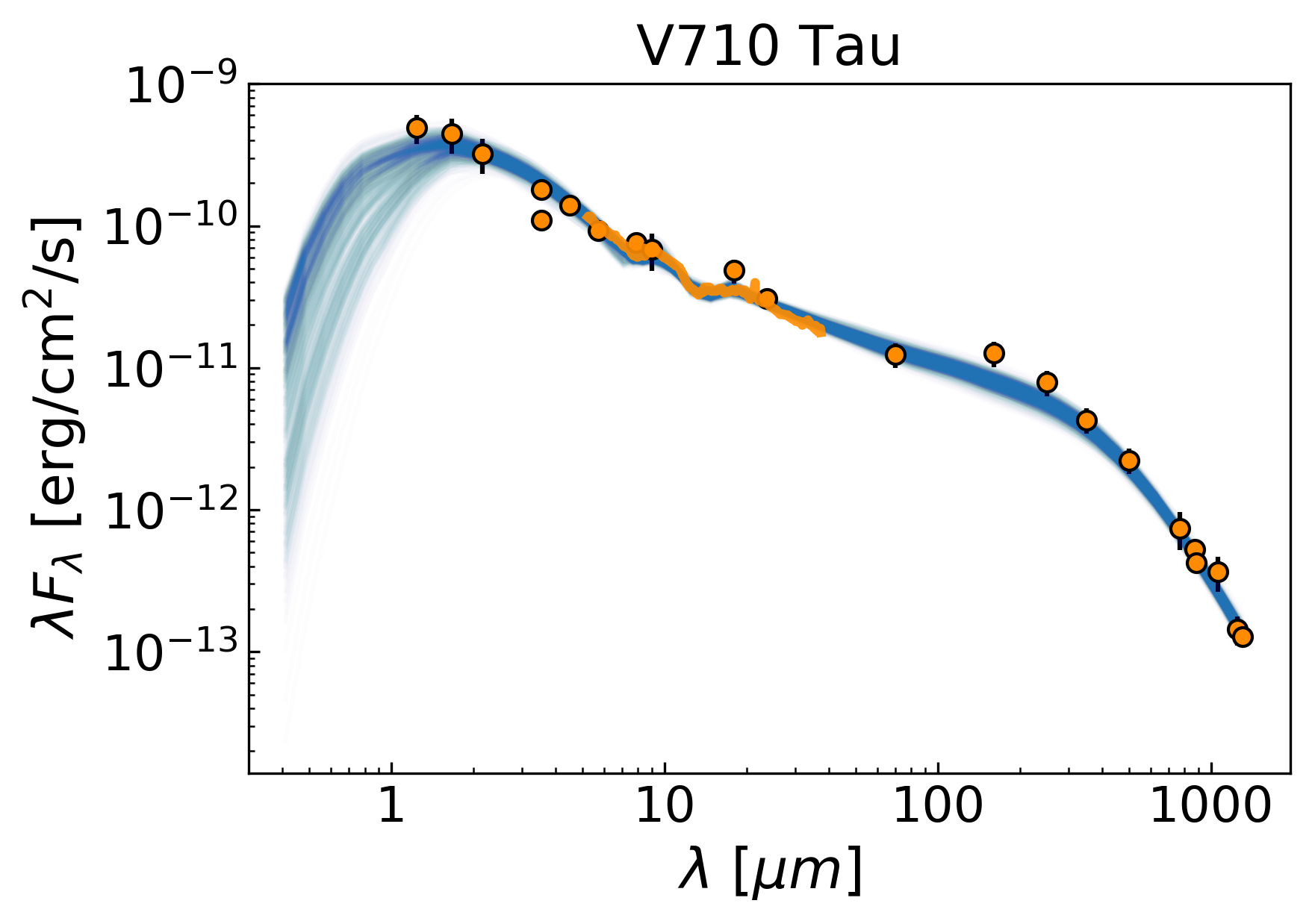}
  \includegraphics[width=0.33\hsize]{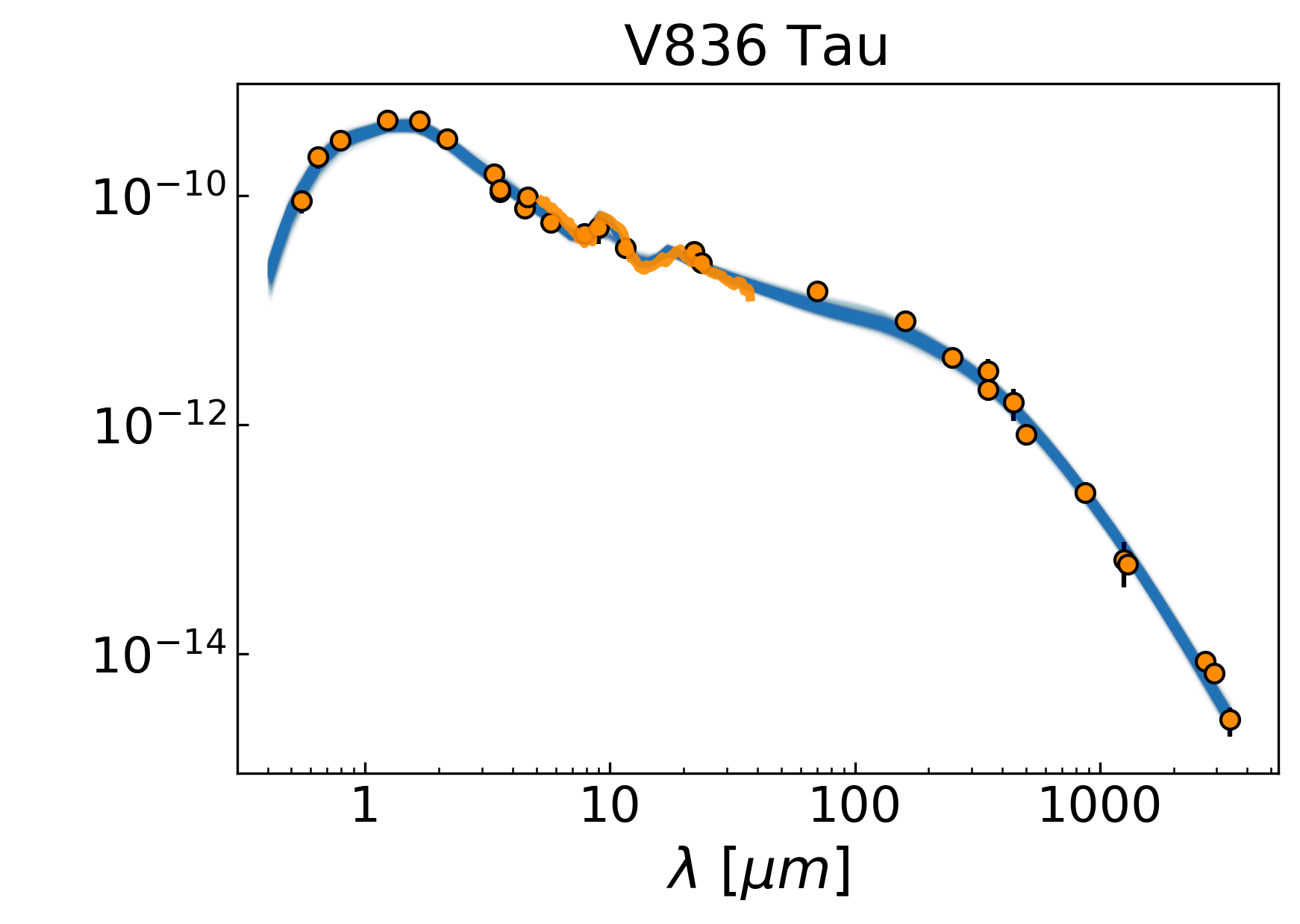}
  \caption{(Continued).}
\end{figure*}

\section{Results}\label{sec:results}

\subsection{Successful fits and discarded sources}\label{sec:fits}

\begin{figure*}
  \centering
  \includegraphics[width=0.33\hsize]{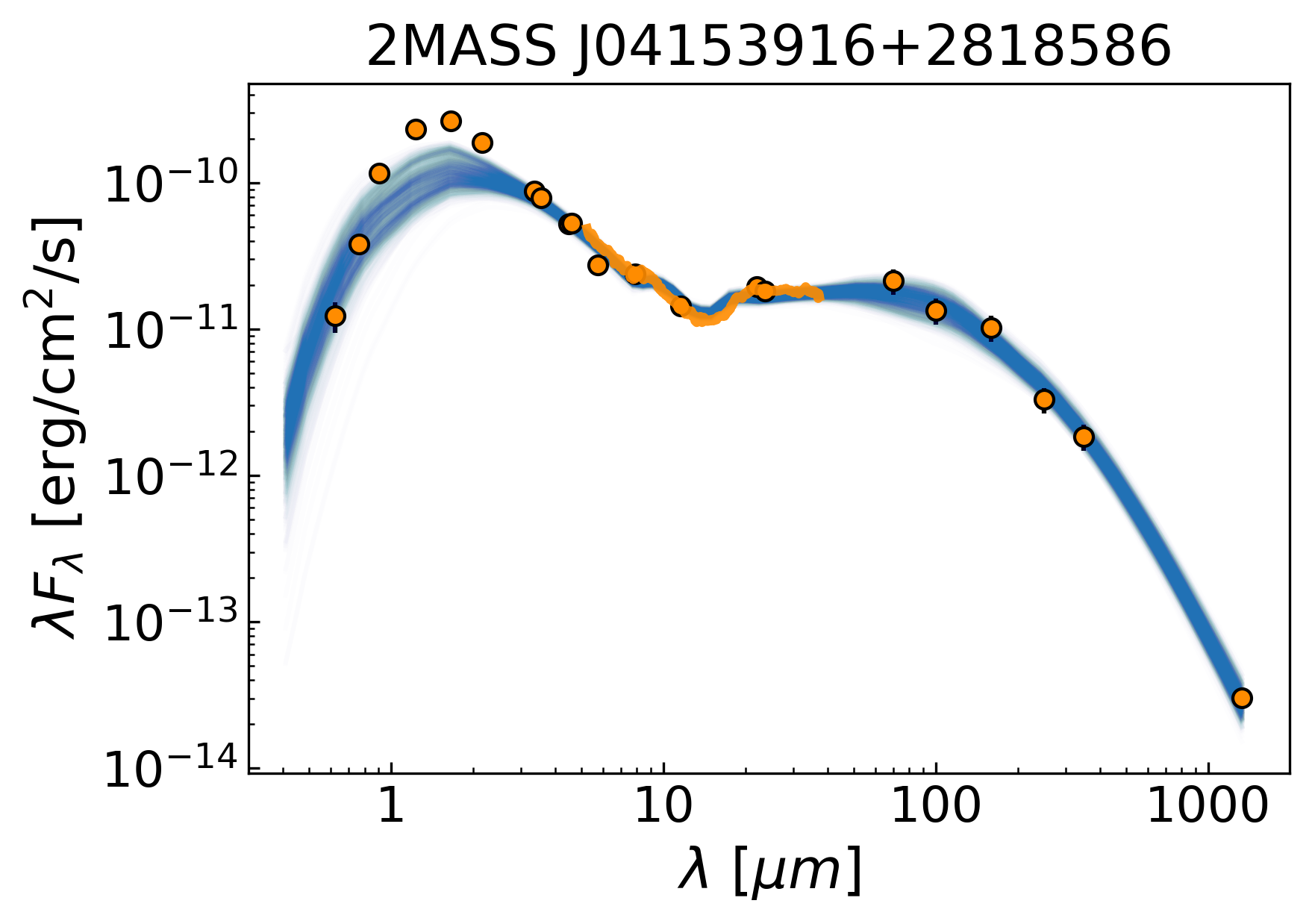}\hfill
  \includegraphics[width=0.33\hsize]{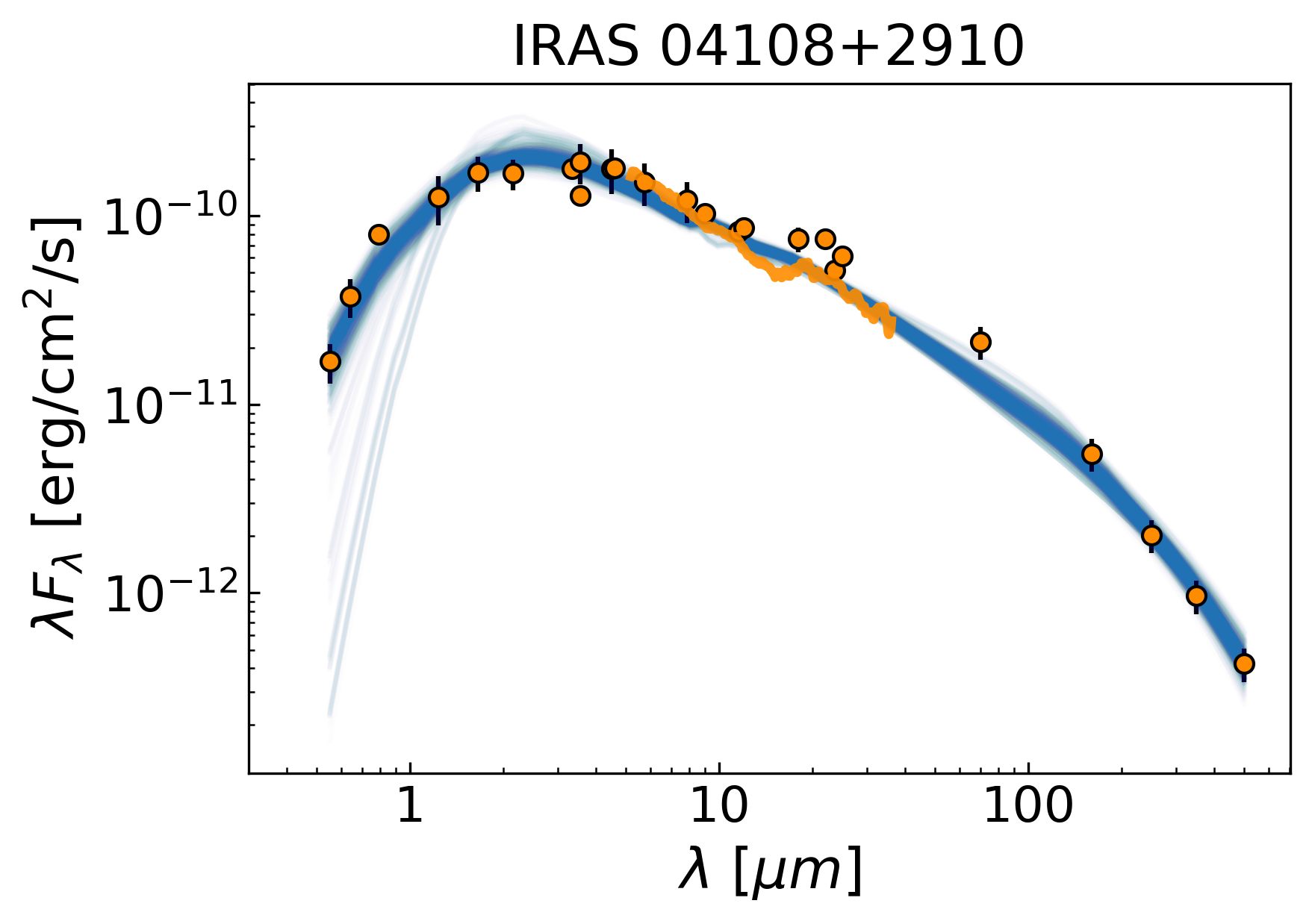}\hfill
  \includegraphics[width=0.33\hsize]{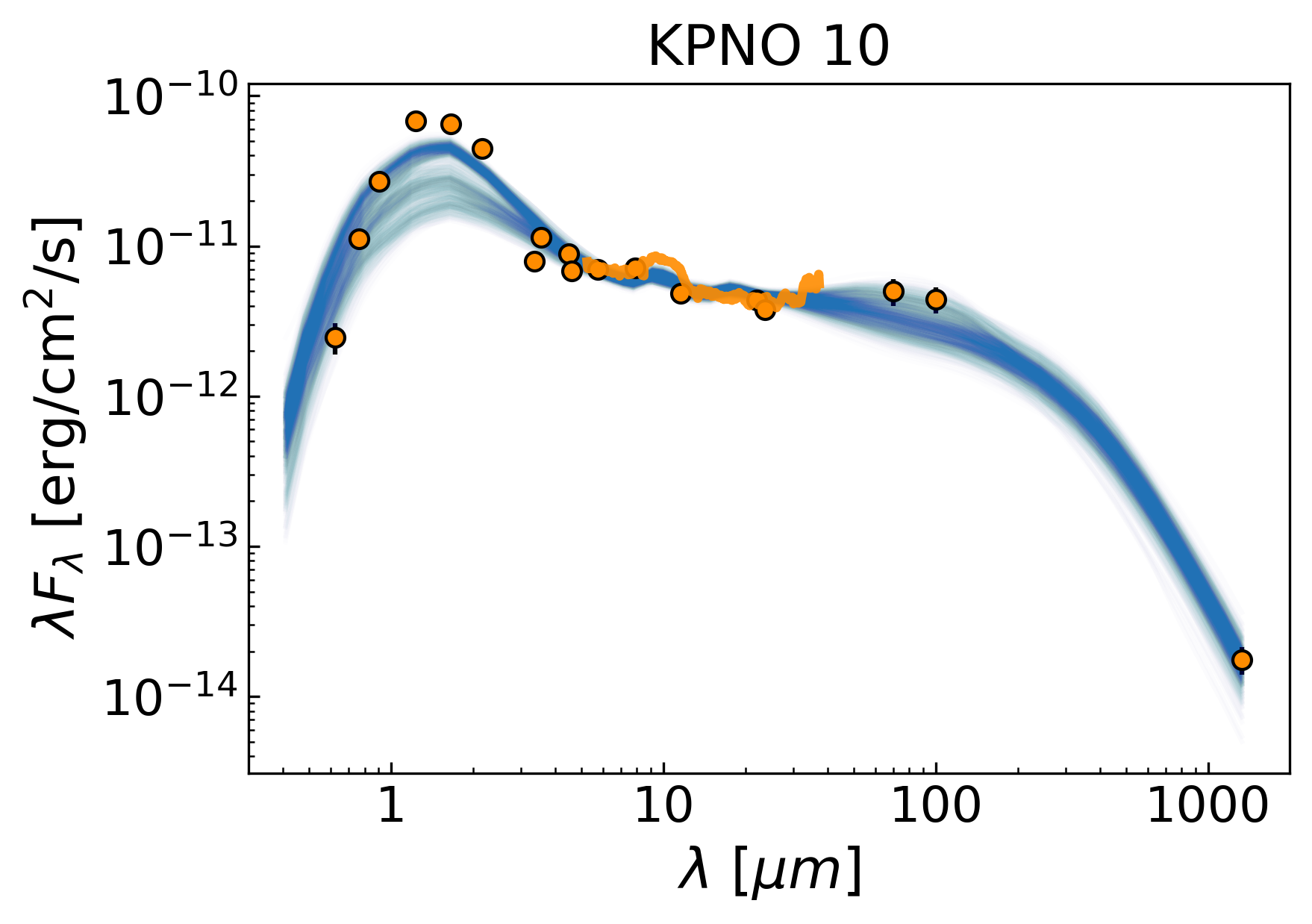}\\
  \caption{Three examples of SEDs that we were not able to properly fit with our adopted model. Symbols are
    the same as in Fig.~\ref{fig:successful_fits}.}\label{fig:failed_fits}
\end{figure*}

It is likely that some disks in the sample have yet unknown substructures or companions, or are affected by
physical processes not considered in our modeling. Thus, some objects may not be properly described by the
DIAD models, and it is expected that we will not be able to fit the whole sample successfully. Before
analyzing our results, we inspected the obtained fits and discarded those sources for which the models were
unable to reproduce the observed SED. Nine objects were discarded during this process:
2MASS~J04153916+2818586, DL~Tau, FN~Tau, IRAS~04108+2910, IRAS~04196+2638, IRAS~04216+2603,
IRAS~04200+2759, KPNO~10, and XEST 13-010. Therefore, the final sample of successfully modeled disks comprises
23 sources. Their SEDs and corresponding models are shown in Fig.~\ref{fig:successful_fits}. For reference, we
also show three discarded cases in Fig.~\ref{fig:failed_fits}.

\subsection{Results for parameters of interest}\label{sec:results_params}

Our goal is to better understand the general properties of disks without focusing on individual objects, since
better fits could probably be achieved by considering the characteristics of each source in detail. For this
purpose, we built the ensemble distribution for each parameter by randomly drawing 1000 values from the
posterior of each source, and combining them in a joint histogram (see
Fig.~\ref{fig:ensemble_distributions}). Nevertheless, the results of the parameters of interest for individual
sources are listed in Table~\ref{tab:modeling_params}, and the corresponding cornerplots can be retrieved from
an online
repository\footnote{\href{https://doi.org/10.5281/zenodo.4011334}{https://doi.org/10.5281/zenodo.4011334}}. Some
parameters in our model are nuisance ones and, in those cases, we marginalized over them (see
Appendix~\ref{appendix:marginalized_params}).

\subsubsection{Disk viscosity and accretion rate}\label{sec:results_alpha_mdot}

As mentioned in Sect.~\ref{sec:DIAD_models}, the ratio of these two parameters determines the scaling of the
surface density profile \citep[$\Sigma \propto \dot{M}/\alpha$, e.g.,][]{DAlessio1998}. Therefore, these two
parameters are correlated. However, their effects on the SED at mid/far-IR wavelengths allow us to constrain
them individually as well: when maintaining a constant ratio between these two parameters, higher
  accretion rates increase the accretion luminosity irradiating the disk, which heats the upper layers and
  increases the mid/far-IR disk emission (see bottom right panel of Fig.~\ref{fig:SED_parameters}). Both
parameters are relatively well constrained for most sources. In the case of $\alpha$, the obtained median
value is 0.003 and a significant fraction of the sample (10 out of 23 objects) is above $\alpha=0.01$, with
individual values distributed throughout the entire explored range ($1 \times 10^{-4}$ to 0.1). These results
are compared with observations in Sect.~\ref{sec:viscosity}. For the accretion rates, the median
($2\times 10^{-8}\,M_\odot$/yr) is in good agreement with previous results \citep{Hartmann1998}. Despite
the accretion rate in the disk and that onto the star not being the same parameter and the significant
variability of the latter, we find an overall agreement (within an order of magnitude) of our derived
accretion rates and those reported in the literature \citep[e.g.,][]{Valenti1993, Hartigan1995, Gullbring1998,
  Ingleby2013, Simon2016}.

\subsubsection{Disk radius}

Disk sizes are poorly constrained by SED fitting alone, and we do not expect the derived radii to be very
accurate. Moreover, dust radial migration causes grains with different sizes to have different radial extents
and, therefore, the observed radii depend on the wavelength \citep[e.g.,][]{Tazzari2016}. Our models do
not account for this effect and, since we fit emission at all wavelengths, our radii estimates should be
either larger or comparable to radii at millimeter wavelengths (for sources with or without significant dust radial
migration, respectively). The ensemble distribution shows a rather unconstrained distribution (with a little
preference for larger radii), except for a family of sources with radii $\leq$50\,au. Interestingly, this
trend seems to reflect a population of compact disks recently discovered with high-spatial resolution ALMA
observations \citep[e.g.,][]{Cieza2019, Long2019}.

\subsubsection{Dust settling}

As discussed in Sect.~\ref{sec:DIAD_models}, the DIAD models include two layers with different dust grain
populations to simulate dust settling (small grains in the upper layers, larger grains in the disk
midplane). The $\epsilon$ parameter describes the depletion of grains in the upper layers with respect to the
standard gas-to-dust ratio. Our results show high levels of settling ($\epsilon < 10^{-2}$) for the whole
sample, indicating that disks in Taurus-Auriga have already undergone significant settling, in agreement with
previous studies \citep[][]{Furlan2011, Grant2018}. Given that this process concentrates large grains in the
disk midplane and enables planet formation, this is in line with the ubiquity of rings and gaps found in
protoplanetary disks if these structures are to be explained by the presence of planets.

\subsubsection{Grain sizes}

The size of grains in the upper layers of disks ($a_{\rm max, upper}$) is mostly determined from mid-IR
spectroscopy and, in particular, from the shape of the silicate features at 10 and 20\,$\mu$m. Given that our
sample includes \emph{Spitzer}/IRS spectra for all objects, we expect to gain some information about
$a_{\rm max, upper}$. Indeed, the results for this parameter suggest that grains of a few $\mu$m are present in
the upper layers of these disks, but the derived values are general estimates only (see
Sect.~\ref{sec:caveats}). On the other hand, information about large grains in the midplane
($a_{\rm max, midplane}$) is usually assumed to come from (sub)mm fluxes, and the slope of the mm emission has
been used for this purpose under some assumptions \citep[e.g.,][]{Ricci2010_Ophiuchus, Ricci2010_Taurus,
  Ribas2017}. However, although many of our sources have well-sampled SEDs in this wavelength range, the
ensemble distribution of $a_{\rm max, midplane}$ is poorly constrained, with only a slight trend toward large
($>1$\,mm) grains. The lack of strong evidence for large grains in our analysis is somewhat surprising, and
could imply that the combined effect of other parameters in grain size uncertainties may have been previously
underestimated when establishing grain growth in disks \citep[e.g.,][]{Ysard2019}. Our results suggest that
very little information about grain sizes in the disk midplane can be gained from SEDs alone.

\subsubsection{Disk masses}

The disk mass ($M_{\rm disk}$) is one of the most relevant quantities in planet formation. Despite
$M_{\rm disk}$ not being a free parameter in the DIAD models, we can estimate the disk mass of each model in
the posterior distributions using ANN$_{\rm disk mass}$. This yielded the corresponding $M_{\rm disk}$
distribution for each source, which is well constrained in all cases. We then computed the ensemble posterior
for the 23 sources as in Sect.~\ref{sec:results_params}. The derived disk masses are listed in
Table~\ref{tab:modeling_params}, and the results for $M_{\rm disk}$ and $M_{\rm disk}/M_{*}$ are shown in
Fig.~\ref{fig:ensemble_distributions}. The ensemble distribution of $M_{\rm disk}$ peaks at $0.02\,M_\odot$,
and is smoothly distributed between 0.001 and 0.1\,$M_\odot$. In the case of $M_{\rm disk}/M_{*}$, we found it
to range from 0.003 to 0.3, with a preference for values between 2-10\,\%. The distribution decreases rapidly
for higher values, in agreement with general disk stability criteria \citep[e.g.,][]{Lodato2005}. Given that
$M_{\rm disk}$ is proportional to the ratio $\dot{M}/\alpha$ in the $\alpha$-disk prescription, we also
searched for possible correlations of $M_{\rm disk}$ with those parameters. Despite $\dot{M}$ and $\alpha$
showing a clear correlation (as expected for $\alpha$-disk models), none of them appear to be significantly
correlated with $M_{\rm disk}$ in our results (see Fig.~\ref{fig:Mdot_alpha_Mdisk_correlations}).

One advantage of the method employed in this study is that it does not require some of the standard
assumptions when estimating disk masses from (sub)mm observations (e.g., emission from isothermal dust,
optically thin emission) and, given the Bayesian framework used, it naturally accounts for the effect and
uncertainties from all the other parameters. However, the caveats in Sect.~\ref{sec:caveats} should be
considered when using these disk mass estimates. In particular, we note that the gas-to-dust value is highly
uncertain and, therefore, the dust content should be more accurate than the total (gas+dust) mass.

 \subsection{Caveats}\label{sec:caveats}

 Although we have tried to account for as many uncertainties and physical phenomena as possible
 during our analysis, it is important to consider the caveats in our models and the implications they may
 have in our results.

 We have not explored different dust compositions, which has a crucial effect on the dust opacity and, therefore,
 on the thermal structure of disks. Including this factor in our models adds a considerable number of free
 parameters, which would significantly increase the complexity of the ANN training. Moreover, a proper
 exploration of dust compositions requires a detailed fitting of the \emph{Spitzer}/IRS mid-IR spectra and, as
 explained in Sect.~\ref{sec:uncertainties}, we have weighted the spectroscopic data heuristically in our
 study. While this makes the overall fitting more consistent, it also decreases the quality of the fit to the
 spectra. Thus, we chose to keep the dust composition fixed. As a result, the derived sizes for grains in the
 upper layers are approximations.

 We have adopted the standard gas-to-dust ratio value of 100 in our models. This parameter affects the gas
 density in the disk and, therefore, its structure. Given the various difficulties in determining this quantity,
 this choice allows us to compare our results with other studies and to test default assumptions about disk
 properties. However, it is very likely that the gas-to-dust ratio changes considerably from source to source
 \citep[e.g.,][]{Bergin2013, Williams2014, Schwarz2018}.

 Our models also assume disks with no substructures, (i.e., no rings in their surface density profiles). We have
 excluded sources with known large cavities from our sample, but this does not guarantee that the modeled
 disks have smooth density distributions since not all the remaining objects have been observed at high-spatial
 resolution and, even then, the presence of smaller, unresolved substructures cannot be ruled out. In fact, a
 few sources in our sample are known to harbor rings in their outer regions. Nevertheless, we note that most disk
 substructures are likely produced by trapping of large particles in pressure bumps \citep[e.g.,][]{Dullemond2018},
 so the distribution of smaller ($<$1 mm) particles, which dominate the SED emission, is probably less affected by
 possible substructures.

 We have not performed any convolution of the models with the corresponding photometric filters when computing
 the likelihoods. This could be relevant mostly at optical and near-IR wavelengths due to stellar absorption
 lines (at longer wavelengths, the flux is dominated by the mostly line-free continuum emission from the
 disk). However, since we enforce that the resulting stellar temperatures match the observed ones and use
 isochrones to guarantee consistent stellar parameters, this should have little to no impact on the
 modeling results, especially for disk parameters.

 We do not account for stellar scattered light, which is usually a negligible contribution with respect to the
 stellar and disk thermal emission in cases where the disks are not highly inclined
 \citep[e.g.,][]{DAlessio2006}. Since we have removed disks with high inclinations, the impact of scattered
 light in our results should be minor. We note that a high polarization fraction has been found in a couple of
 sources in our sample \citep[DG~Tau and Haro~6-13 show polarization fractions of $\sim$4.4\,\% at J
 band,][]{Moneti1984}, which indicates that they could have a more relevant contribution from stellar
 scattered light. However, as in the previous case, the observed stellar temperatures and parallaxes from the
 \emph{Gaia} DR2 provide strong constraints for the stellar parameters and they should not be significantly
 affected by this issue.
  
\begin{figure*}[h]
    \centering
  \includegraphics[height=5.3cm]{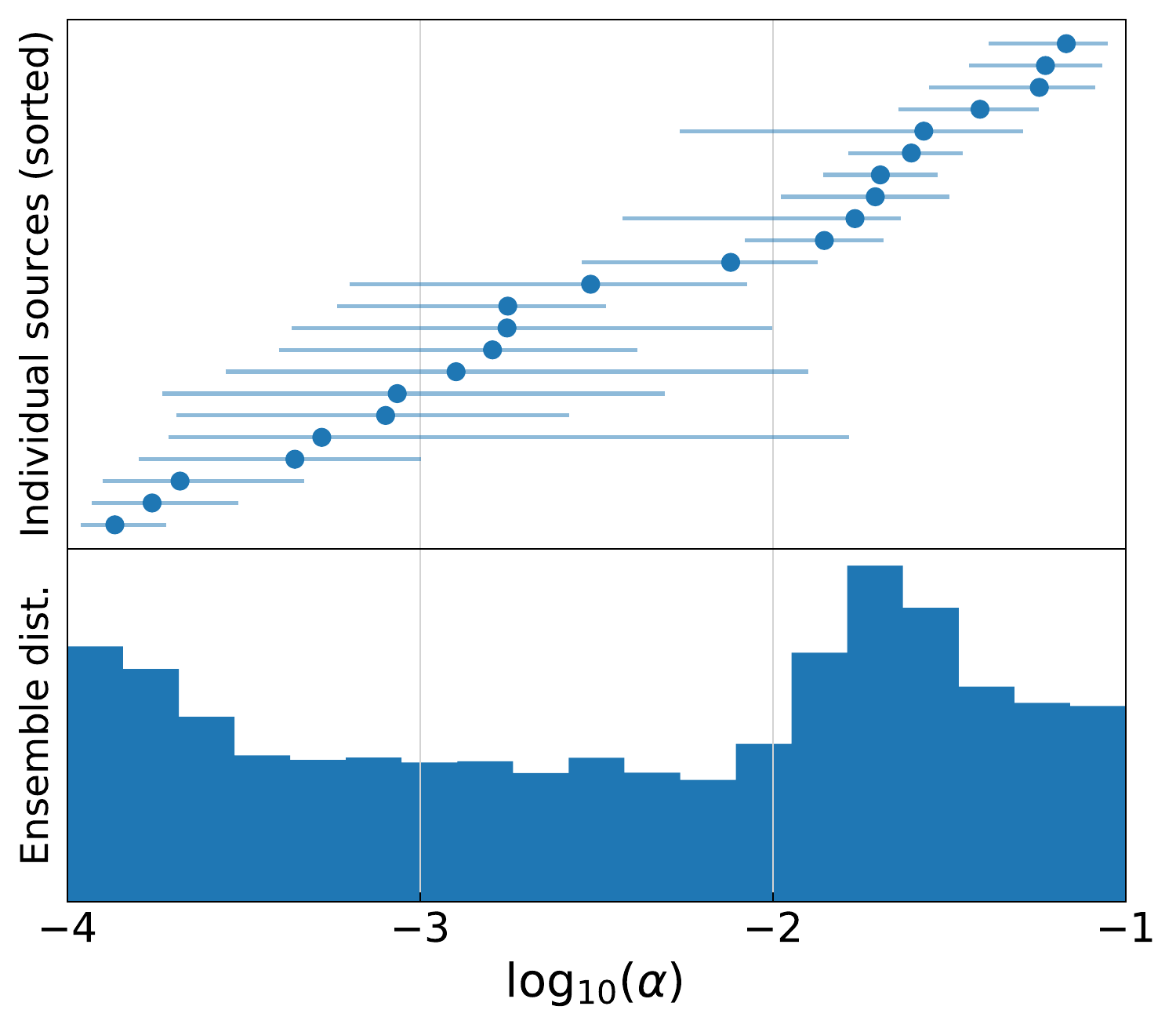}
  \includegraphics[height=5.3cm]{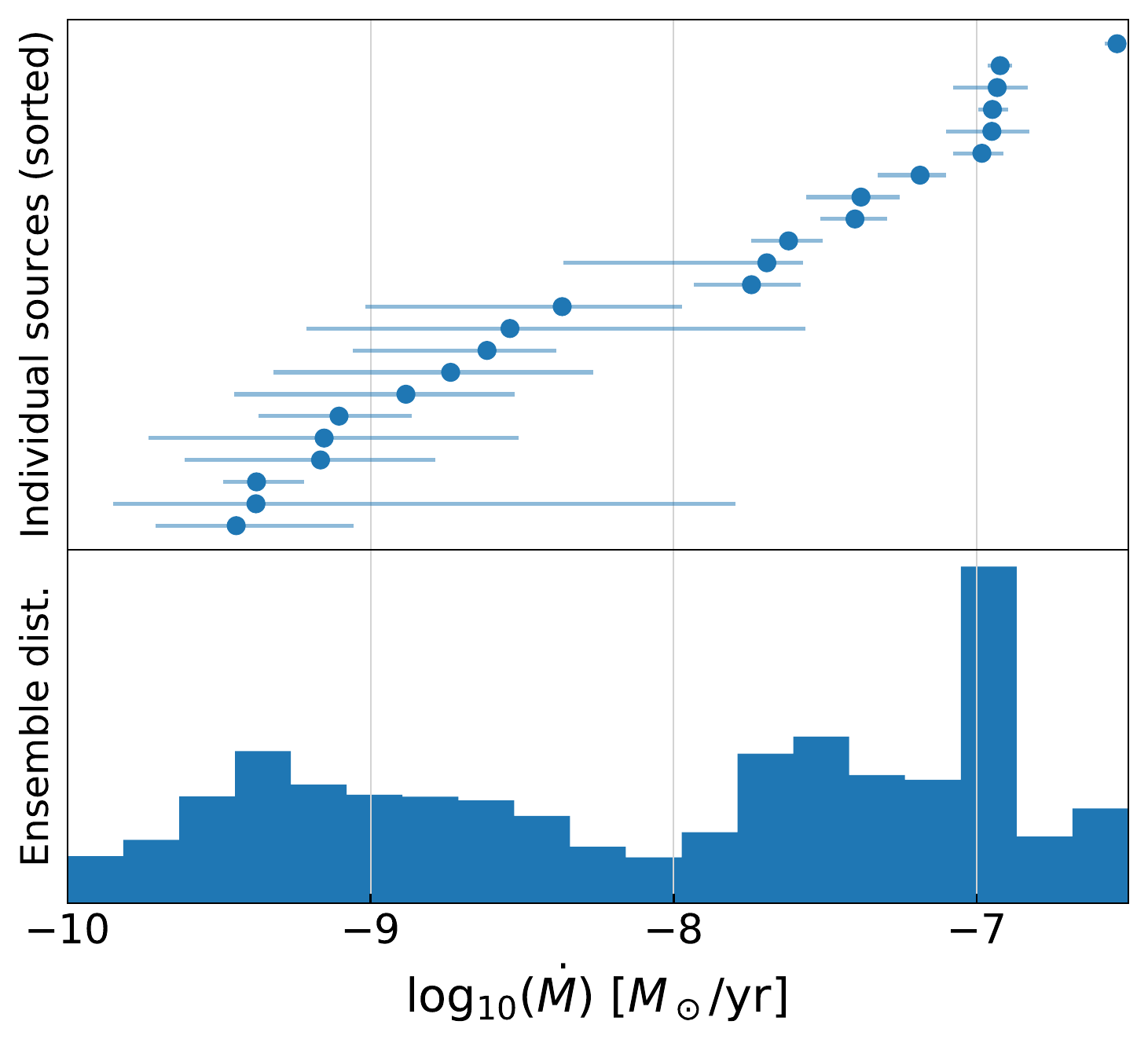}
  \includegraphics[height=5.3cm]{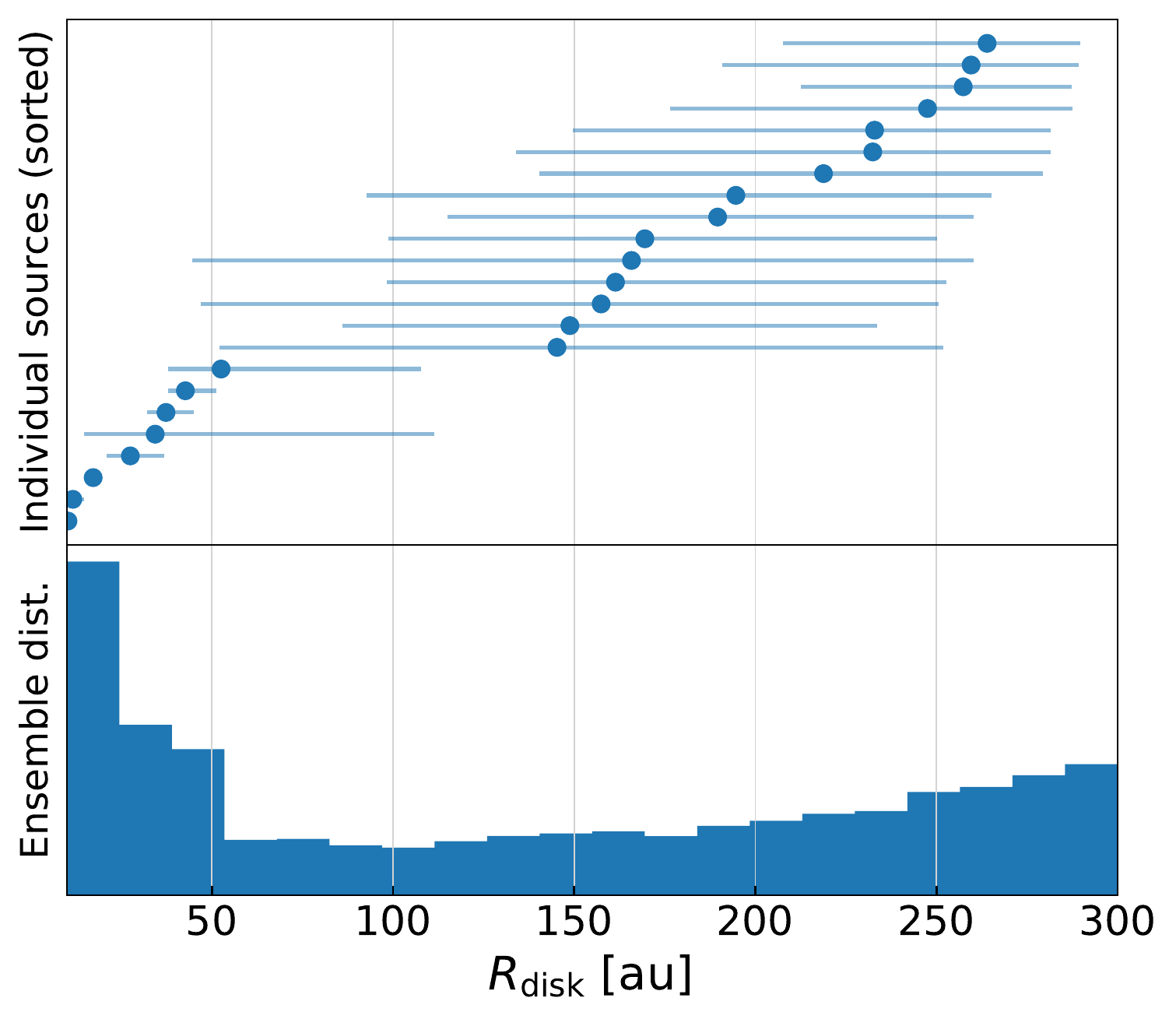}\\
  \includegraphics[height=5.3cm]{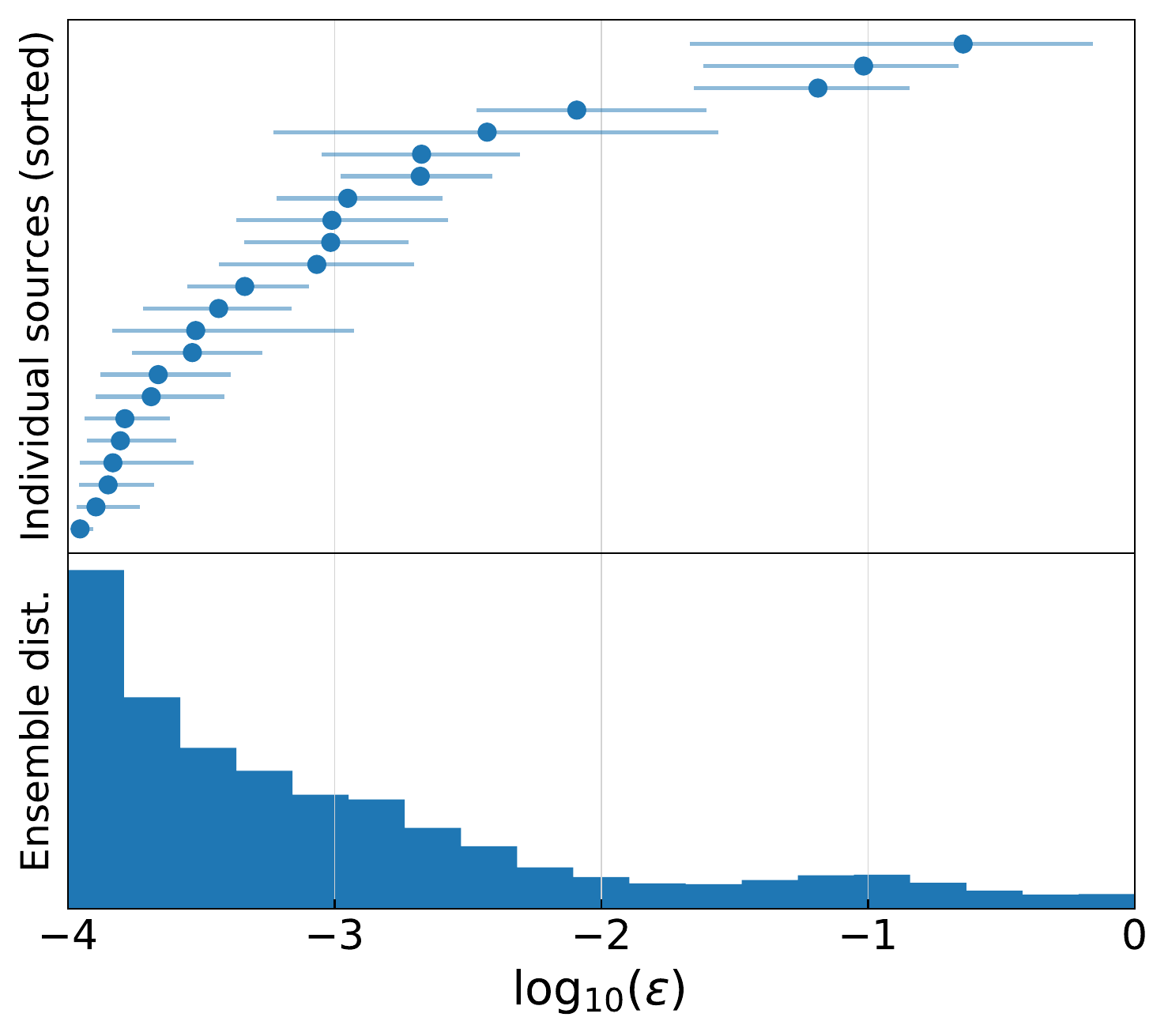}
  \includegraphics[height=5.3cm]{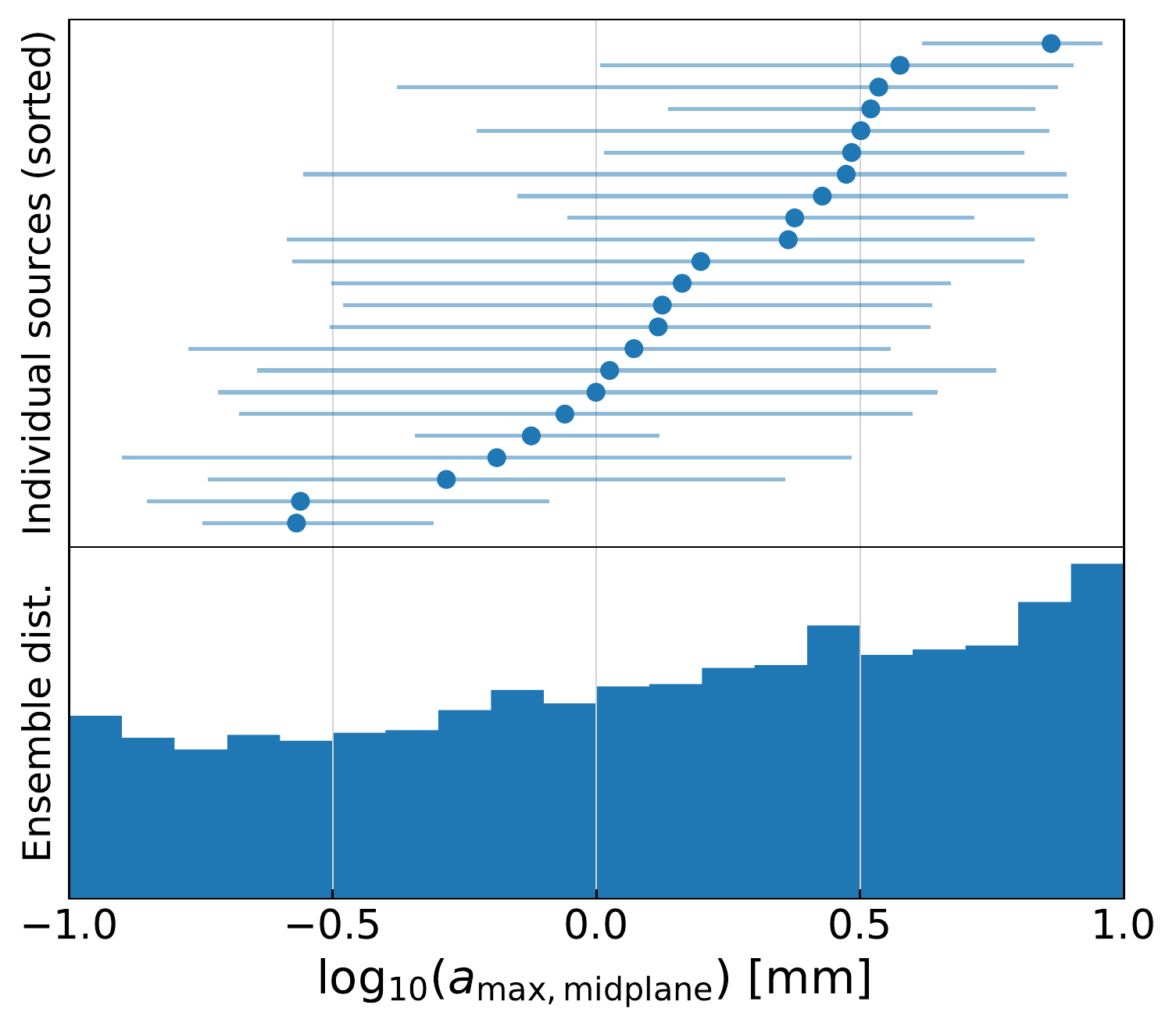}
  \includegraphics[height=5.3cm]{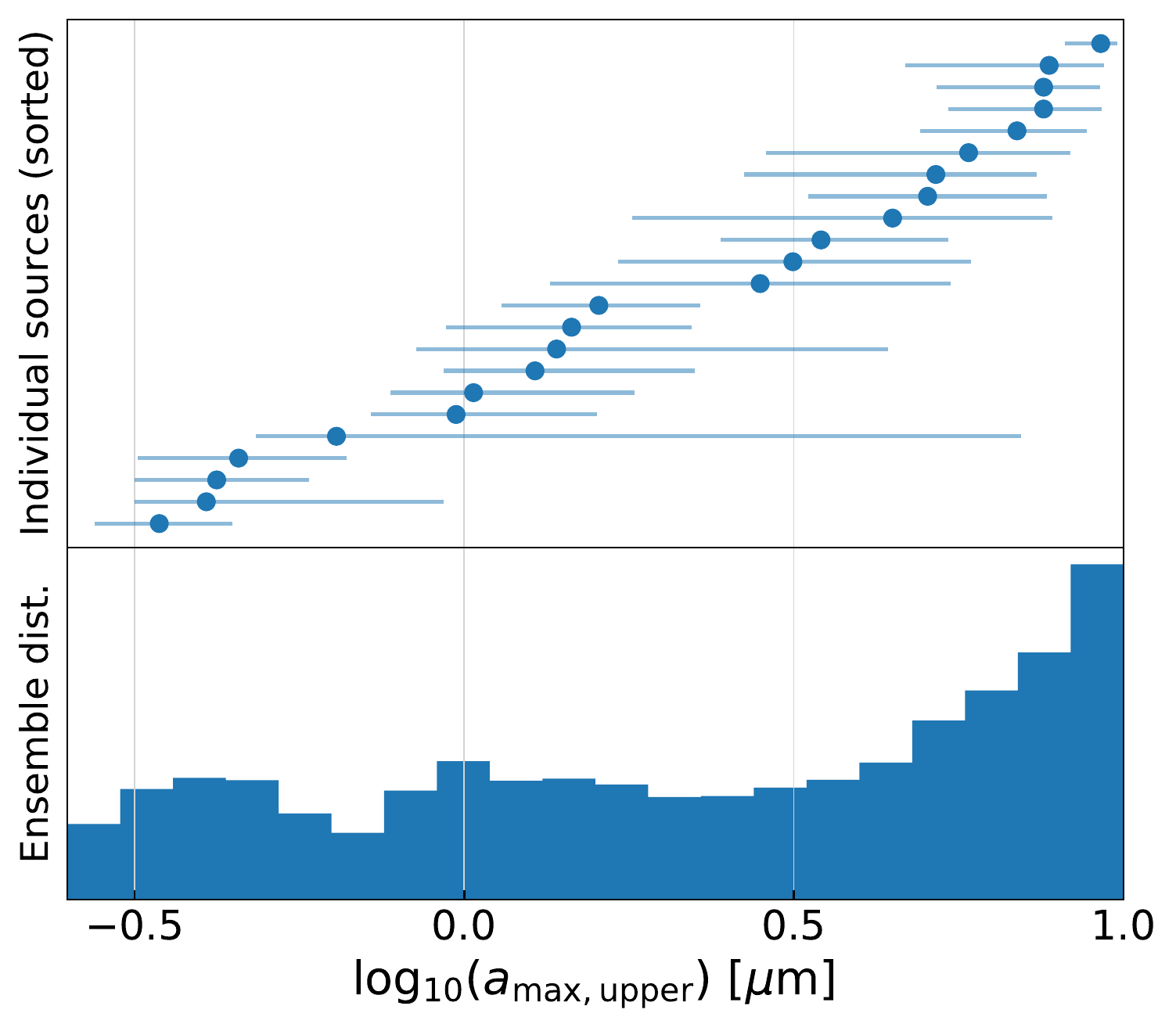}\\
  \includegraphics[height=5.3cm]{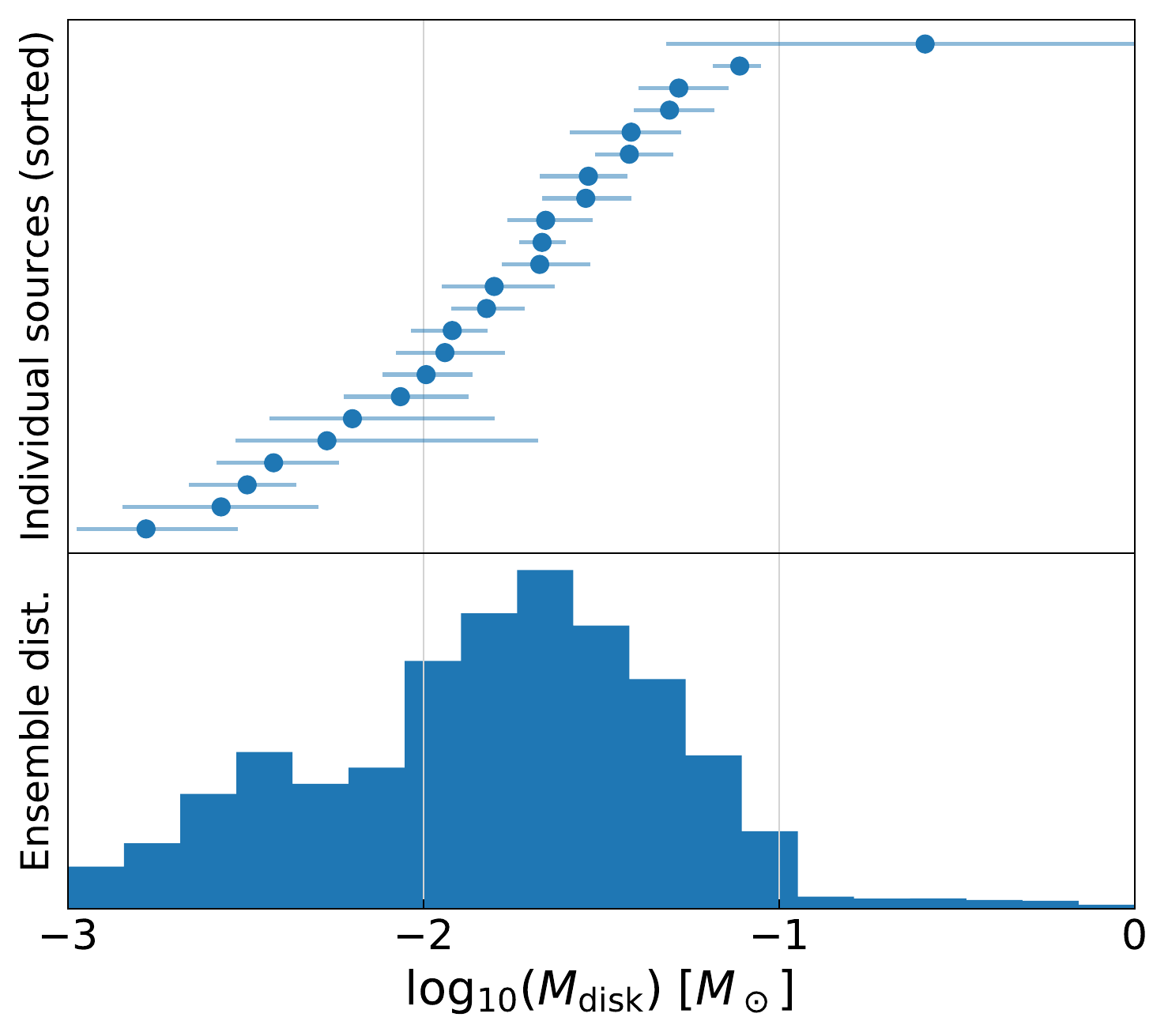}
  \includegraphics[height=5.3cm]{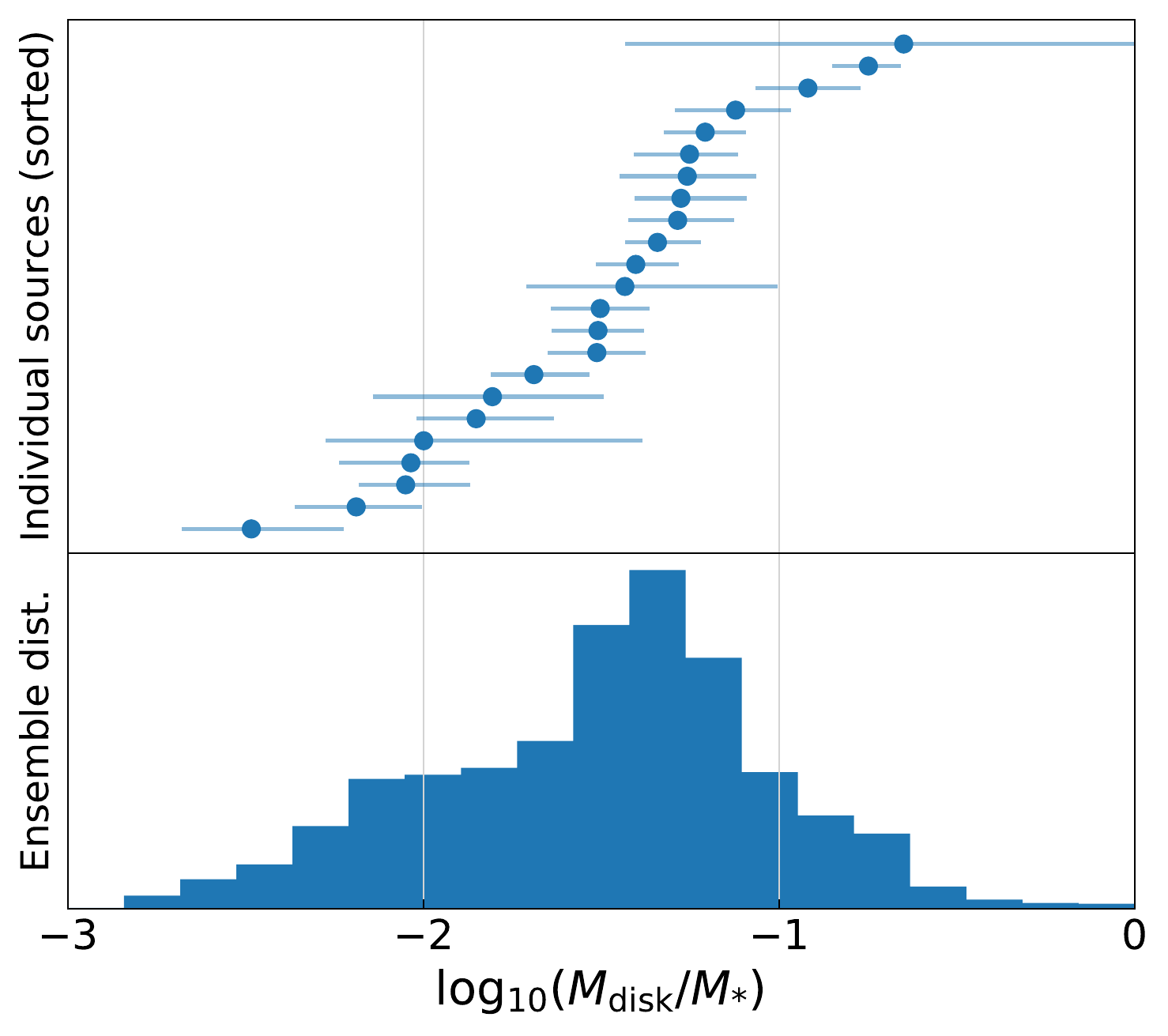}\\  
  \caption{Results for individual sources (top panels) and ensemble distribution (bottom panels) of
      each relevant parameter for the sample of 23 disks modeled in our study. The top
      panels show the parameter value (blue dot) for each source, sorted by increasing value. The errorbars
      are the corresponding 16\,\% and 84\,\% percentiles. The ensembles in the bottom panels show the overall
      distribution of the sample, and are produced by randomly selecting 1000 values from the posterior
      distribution of each object, and then combining all of them.}\label{fig:ensemble_distributions}
  \end{figure*}

 Finally, our sample represents only a subset of disks taken from a single star-forming region. Therefore,
 these results are likely representative of disks with no large gaps around single T~Tauri stars only, and may
 not be applicable to all disks; those with significant substructures or in multiple systems could
 have different physical conditions. Moreover, the details of the ensemble distributions in
 Fig.~\ref{fig:ensemble_distributions} are probably caused by the moderate sample size, and only the overall
 trends in them are likely real. Extending this analysis to other regions will allow us to increase the sample
 size and to determine which structures in the ensemble distributions are in fact real.

\section{Discussion}\label{sec:discussion}

\subsection{Implications for the $\alpha$-disk prescription}\label{sec:viscosity}

One of the most important open problems in planet formation is the transport of angular momentum in disks,
which plays a major role in this process and in disk evolution. In the last couple of decades, the leading
explanation for such transport has been disk viscosity. A very common prescription of this model is the
$\alpha$-disk \citep{Shakura1973}, in which the viscosity $\nu$ is assumed to follow $\nu=\alpha c_s H$
\citep{Lynden-Bell1974, Pringle1981}, where $c_s$ is the local sound speed, $H$ is the pressure scale height,
and $\alpha$ is a constant that encompasses the unclear nature of the viscosity. Molecular viscosity in disks
is too low to account for the measured accretion rates and disk lifetimes, and thus turbulence induced by
magnetorotational instability \citep[MRI, e.g.,][]{Balbus1991, Gammie1996} is believed to be the main source
of viscosity. However, the use of the $\alpha$-disk prescription for protoplanetary disks has also been questioned
by theoretical studies, as ionization levels are expected to be too low in most regions of the disk for MRI to
be effective: the midplane is largely shielded from ionizing stellar radiation, and non-ideal
magnetohydrodynamic effects can suppress MRI even in the upper layers of disks \citep[e.g.,][]{Gammie1996,
  BaiStone2013_winds}. As a result, disk winds \citep[e.g.,][]{Blandford1982} have regained substantial
attention lately as the main drivers of angular momentum transport \citep[e.g.,][]{BaiStone2013_winds,
  Gressel2015, Bai2016}. In this scenario, magnetothermal winds remove material (and thus angular momentum)
from the disk while producing sufficient accretion onto the central source.

Some observational evidence exists in favor of the $\alpha$-disk prescription. Early works using the
$\alpha$-disk model found that the properties of protoplanetary disks (e.g., accretion rates, sizes) can be
successfully explained with $\alpha\sim10^{-2}$ \citep{Hartmann1998, Calvet2000}. Observations of CO overtone
emission in disks yielded non-thermal velocities consistent with turbulence caused by MRI in their inner
regions \citep{Carr2004, Najita2009}. More recently, \citet{Manara2016} also found a correlation between mass
accretion rates and disk dust masses derived from 890\,$\mu$m fluxes for sources in Lupus, in agreement
with predictions from viscous accretion theory. A similar result was later reached by \citet{Mulders2017} in
Chamaeleon~I using the same method. \citet{Najita2018} compared the observed radii of Class I and Class II
disks and found the latter to be generally larger, which suggests viscous spreading of disks with
time. \citet{Sellek2020} also explored the effect of external far ultraviolet photoevaporation in disk
evolution and found that the similar lifetimes of dust and gas in disks are consistent with
$\alpha\sim10^{-2}$, while the observed relationship between their 890\,$\mu$m fluxes and disk radii is better
characterized by $\alpha\sim 10^{-3}$.

On the other hand, new ALMA observations support low turbulence levels in several
disks. \citet{deGregorio-Monsalvo2013} estimated a low ($\sim$0.1\,km/s) non-thermal velocity in the disk
around HD~163296 when modeling its CO(3-2) emission. \citet{Flaherty2015, Flaherty2017} used additional
CO(2-1), C$^{18}$O(2-1), and DCO+(2-1) observations of this source to place an upper limit of
$\alpha < 3\times10^{-3}$ in the disk. CO observations of TW~Hya also constrained its $\alpha$ value to be
lower than 0.007 \citep{Flaherty2018}, even when accounting for possible CO depletion over time
\citep{Yu2017}.  Recently, \citet{Flaherty2020} employed a similar method to study the CO(2-1) emission of
MWC~480, V4046~Sgr, and DM~Tau, measuring significant turbulence only for the latter ($\alpha \sim 0.08$) and
upper limits for the remaining two objects ($\alpha < 0.006$ and $\alpha < 0.01$ for MWC~480 and V4046~Sgr,
respectively). From the continuum side, \citet{Pinte2016} found that high levels of dust settling are required
to reproduce the shape and contrast of dust rings in HL~Tau, predicting $\alpha\sim3\times10^{-4}$. Indirect
evidence for low $\alpha$ also comes from the increasing number of known multi-ringed disks
\citep[e.g.,][]{HLTau, Dipierro2018, Andrews2016, Andrews2018_DSHARP}: An emerging explanation of multiple
gaps does not require the presence of several planets but, instead, they can be produced by a single low-mass
planet (e.g., a super-Earth or mini-Neptune) in a low viscosity disk
\citep[$\alpha \lesssim 1\times10^{-4}$,][]{Dong2017, Bae2017, Dong2018}. In fact, some disks observed by the
DSHARP program show substructures that may be in mean-motion resonance, favoring this interpretation
\citep[e.g.,][]{Huang2018_DSHARP2}.

Two studies \citep{Rafikov2017, Ansdell2018} used a different approach to derive $\alpha$ values for disks in
Lupus. These works combined measured accretion rates and dust and/or gas disk sizes of sources in the region to
estimate their $\alpha$ values, both obtaining a wide range of values ranging from $10^{-4}$ to 0.4. In both
cases, no correlation was found between global disk parameters (e.g., $M_{\rm disk}$, $R_{\rm disk}$, or surface density
profiles) and the derived viscosities, which could be due to the moderate sample sizes ($\sim$25
sources). \citet{Rafikov2017} proposed that this lack of correlation could also be explained if winds, and not
viscosity, are driving disk evolution.

To our knowledge, this is the first study that models several sources using the $\alpha$-disk prescription and
a Bayesian approach, and the obtained posteriors for $\alpha$ offer an indirect look at viscosity in disks.
Our results show that $\alpha$ estimates derived from SEDs with these type of models are, in many cases, at
least one order of magnitude higher than those suggested by recent turbulence measurements using ALMA, as well
as by the large number of multi-ringed disks (if they are to be explained by resonances).  Additionally, we
find significant levels of settling in our modeling, which should imply that turbulence levels in disks are
generally low\footnote{The dust settling prescription in the DIAD models is independent of the $\alpha$
  value.}. We also find no significant correlation between $\alpha$ and dust settling in our results (Pearson
correlation coefficient $r$=0.18). The need for high $\alpha$ values in our modeling, in contrast with the new
observational estimates, together with the required high levels of dust settling, suggests that the $\alpha$-disk
prescription may not fully explain the surface density profile in protoplanetary disks, and physical
mechanisms other than viscosity play an important role in the angular momentum transport in the disk.
  
We note that our model assumes a constant value of $\alpha$ in the disk.  In principle, the discrepancy
between our results and the recent estimates of low viscosity values in disks could be reduced with a
radially-changing $\alpha$, with lower values at radii $>$50 au (the regions probed by the observational
studies that have constrained this parameter), and an increasing $\alpha$ toward inner radii. However, such a
radial dependence would be contrary to theoretical expectations of MRI-induced turbulence: The decrease in
optical depth in the outer regions of disks (beyond the dead-zone) yields higher ionization fractions, and
should produce a radial increase of $\alpha$ \citep[e.g.,][]{Liu2018}. Although recent works have shown that
cosmic rays accelerated by accretion shocks could increase the ionization fraction in the inner disk, this
would mostly affect the upper layers in the inner regions ($<10$\,au) only \citep{Offner2019}. Likewise,
accretion is known to be highly variable in young stellar objects \citep[e.g.,][]{Robinson2019}, which
suggests that the mass accretion rate may not be constant throughout the disk and so we cannot rule out that
some radial variation in $\dot{M}$ currently exists in these disks.

If viscosity in disks is generally low as suggested by recent observations, the outer regions of large disks
may not have reached a steady state at the age of Taurus given the increasing viscous timescales at larger
radii. Therefore, our derived $\alpha$ values for disks with large radii might not be accurate estimates of
the true viscosity in these systems. However, most of the high $\alpha$ values in our analysis appear for disks
with derived small radii: Seven out of the eight disks with derived radii $\lesssim$\,50\,au have
$\alpha \gtrsim 0.01$. Although there is limited information on gas radii for these sources, (sub)mm continuum
radii measurements exist for all these disks and in all cases show $R_{\rm dust}\lesssim$50\,au \citep{Pietu2014,
  Andrews2018, Bacciotti2018, Long2019}, in very good agreement with our results. Thus, these systems are
probably compact, and it is unlikely that their derived high viscosities are due to their outer regions not
being in a steady state.

The reason why the SED fits are able to constrain $\alpha$ in our model is because of the relation that arises
between this parameter and the mass accretion rate $\dot{M}$ within the $\alpha$-disk prescription. As
mentioned in Sect.~\ref{sec:results_alpha_mdot} (see also Fig.~\ref{fig:SED_parameters}), $\dot{M}$ has a
significant impact in the mid/far-IR emission by changing the accretion luminosity, and the $\alpha$ parameter
changes accordingly to produce a surface density profile that can reproduce the observed SED
\citep[$\Sigma \propto \dot{M}/\alpha$, e.g.,][]{DAlessio1998}. However, if other processes affect the
transport of angular momentum, the adopted relationship between these two quantities in the $\alpha$-disk
model may not be accurate. As an example, \citet{Manara2020} studied the mass accretion rates of disks
in the 5-10\,Myr-old Upper Scorpius association and found values similar to those of sources in the younger
Lupus and Chamaeleon~I regions ($\sim$1-3\,Myr), in contrast with theoretical expectations of viscous evolution
(which predicts decreasing accretion rates with age).  Future observations of turbulence levels (both in the inner regions with
high-resolution near-IR spectroscopy and in the outer radii with ALMA), disk sizes, mass accretion rates, and
disk masses for large samples of disks, together with a better theoretical understanding of disk winds, will
be crucial in determining the underlying mechanism behind angular momentum transport in these systems.

Finally, we note that wind-dominated models can produce surface density profiles similar to that of the $\alpha$-disk
\citep[e.g.,][]{Bai2016}. Observationally, surface density profiles are broadly consistent with the
$\Sigma(r) \propto r^{-1}$ dependence in the $\alpha$-disk model, and thus the radial surface density profiles
from DIAD are likely a good approximation of the real ones. If winds are in fact responsible for the structure
of disks, then the $\alpha$ parameter in the DIAD models does not have a physical interpretation and simply
acts as a scaling factor of the surface density ($\Sigma \propto \alpha^{-1}$).  Likewise, if the large disks
have not reached a steady-state yet as discussed before, the derived viscosities would not have a physical
interpretation, but would only act to produce surface density profiles that fit the observed SEDs.  Regardless
of this, estimates of other parameters such as disk masses should still be reliable, since the dust emission
probed by the SED is largely sensitive to the dust surface density.

\subsection{Improved disk masses and a possible solution to the missing mass problem}

Because of its relevance for planet formation theories, the mass of protoplanetary disks is probably the most
studied of their parameters. Recent (sub)mm surveys have yielded censuses of disk masses in different
star-forming regions \citep[e.g.,][]{Andrews2013, Ansdell2016, Pascucci2016, Cieza2019}. When combined with
the also increasing statistics of known exoplanets, however, protoplanetary disks do not appear be massive
enough to explain the observed planetary systems \citep[e.g.,][]{Pascucci2016, Manara2018}. Proposed
explanations for this discrepancy include underestimated disk masses (for various reasons), very early planet
formation, and/or continued mass accretion from the ISM onto the disk \citep[see e.g.,][]{Manara2018}.

The standard approach to derive disk masses from (sub)mm fluxes is to assume that the disk emission at those
wavelengths is isothermal and optically thin. In that case, the total continuum emission is directly
proportional to the mass of dust in the disk \citep[e.g., see][]{Beckwith1990}:

\begin{equation}
M_{\rm dust} = \frac{F_\nu \, d^2}{\kappa_\nu  \, B_\nu(T_{\rm dust})},
\end{equation}

where $M_{\rm dust}$ is the mass of dust in the disk, $F_\nu$ is the flux at the observed frequency $\nu$, $d$
is the distance to the source, $\kappa_\nu$ is the dust opacity at the frequency $\nu$, and
$B_\nu(T_{\rm dust})$ is the blackbody radiation at the corresponding frequency and dust temperature $T_{\rm
  dust}$. However, despite its usefulness, mass estimates obtained with this method require that one adopts a
grain opacity (which is largely uncertain) and a single dust temperature \citep[the appropriate temperature
value and its dependence with stellar luminosity are also uncertain, e.g.,][]{Andrews2013,
  Pascucci2016}. Moreover, \citet{Tripathi2017} and \citet{Andrews2018} found a strong correlation between
disk sizes and their millimeter luminosity, and \citet{Zhu2019} showed that neglecting dust scattering can
result in significant underestimates of the disk optical depth. These results indicate that the optically thin
assumption may not hold at $\lambda\lesssim$1\,mm.

\begin{figure}
  \centering
  \includegraphics[width=\hsize]{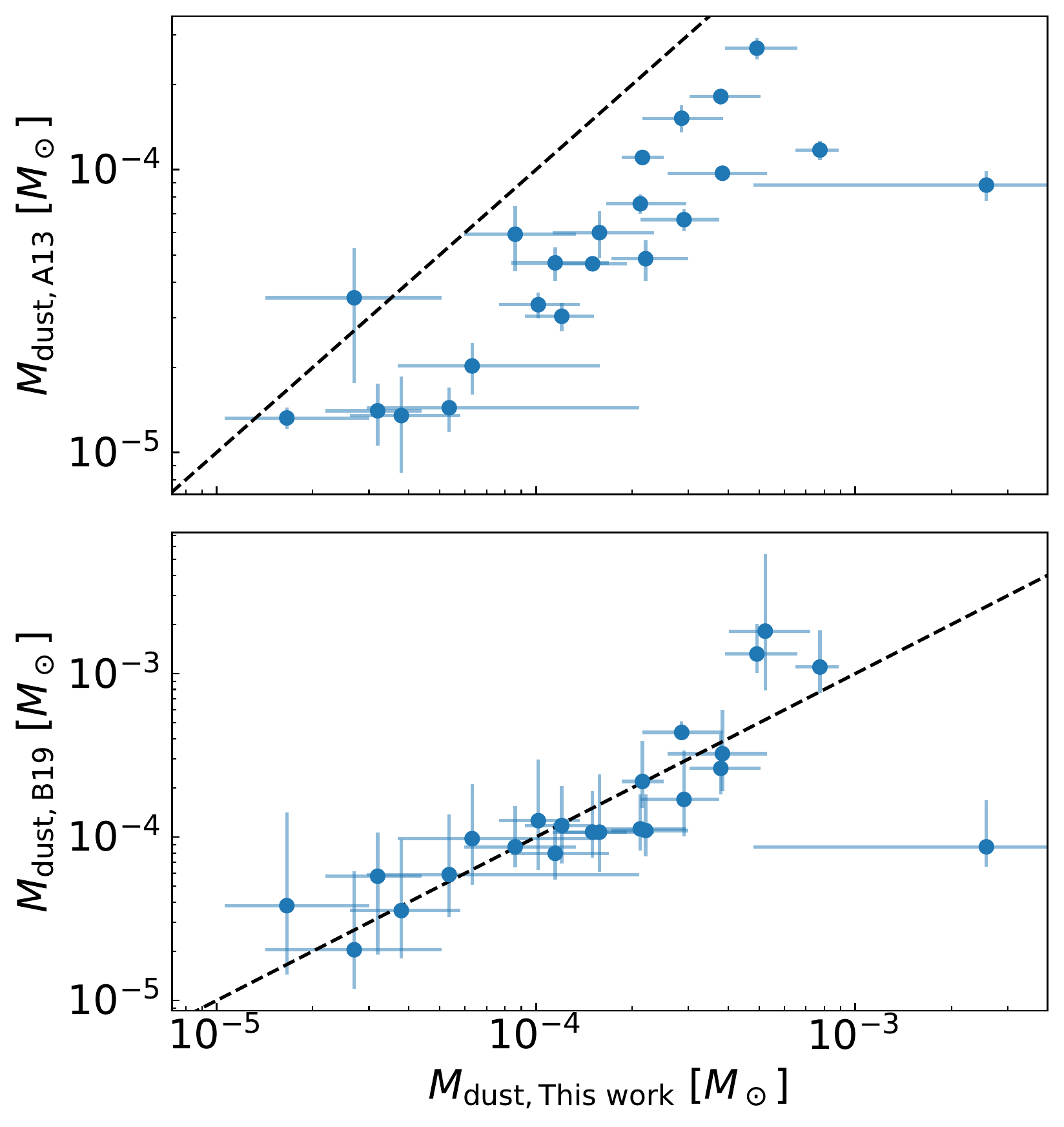}
  \caption{Comparison of the disk dust masses derived in this study with those from \citet{Andrews2013} based
    on 1.3\,mm fluxes (top) and from \citet{Ballering2019} using radiative transfer models (bottom). The one-to-one
    relation is shown as a dashed line.}\label{fig:mdisk_comparison}
\end{figure}

On the other hand, because DIAD solves the temperature and density structure of the disk, the dust masses
derived in this study do not require the assumptions of isothermal and/or optically thin emission. The adopted
Bayesian approach also accounts for the effect that other parameters have in the uncertainties of disk
masses. In Fig.~\ref{fig:mdisk_comparison} we compare our disk dust mass estimates with those of
\citet{Andrews2013} based on 1.3\,mm fluxes (after correcting them for the new \emph{Gaia} DR2 distances), as well as
with those of \citet{Ballering2019} from radiative transfer models. Our mass values are systematically higher
than those from \citet{Andrews2013}, with the median being $\sim$3 times higher. In contrast, the comparison with
\citet{Ballering2019} yields much more compatible results. These authors computed two sets of dust masses
using radiative transfer models: Their estimates enforcing a $L_{\rm mm}-R_{\rm disk}$ relation are still
$\sim$25\,\% lower than ours, but their results when not including that relation show a very good agreement with ours
(median difference of $\sim$2\,\%). These comparisons clearly indicate that, as discussed in
\citet{Ballering2019}, dust mass estimates from (sub)mm fluxes are very likely underestimated by factors of a
few as a considerable fraction of the emission at these wavelengths remains optically thick. A similar
conclusion was also reached by \citet{Woitke2019} when modeling SEDs, gas lines, and spatially resolved
observations of disks. These findings imply that results based on disk dust masses measured from (sub)mm
fluxes alone should be considered with caution, because they may be reflecting properties other than mass. A
full radiative transfer modeling considering the physical structure of the disk is therefore needed to obtain
reliable mass estimates.

The higher disk masses found when using full radiative transfer models alleviate the apparent discrepancy
between the measured masses of exoplanetary systems and those of protoplanetary disks. \citet{Manara2018}
compared dust masses for disks in Lupus and Chamaeleon~I derived from ALMA observations at 890\,$\mu$m with
those of core masses in planets and planetary systems, and found that the core masses were 3-5 times more
massive (median value) than the dust disks. Since the disk masses measured in this work are $\sim$3 times
higher than those measured with (sub)mm data alone, this could explain the aforementioned
discrepancy. Nevertheless, we caution that our sample is still small in size and mostly includes disks without
substructures, so we cannot exclude the possibility that a substantial mass reservoir is already locked in
planetesimals in these disks.

Finally, we also searched for the $M_{*}-M_{\rm disk}$ relation, which was identified in the past using disk
masses derived from (sub)mm fluxes \citep[e.g.,][]{Andrews2013, Pascucci2016}. Recently, \citet{Ballering2019}
found tentative evidence for such a correlation (at a 2-$\sigma$ confidence level) using their mass estimates
from SED fitting when enforcing a $L_{\rm mm}-R_{\rm disk}$. However, given the relatively low significance of the
correlation and that the influence of the $L_{\rm mm}-R_{\rm disk}$ relation in the $\chi^2$ value had to be
introduced in an ad hoc manner, further evidence is required to unambiguously confirm the $M_{*}-M_{\rm disk}$
relation.  Using our results and the approach in \citet{Andrews2013} based on the Bayesian analysis of
\citet{Kelly2007}, we found no evidence of a correlation between these two quantities (see also
Fig.~\ref{fig:Mdot_alpha_Mdisk_correlations}), possibly due to the small size of our sample. Nevertheless, we
note that if (sub)mm fluxes are in fact optically thick, then this previously claimed correlation may not
reflect a true connection between $M_{*}$ and $M_{\rm disk}$, and may instead be probing changes the disk
structure or dust properties. Disk mass estimates from radiative transfer models accounting for optical depth
effects are needed for larger samples of disks to revisit this correlation.

\section{Summary}\label{sec:summary}

We have used the physically motivated DIAD models to fit the complete SEDs of 23 protoplanetary disks in the
Taurus-Auriga star-forming region using a Bayesian framework. This analysis was possible thanks to the vast
increase in computational performance achieved by combining these models with artificial neural networks. Our
main results are:

\begin{itemize}

  \item Several of the modeled disks require high viscosities and accretion rates, in contradiction with recent
    observational estimates of low turbulence in disks. We also derive high levels of dust settling in the
    sample, which the DIAD models treat independently of the viscosity and can thus be used as
    another indirect indicator of low turbulence in disks. Combined with theoretical predictions of low
    ionization levels in most regions in the disk and other observational results, these findings support the idea
    that disk winds could play an important role in angular momentum transport in disks.
  
  \item We find evidence of a population of very compact disks, also in agreement with recent findings of ALMA
  high-resolution observations.

\item The posterior distributions of grain sizes in the disk midplane are largely unconstrained, with only mild
  evidence of large grains in disks. This suggests that little to no evidence about grain growth
    can be gained by analyzing SEDs or (sub)mm spectral indices alone.

  \item The derived disk masses are systematically higher than those obtained using the standard conversion
    from (sub)mm fluxes, probably because the emission at these wavelengths is still (partially) optically
    thick. Disk dust masses computed directly from (sub)mm fluxes should therefore be considered with
    caution. The higher disk masses derived in this study decrease previous tensions between disk mass
    measurements and those of exoplanetary systems.

\end{itemize}

\vspace{0.4cm}

\begin{acknowledgements}
  We thank the anonymous referee for their detailed revision of our work, which helped to improve the quality
  of this study. We also thank Connor Robinson for his work developing software to process the output of the
  DIAD models. AR acknowledges financial support from the European Southern Observatory (ESO). AR and CCE
  acknowledge support from the National Science Foundation under CAREER grant AST-1455042.  The data
  processing and analysis in this manuscript made extensive use of the following open source software:
  \texttt{Matplotlib} \citep{Matplotlib}, \texttt{SciPy} \citep{Scipy}, \texttt{Numpy} \citep{Scipy},
  \texttt{pandas} \citep{pandas}, \texttt{Astropy} \citep{Astropy2013}, and \texttt{Scikit-learn}
  \citep{scikit-learn}. We thank the developers of these software for enabling this and many other studies.
  This paper utilizes the D'Alessio Irradiated Accretion Disk (DIAD) code. We wish to recognize the work of
  Paola D'Alessio, who passed away in 2013. Her legacy and pioneering work live on through her substantial
  contributions to the field.
\end{acknowledgements}


\bibliographystyle{aa}
\bibliography{biblio}

\begin{appendix}

\section{Effect of different parameters on the SED} \label{appendix:parameters}

\begin{figure*}[h]
  \centering
  \includegraphics[width=0.33\hsize]{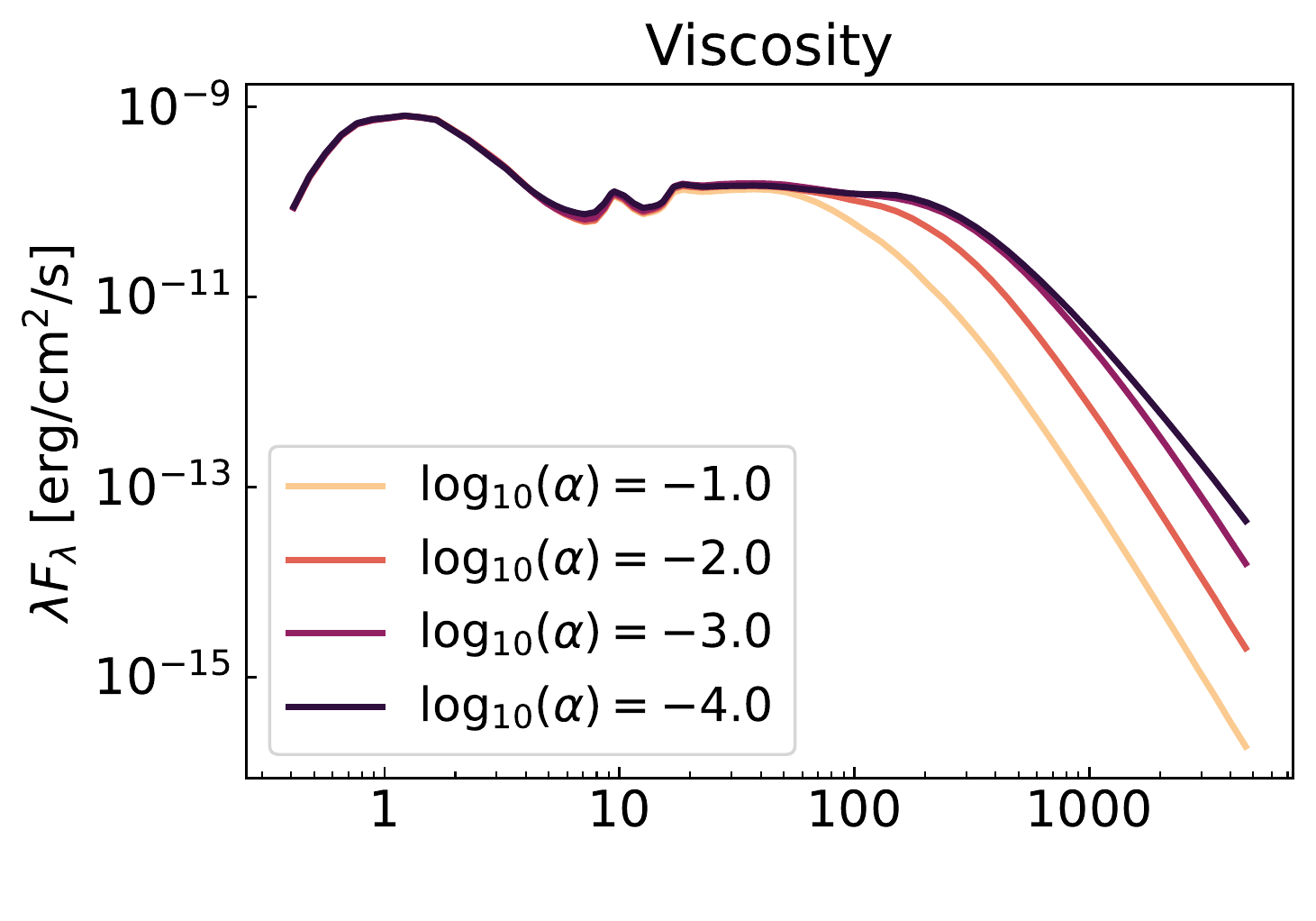}\hfill
  \includegraphics[width=0.33\hsize]{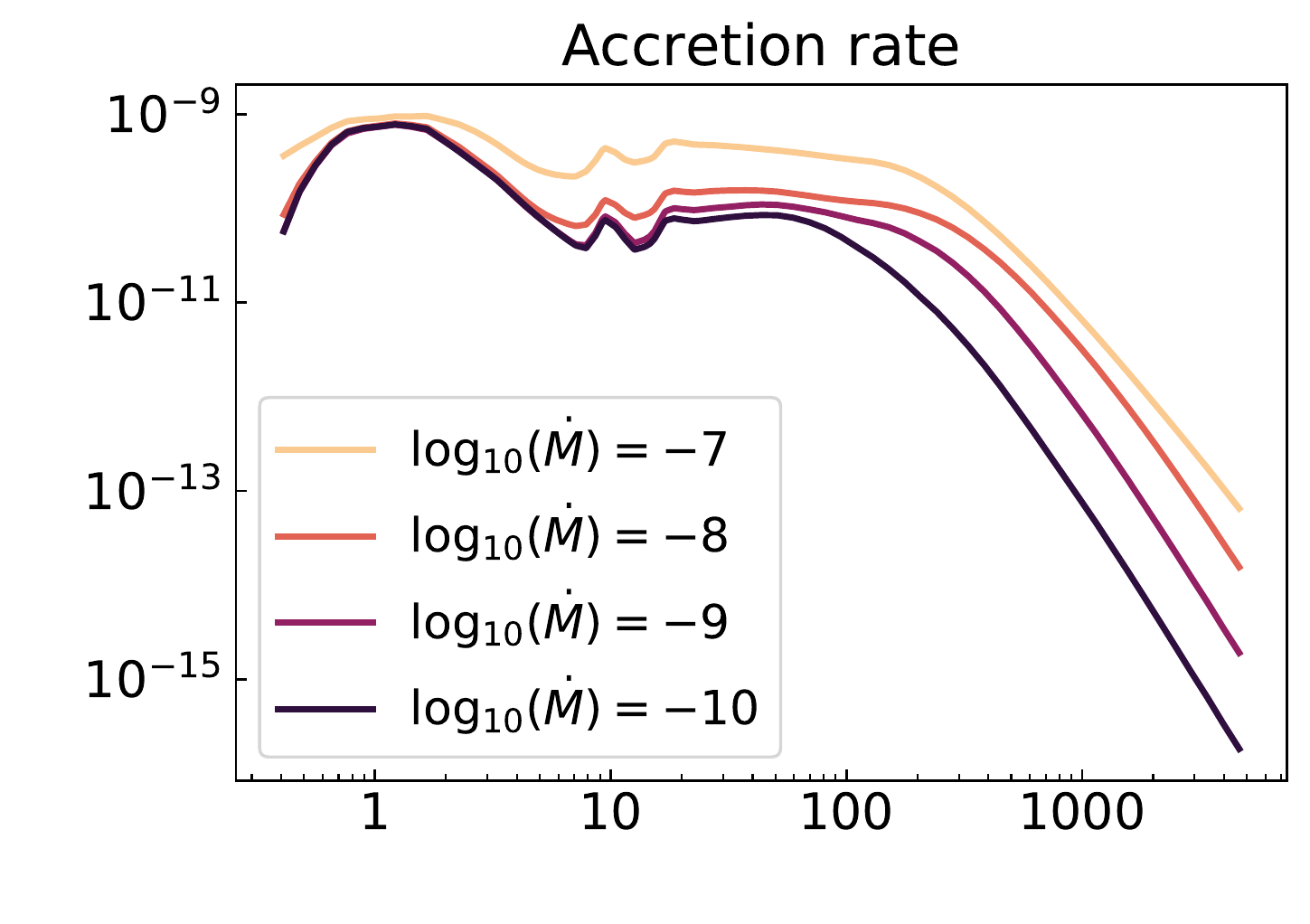}\hfill
  \includegraphics[width=0.33\hsize]{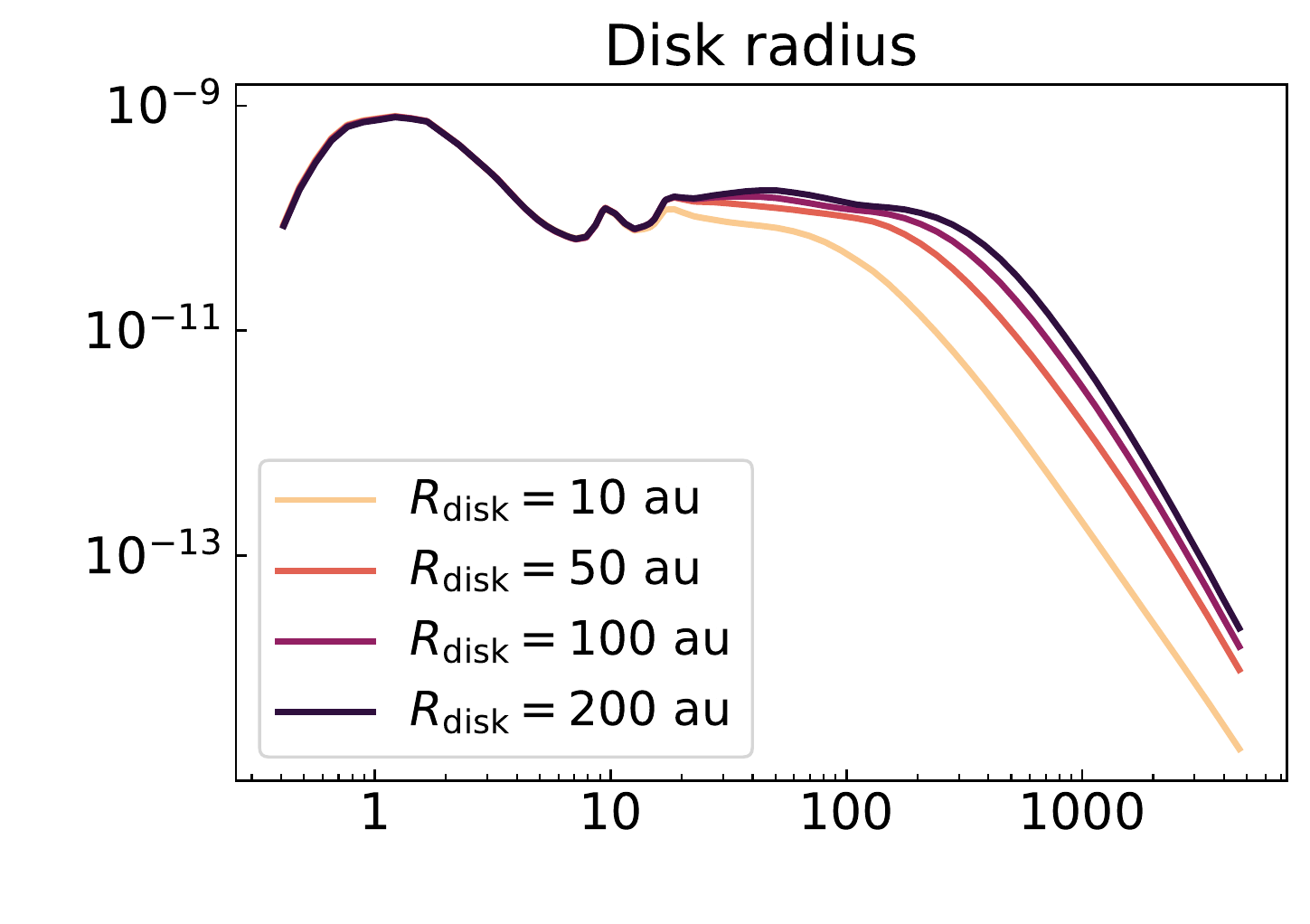}\\
  \includegraphics[width=0.33\hsize]{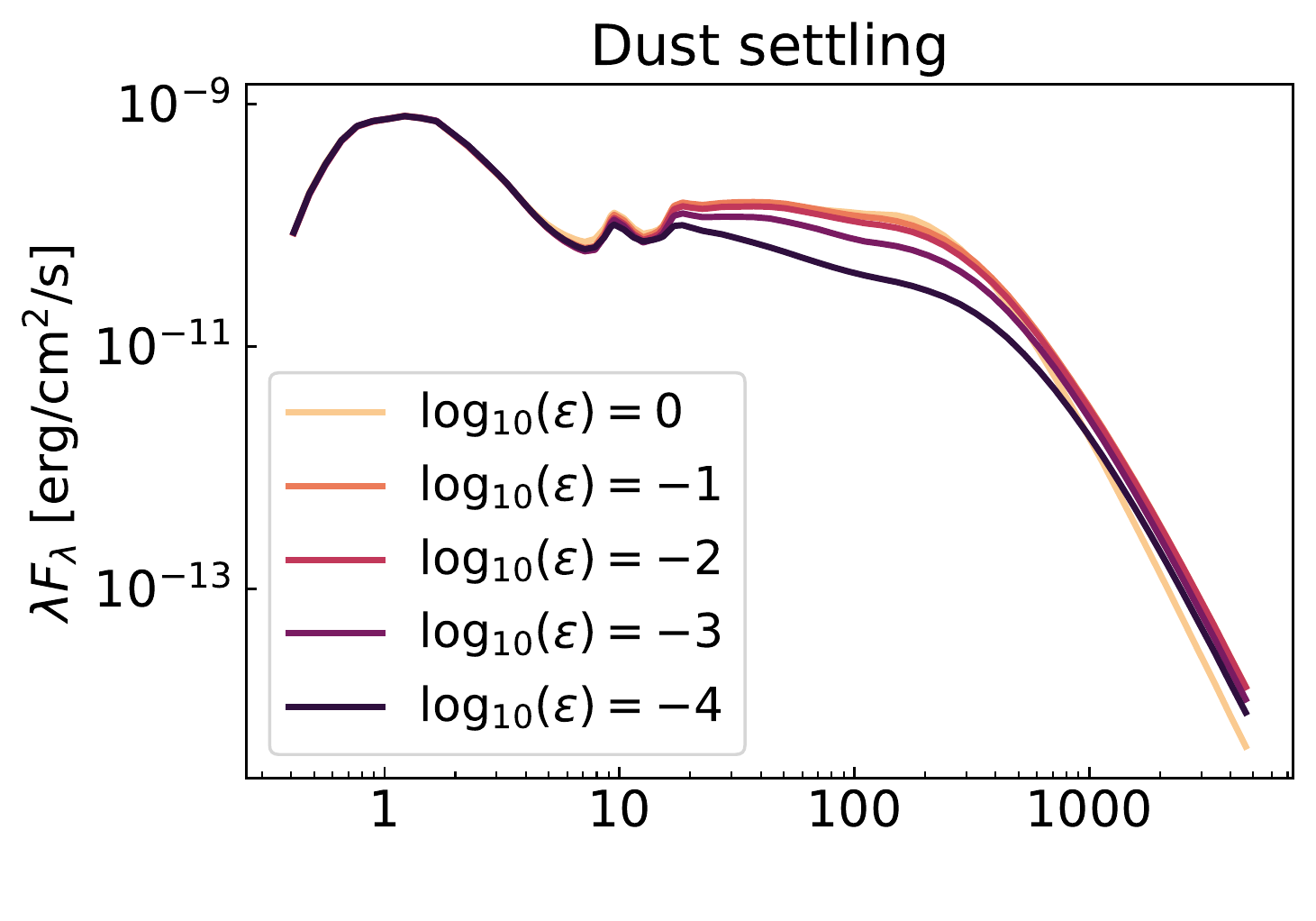}\hfill
  \includegraphics[width=0.33\hsize]{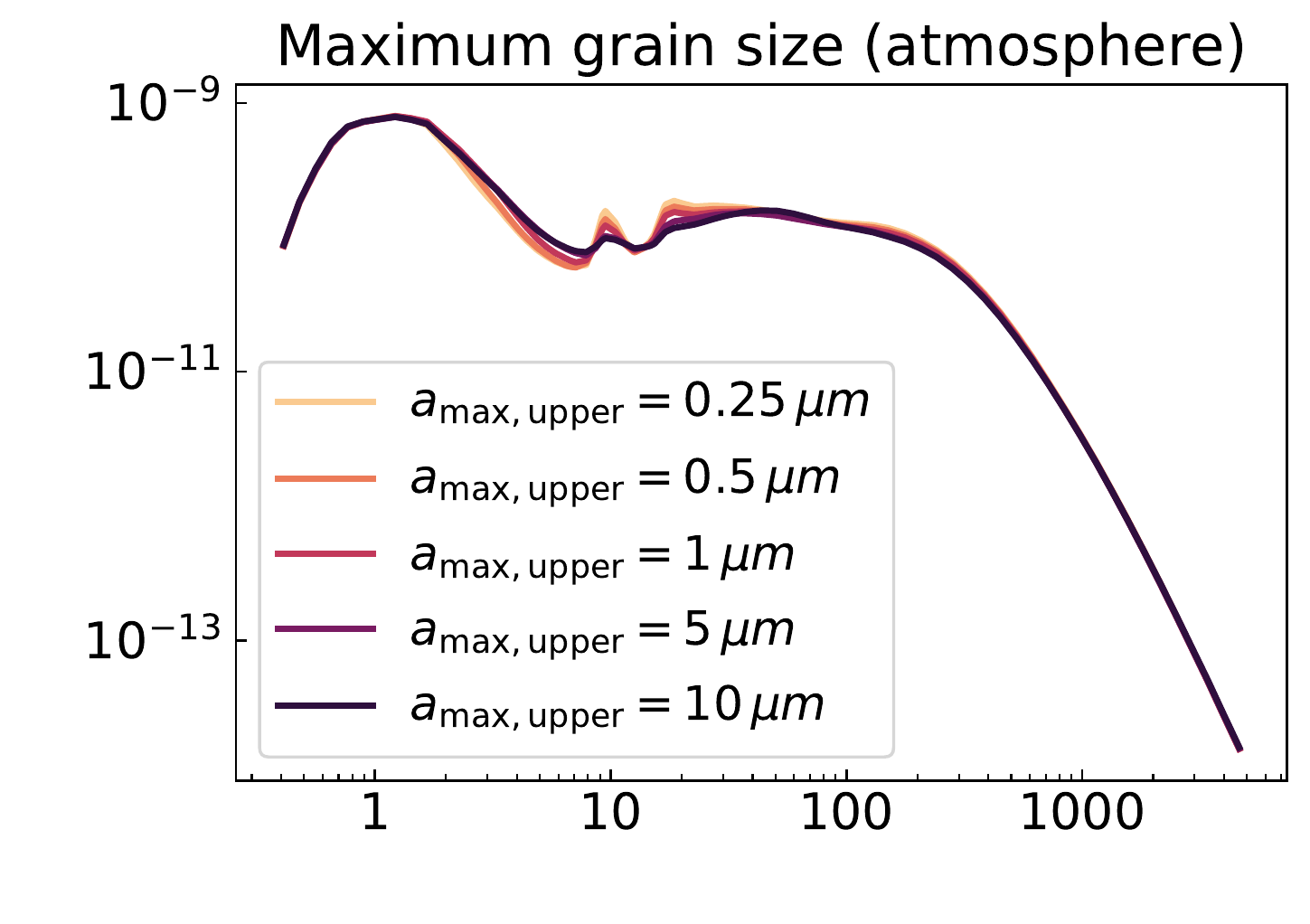}\hfill
  \includegraphics[width=0.33\hsize]{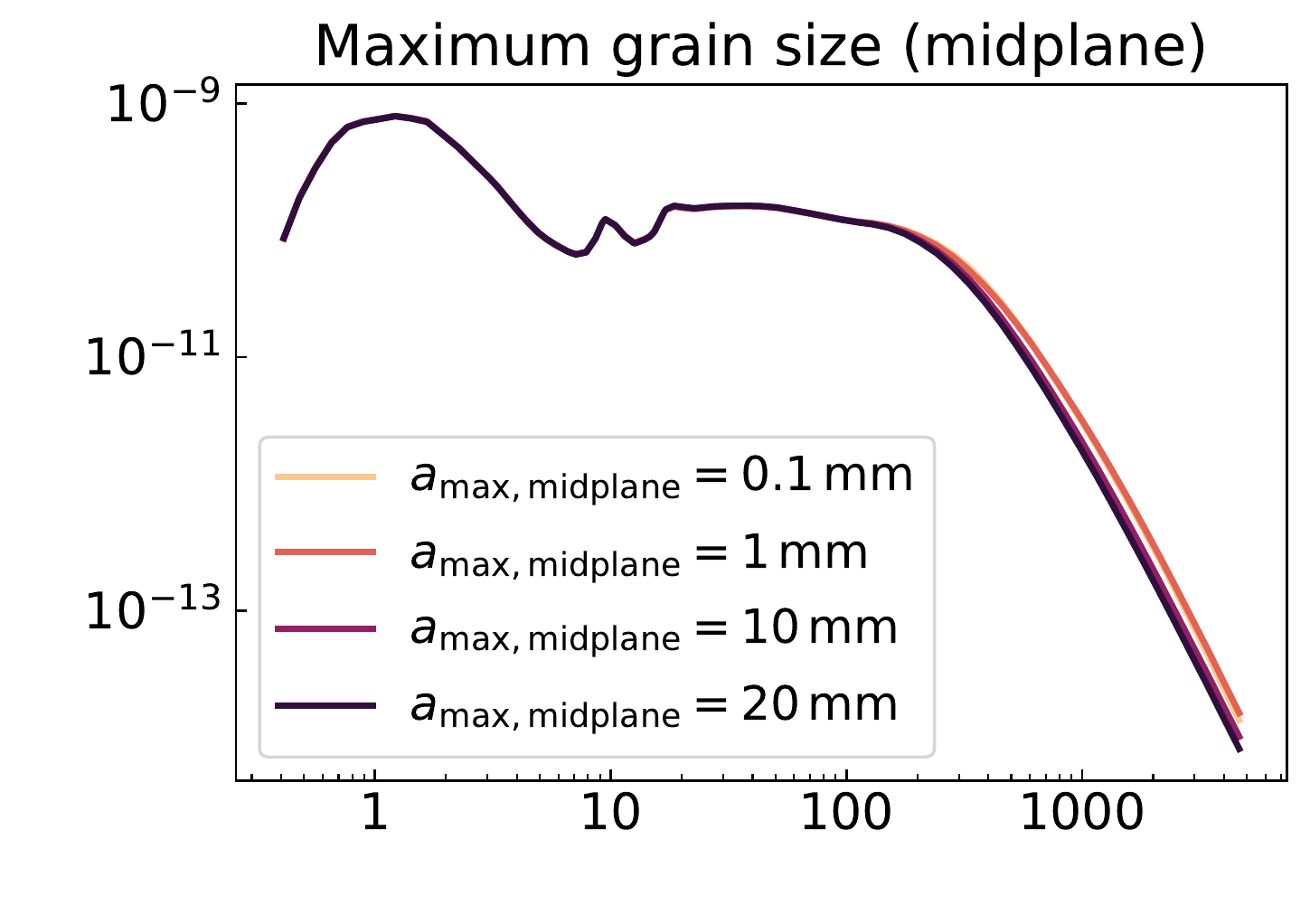}\\  
  \includegraphics[width=0.33\hsize]{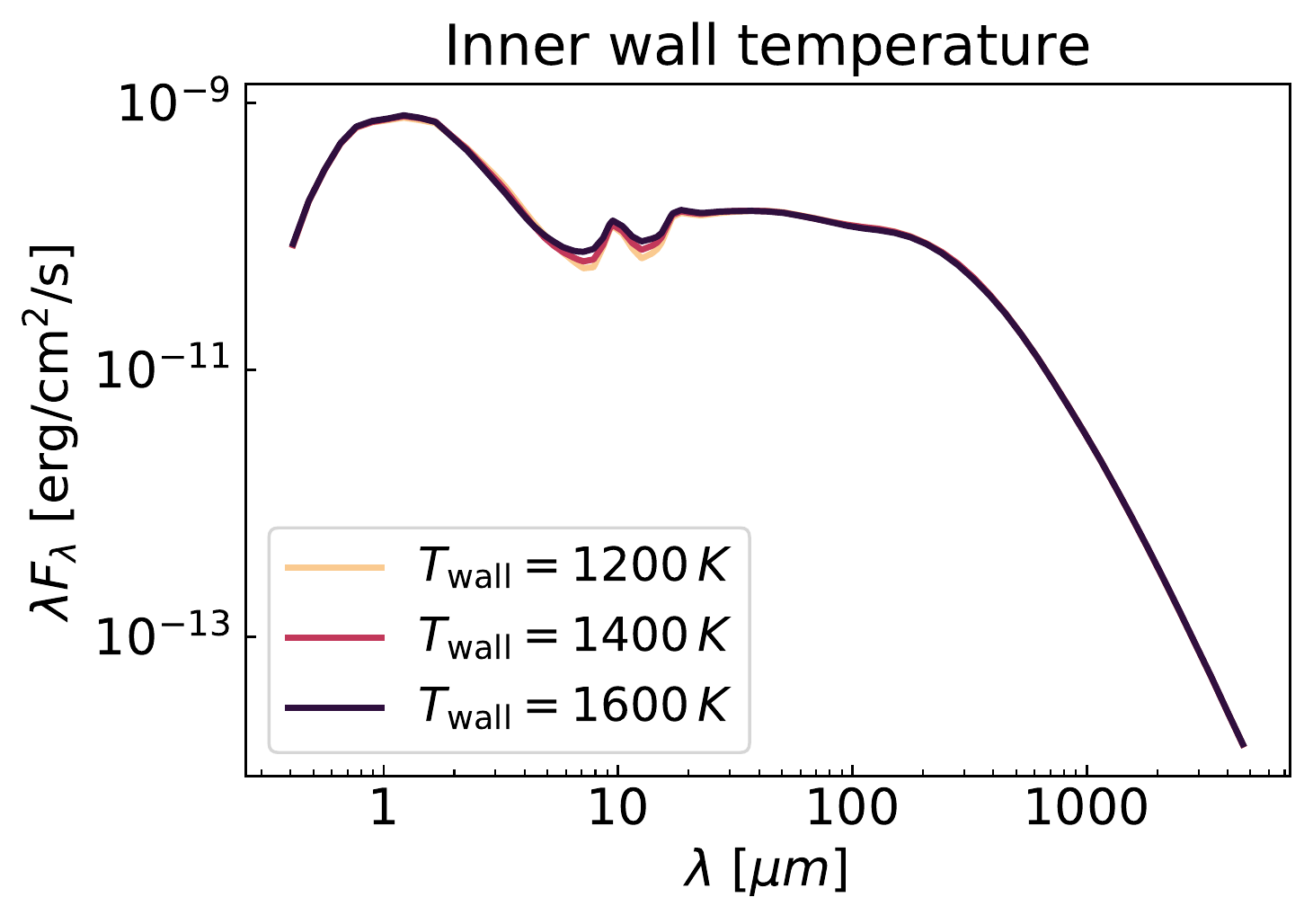}\hfill
  \includegraphics[width=0.33\hsize]{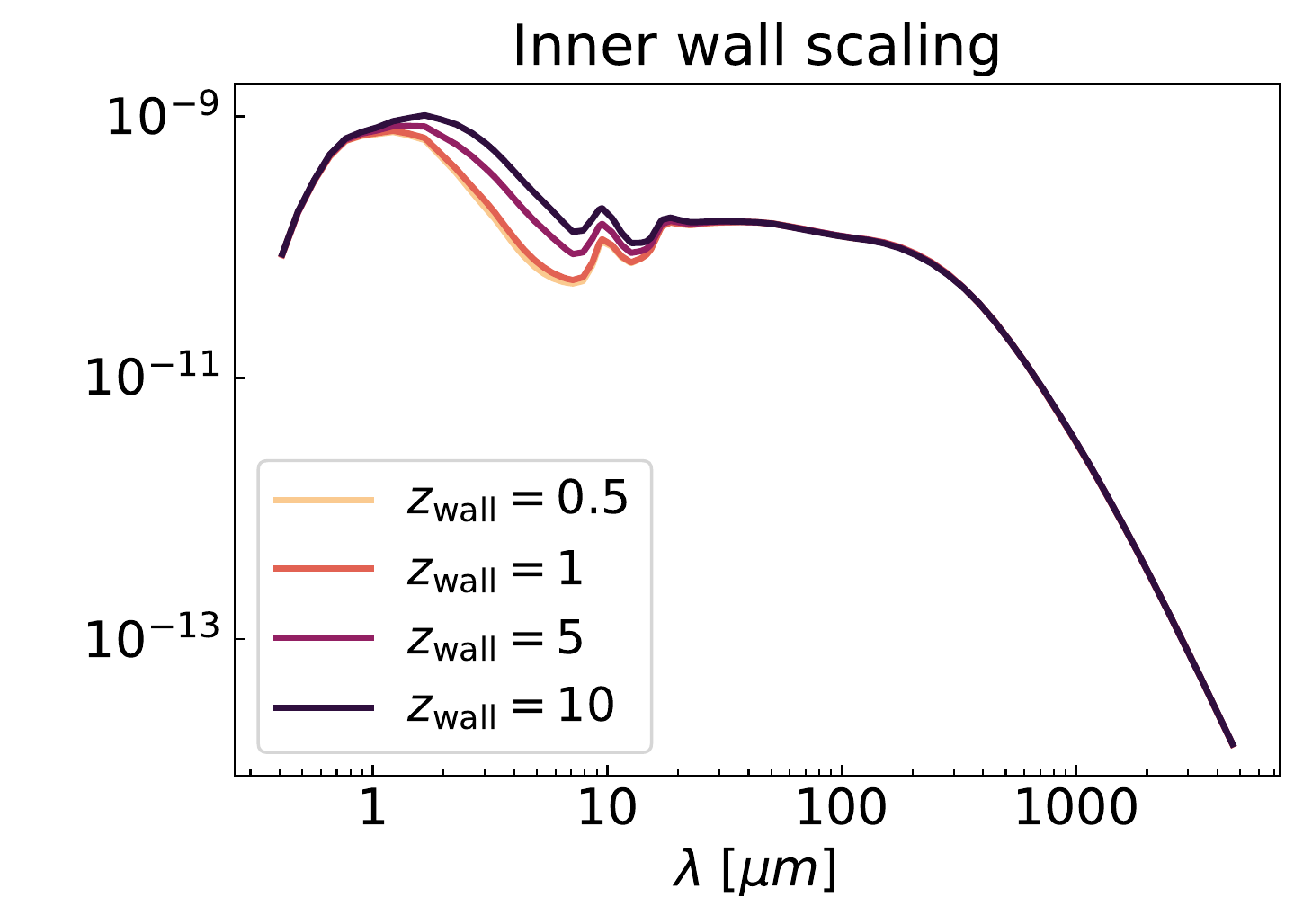}\hfill
  \includegraphics[width=0.33\hsize]{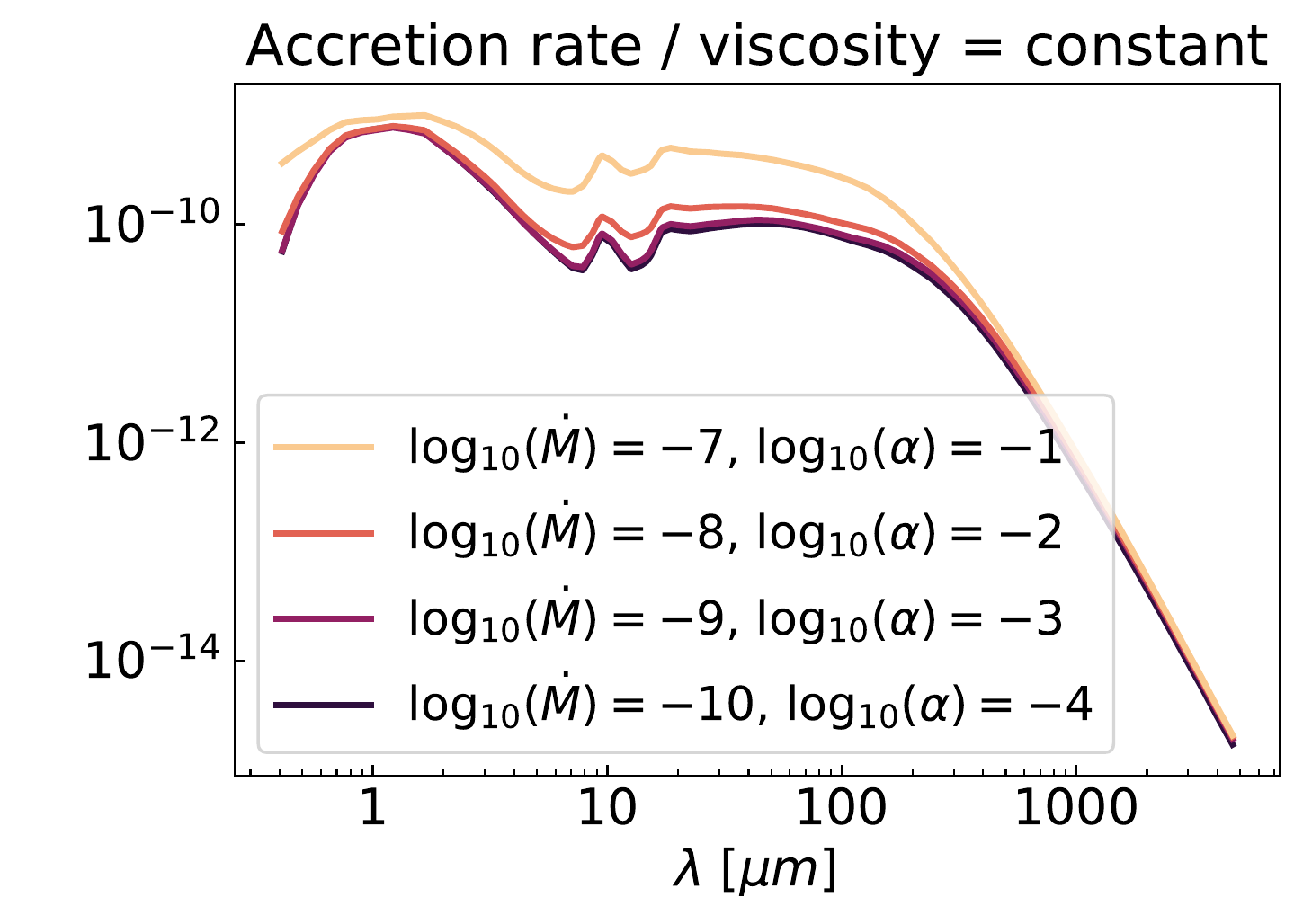}\\
  \caption{Effect of changing different parameters on the SED. Parameters are modified one at a time, the
    others are fixed to the values described in Sect.~\ref{appendix:parameters}. The bottom right panel also shows
    the change in the SED produced by varying both the viscosity and accretion rate while keeping their ratio
    constant.}\label{fig:SED_parameters}
\end{figure*}

Here we show the impact of different parameters on the SEDs from the DIAD models. The default disk has the
following parameters: age=2\,Myr and $M_*$=0.5\,M$_\odot$ (corresponding to $T_*=3725$\,K and
$R_*$=1.5\,R$_\odot$ according to the MIST isochrones used), viscosity parameter $\alpha=10^{-3}$, mass accretion rate
$\dot{M}=10^{-8} M_{\odot}$/yr, dust settling $\epsilon=10^{-1}$, disk radius $R_{\rm disk}=$100\,AU,
inner wall temperature and scaling $T_{\rm wall}$=1400\,K and $z_{\rm wall}$=2, grain sizes in the upper
layers and midplane $a_{\rm max, upper}=1\,\mu$m and $a_{\rm max, midplane}$=1\,mm, and distance
$d=100$\,pc. Figure.~\ref{fig:SED_parameters} shows how modifying these parameters affects the corresponding
SED. We also show the effect of varying both $\alpha$ and $\dot{M}$ while keeping their ratio constant (at
$10^{-6} M_{\odot}$/yr) to separate the effects of the individual parameters (accretion
luminosity and viscous heating) from their effect on the dust surface density.

\section{The artificial neural network} \label{appendix:ANN}

\subsection{Architecture of the artificial neural networks}

\begin{figure*}[h]
  \centering
  \includegraphics[width=\hsize]{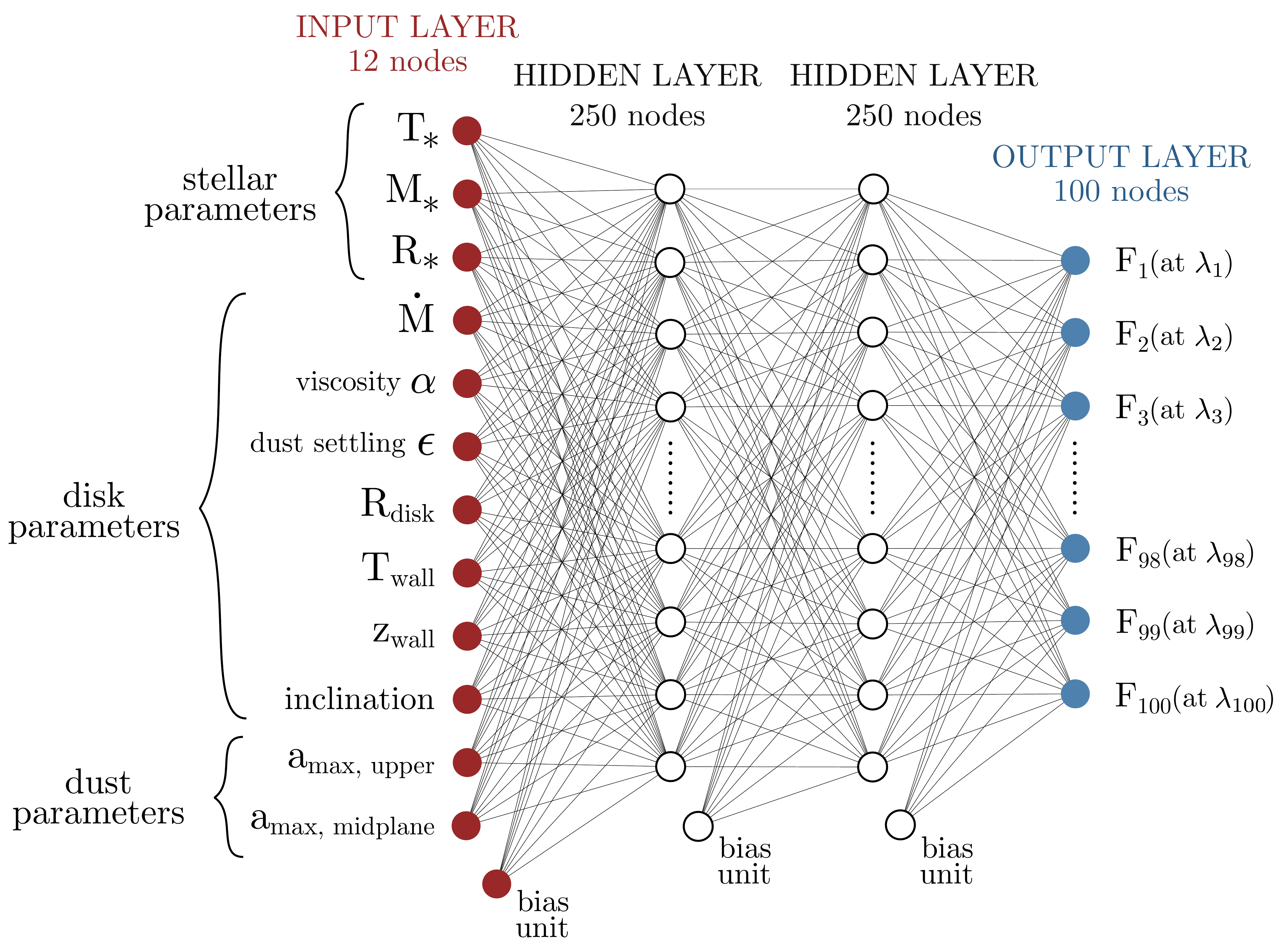}
  \caption{Architecture of the ANN used to mimic the DIAD models. The input layer (red circles) contains 12
    nodes, one per free parameter considered in DIAD. The input layer is connected to two hidden layers (empty
    circles), with 250 nodes each. Finally, the output layer contains 100 nodes, each being the flux at one of
    the considered 100 wavelengths. The bias units of the input and hidden layers are also
    shown.}\label{fig:ANN_architecture}
\end{figure*}

\begin{figure*}[h]
  \centering
  \includegraphics[width=\hsize]{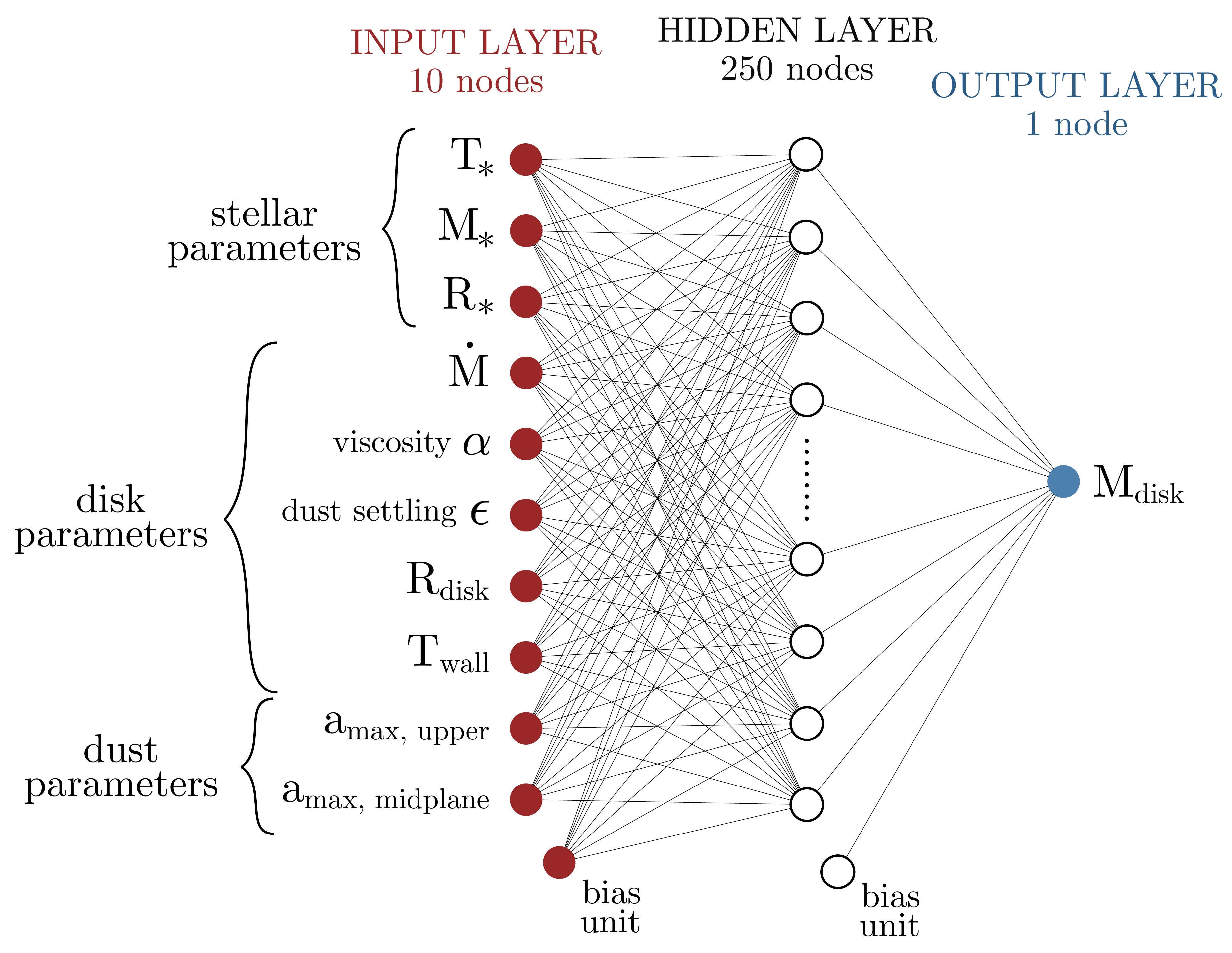}
  \caption{Architecture of the ANN used to compute disk masses based on the DIAD
    input parameters. In this case, the disk mass depends on 10 free parameters, which is the number of nodes
    in the input layer (red circles). The input layer is connected to a single hidden layer (empty circles),
    with 250 nodes. The output layer has a single node (the disk mass). The bias units of
    the input and hidden layer are also shown.}\label{fig:ANN_diskmass_architecture}
\end{figure*}

ANNs can have different architectures based on the arrangement of their layers. For our study, we have chosen
a feedforward configuration (i.e., the connections between the nodes do not go backward or form cycles).
Figs.~\ref{fig:ANN_architecture} and ~\ref{fig:ANN_diskmass_architecture} show the chosen architectures for
the ANNs used to estimate SEDs and disk masses, respectively. These configurations were selected by trial and
error by increasing the number of nodes in each layer until the validation error did not improve
significantly. Given that this is a regression problem, we used a rectified linear unit function for the
activation function.

\subsection{Generation of the training sample}\label{appendix:training_sample}

A sample of DIAD models is needed in order to train the ANN (i.e., to determine the correct weight for each
connection). This training sample should cover the desired range of the parameter space.  We chose to
distribute the DIAD models randomly (although not uniformly) in this parameter space instead of using a
grid. This guarantees that each model has a unique value of each parameter (except for grain sizes, which can
only have values from a discrete list, see below), maximizing the available information for training. The
models were computed as follows:

\begin{itemize}

\item Stellar temperature, radius, and mass ($T_*$, $R_*$, and $M_*$): These are the stellar parameters
  in the DIAD models, and are tightly connected through stellar evolution. We therefore relied on the MIST
  isochrones \citep{Dotter2016, Choi2016, Paxton2011, Paxton2013, Paxton2015} to avoid sampling non-physical
  combinations of these parameters. For this purpose, we randomly chose a value for $M_*$ between 0.1
  and 2.5\,$M_\odot$ and an age between 0.1 and 10 \,Myr. We then computed the corresponding $T_*$ and $R_*$
  according to MIST, and added Gaussian noise to these values (with a standard deviation of 250\,K and
  25\,\%, of the stellar radius, respectively). If the resulting values met 3000\,K $< T_* <$
  7000\,K, the combination of $T_*$, $R_*$, and $M_*$ was accepted. We note that the only effect of this process
  is to generate more models in areas consistent with stellar evolution, thus improving the accuracy of the
  ANN in these regions.

\item Mass accretion rate ($\dot{M}$): Typical accretion rates for T Tauri stars are between
  $10^{-7}-10^{-9}$\,$M_\odot$/yr, although both higher and lower values
  are also found \citep[e.g.,][]{Ingleby2013}. Therefore, we selected random $\dot{M}$ values in logarithmic
  scale, ranging from $10^{-10}$ to $10^{-6.5} M_\odot$/yr.

\item Disk viscosity parameter ($\alpha$): This parameter was explored in logarithmic scale from $10^{-4}$ to 0.1.

\item Dust settling ($\epsilon$): We explored this parameter in logarithmic scale from $10^{-4}$ (very settled disk) to
  0 (no settling). Values below this range have no impact on the SED (at that value the upper layers of the
  disk are mostly devoid of dust), and $\epsilon>1$ values have no physical meaning (they correspond to disks with
  more dust in the upper layers than in the midplane).

\item Disk radius ($R_{\rm disk}$): This parameter was explored uniformly from 10\,au to 300\,au, a broad
  range covering typical disk size estimates from resolved observations.
 
\item The dust sublimation temperature ($T_{\rm wall}$): A commonly adopted sublimation temperature is
  1400\,K. Therefore, we uniformly explored temperatures in a 500\,K range centered around this value (from 1150 to
  1650\,K).

\item The height of the disk wall ($z_{\rm wall}$): The inner wall may be curved and/or puffed-up due to the vertical
  structure of the disk \citep[see e.g.,][]{Natta2001, Dullemond2010}, which increases its surface area. Since the true
  shape of the wall is still uncertain, we scaled the wall height by a factor of $z_{\rm wall}$ times
  the local hydrostatic scale height to account for this uncertainty.  $z_{\rm wall}$ was uniformly probed
  from 0.5 to 15.

\item Inclination ($i$): Following the observational distribution of inclinations, $i$ was sampled uniformly in
  $\cos{(i)}$ space. For inclinations above 70\,\degr, self-extinction from the disk becomes important and
  small variations of the inclination produce drastic changes in the photospheric emission. This adds a
  significant complication for training an ANN to accurately predict SEDs with high inclinations. Therefore,
  we sampled the inclination using the observational distribution from 0 to 70\,\degr, and removed sources
  with $i \geq 70$ from our sample (see Sect.~\ref{sec:sample}).

\item Maximum grain size in the disk atmosphere ($a_{\rm max, upper}$): It is usually assumed that the size
  distribution of grains in the upper layers of the disk follows that of the ISM, with $a_{\rm max, upper}$
  values of 0.25\,$\mu$m. We thus randomly sampled from a list of discrete values of $a_{\rm max, upper}$ from
  0.25\,$\mu$m to 10\,$\mu$m to account for the possibility of larger grains in the disk atmosphere that have
  not settled yet. The list of explored values starts at 0.25\,$\mu$m and increases in steps of 0.25\,$\mu$m
  up to 2.5\,$\mu$m, then includes 3, 4, 5, and 10\,$\mu$m.

\item Maximum grain size in the disk midplane ($a_{\rm max, midplane}$): Several studies have found evidence of
  grain growth in protoplanetary disks, with most sources harboring dust grains of a few mm-cm
  \citep[e.g.,][]{DAlessio2006, Ricci2010_Ophiuchus, Ricci2010_Taurus, Ribas2017}. Therefore, we considered
  $a_{\rm max, midplane}$ values from 100\,$\mu$m to 2\,cm to cover this range. This parameter was also explored in
  a discrete manner by randomly choosing from a list of discrete values, from 100\,$\mu$m to 1\,mm in steps of
  100\,$\mu$m, and from 1\,mm to 2\,cm in steps of 1\,mm.

\end{itemize}

\subsection{ANN training, committees, and accuracies}\label{appendix:ANN_details}

As mentioned in Section~\ref{subsec:ann}, we used the multilayer perceptron regressor in the scikit-learn Python
package \citep{scikit-learn}. Some input parameters of the DIAD models, as well as their outputs, cover
several orders of magnitude, and the accretion rate, the alpha viscosity parameter, the dust settling, the SED
fluxes, and the disk mass were all trained in logarithmic space. The range of each input and output parameters were
re-normalized between 0 and 1 before training.

\begin{figure*}[h]
  \centering
  \includegraphics[width=0.7\hsize]{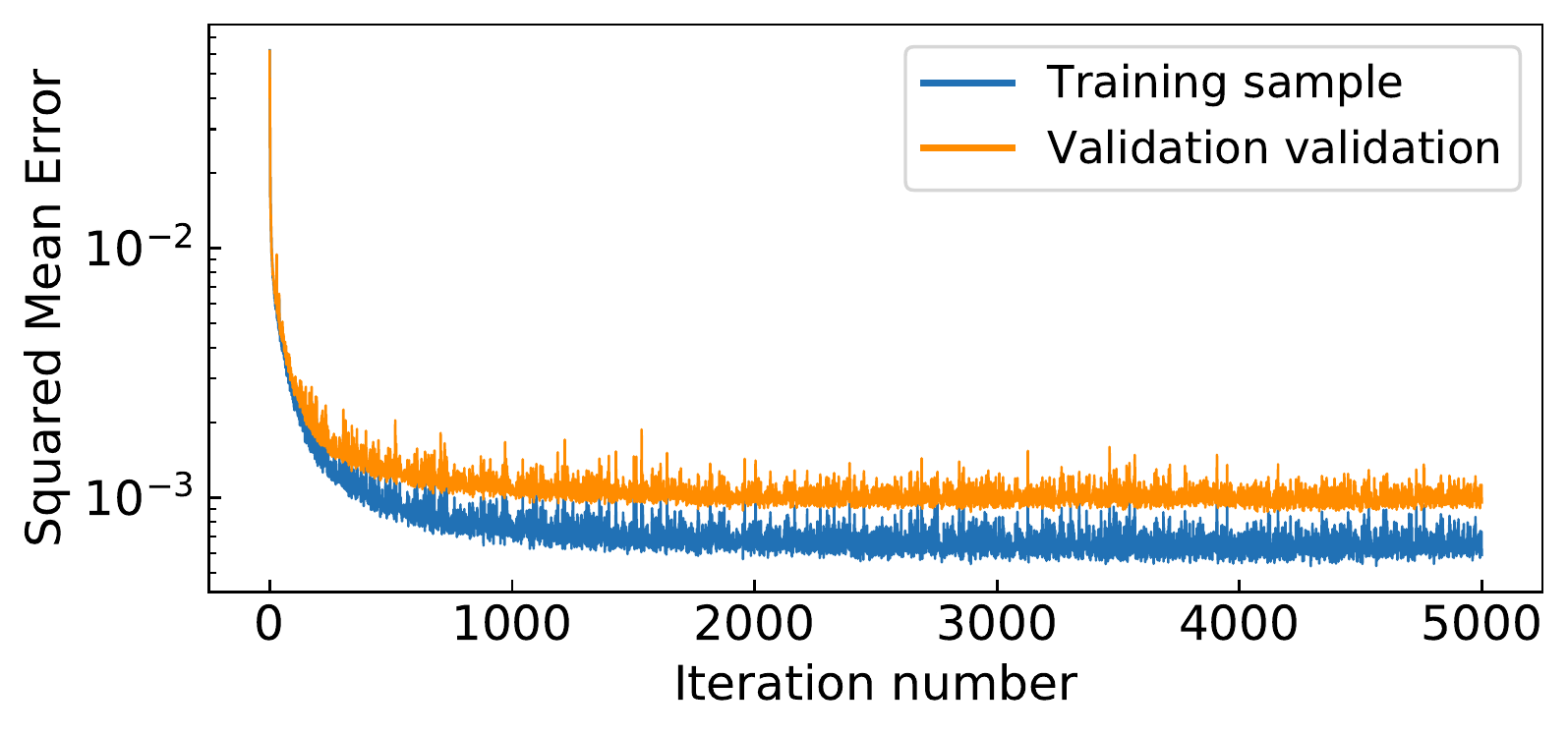}
  \caption{Evolution of the SME for the training and validation samples during the training of
    the ANN. The SME decreases rapidly during the first iterations until convergence is reached. Since the
    training sample used (the DIAD models) do not have noise, no overfitting occurs (no increase is
    seen in the validation error with increasing iterations).}
\end{figure*}\label{fig:fitting_evolution}

The training was performed with the scikit-learn implementation of the stochastic gradient-based optimizer
Adam algorithm \citep{Adam}. The training sample comprises 70000 DIAD models, and another 17000 are used for
validation purposes ($\sim$80\% and $\sim$20\%, respectively). We trained the ANN for 5000 iterations and
monitored the squared mean error (SME) of both the training and validation samples. The ANN with the lowest
validation SME was used. The models do not contain any noise (i.e., the same input parameters will always
yield the same SED) so we do not expect overfitting to occur. This is confirmed by the approximately constant
SME of the validation sample with increasing iterations once convergence has been reached, which in all cases
happened before the 5000 iterations used (see Fig.~\ref{fig:fitting_evolution}).

Training a neural network (almost) never reaches absolute convergence but, instead, it yields weights that are
close to their optimal values \citep[e.g., see][]{Bailer-Jones2002}. For this reason, predictions from neural
networks include some error. In our case, this implies that the output SED from the ANN is not exactly the
same as the SED from the DIAD models given the same input parameters. To mitigate this effect, we trained a
total of five different ANNs using the process formerly described, and use them jointly as a committee: The
final prediction of the flux at each wavelength is the median value of the predictions from each of the five
ANN. This was performed for both the ANN predicting the SED and the one predicting disk masses.

\begin{figure*}[h]
  \centering
  \includegraphics[width=0.8\hsize]{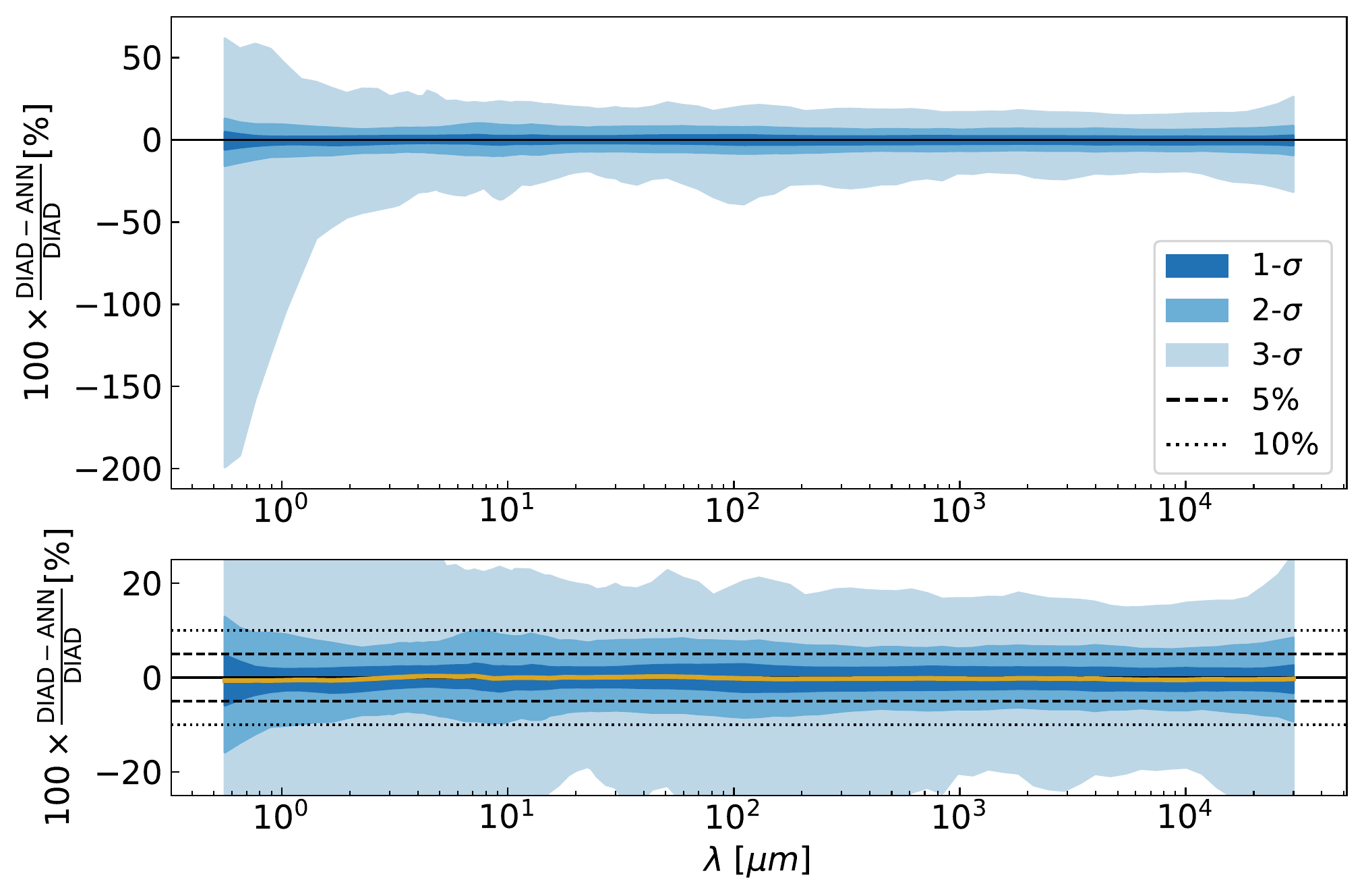}
  \caption{Accuracy of the ANN. For each object in the blind sample (5000 models), we estimated the difference
    between the SED from DIAD and the SED predicted by the ANN at each wavelength. We then estimated the 68\,\% (1-$\sigma$),
    95\,\% (2-$\sigma$), and 99.9\,\% (3-$\sigma$) percentiles. The yellow line corresponds to the median
    difference between DIAD and the ANN at each wavelength. The bottom plot shows a zoom to the -25\,\% -
    25\,\% range.
  }
\end{figure*}\label{fig:ANN_accuracy}

The accuracy of the ANNs was tested with an additional blind sample of 5000 models that were randomly
generated following the same procedure as for the training and validation samples. These models were not used
during the training. Once the ANN was trained, we used it to predict the SEDs of these 5000 models and estimated
the resulting residuals at each wavelength: we found that, in general, 1-$\sigma$ deviations correspond to
a~5\,\% discrepancy between the ANN prediction and the true DIAD model (see Fig.~\ref{fig:ANN_accuracy}), with
its median value centered at 0 (i.e., the ANN does not have a tendency to over- or underpredict). At high
inclinations, self-extinction of the stellar and inner wall emission by the disk becomes important, producing
a strong decrease in the flux at optical and near-IR wavelengths. These changes are extremely sensitive to
small variations of the relevant disk parameters (e.g., inclination, settling), and the ANN weights need to be
very accurate in order to account for these effects. As a result, the uncertainty at wavelengths
$\lesssim$10\,$\mu$m increases: while the 1-$\sigma$ and 2-$\sigma$ deviation levels are still within 5\,\%
and 10\,\%, respectively, the 3-$\sigma$ level reaches up to 200\,\%. Performing the same process with
inclinations $\leq$\,60\,\degr yields a 3-$\sigma$ uncertainty within 50\,\%, clearly showing that this
problem is only important for cases with considerable self-extinction. Since this affects mostly disks that
are close to edge-on (which we do not attempt to fit in this study), we do not expect this to have any
significant impact on our results. Nevertheless, we adopted a conservative 1-$\sigma$ uncertainty of 10\,\% in
the ANN prediction of the SED. We also used the same procedure to quantify the accuracy of the ANN estimating
disk masses, finding a 1-$\sigma$ difference between the ANN prediction and the correct value from DIAD of
$<$\,5\,\%.

\section{Priors} \label{appendix:priors}

\subsection{Priors used in this study}

Bayesian analysis requires the use of priors that encompass previous knowledge about parameters. The adopted
priors in this study are listed in Table~\ref{tab:priors}.

\begin{table*}
  \caption{Priors used for the Bayesian analysis in this study}
  \label{tab:priors}
  \centering
  \begin{tabular}{l c l}
    \hline \hline
    Parameter &  Prior & [Range] or mean, standard deviation \\
    \hline
    Age & ${\mathcal U}$(min, max) & [0.1, 10] Myr\\
    Disk viscosity, $\alpha$ & ${\mathcal J}$(min, max) & [$10^{-4}$, $10^{-1}$]\\
    Accretion rate, $\dot{M}$ & ${\mathcal J}$(min, max) & [$10^{-10}$, $10^{-6.5}$] $M_\odot$/yr\\
    Disk radius, $R_{\rm disk}$ & ${\mathcal U}$(min, max) & [10, 300] au\\
    Dust settling, $\epsilon$ & ${\mathcal J}$(min, max) & [$10^{-4}$, 1]\\
    Temperature of inner wall, $T_{\rm wall}$ & ${\mathcal N}$($\mu$, $\sigma$)& $\mu$=1400\,K, $\sigma$=50\,K\\
    Height of inner wall, $z_{\rm wall}$ & ${\mathcal U}$(min, max) & [0.5, 15]\\
    Inclination, $\cos{(i)}$ (no observational constraint) & ${\mathcal U}$(min, max) & [0, 70] \degr\\
    Inclination, $i$ (observational constraint) & ${\mathcal N}$($\mu$, $\sigma$) & $\mu=i_{\rm observed}$,
                                                                                    $\sigma=i_{\rm unc}$\\
    Max. grain size upper layers, $a_{\rm max, upper}$ & ${\mathcal J}$(min, max) & [0.25, 10] $\mu$m\\
    Max. grain size midplane, $a_{\rm max, midplane}$ & ${\mathcal J}$(min, max) & [0.01, 1] cm\\
    Distance, $d$ & $ \propto d^{2}$(min, max) & [100, 200] pc\\
    Interstellar extinction, $A_V$ &  ${\mathcal U}$(min, max & [0, 8] mag\\
    Outliers mean, $y_{\rm out}$ & ${\mathcal U}$(min, max) & [$10^{-15}$, $10^{-8}$] $\lambda$F$_\lambda$\\
    Outliers standard deviation, $\sigma_{\rm out}$ & ${\mathcal U}$(min, max) & [$10^{-15}$, $10^{-8}$] $\lambda$F$_\lambda$\\
    Outliers fraction, $P_{\rm out}$ & ${\mathcal U}$(min, max) & [0, 0.2]\\
    \hline
  \end{tabular}
  \tablefoot{${\mathcal U:}$ uniform distribution, ${\mathcal N}$: Normal distribution, ${\mathcal J}$: Jeffreys prior.}
\end{table*}

\subsection{Origin of observational inclination measurements}\label{appendix:priors_incl}

Several sources in this work have been observed with high-spatial resolution, which can provide prior
information about disk inclinations. Inclination values for BP~Tau, CI~Tau, DH~Tau, DK~Tau, DL~Tau, DN~Tau,
DO~Tau, DQ~Tau, DR~Tau, DS~Tau, FT~Tau, GI~Tau, GO~Tau, Haro~6-13, HK~Tau, HO~Tau, HP~Tau, IQ~Tau, V710~Tau,
and V836~Tau were compiled from \citet{Long2019}. For CW~Tau, we used the inclination derived from continuum
observations in \citet{Pietu2014}, since it has a smaller uncertainty than the corresponding estimate from CO
emission. Inclination values from gas observations in \citet{Simon2017} were used for CX~Tau, CY~Tau, FM~Tau,
and FP~Tau (in this case, uncertainties derived from the gas emission were smaller than those from the continuum
data). Finally, inclination values for FN~Tau and DG~Tau were compiled from \citet{Kudo2008} and
\citet{Bacciotti2018}, respectively. No uncertainties are reported for these two objects, and we adopted a value
of 10\degr. In all cases, an additional uncertainty of 3\degr\,  was added in quadrature to the ones
listed in the literature (or adopted) to account for possible systematics or underestimated uncertainties.

\section{Likelihoods} \label{appendix:likelihoods}

As mentioned in Sect.~\ref{sec:likelihoods}, we used a likelihood function that includes up to four different terms,
namely: a photometric likelihood, a spectroscopic likelihood, a likelihood for the stellar temperature, and a
likelihood for the observed parallax (when available). Here we describe each of these terms.

{\bf Photometric likelihood}: $\mathcal{L}_{\rm phot}$ evaluates the likelihood of the photometric data. Due
to their heterogeneity, it is likely that some data do not agree with the overall SEDs for different reasons
(e.g., variability, confusion with companions in the case of low-spatial resolution), and we expect some
outliers. Therefore, we used a mixture model to include the possibility that a fraction $P_{\rm out}$ of the
observations arise from a different (outlier) model: a Normal distribution centered at $y_{\rm out}$ and with
a standard deviation of $\sigma_{\rm out}$ \citep[for a description of this method, see][]{Hogg2010}. This
adds three additional parameters to the modeling process. While we marginalized over them after the fitting,
these parameters also require priors: We chose a uniform prior for $P_{\rm out}$ from 0 to 0.2 (i.e., the
maximum number of possible outliers in an SED is 20\,\%), and uniform priors for $y_{\rm out}$ and
$\sigma_{\rm out}$, both of them from $10^{-15}$ to $10^{-8}$ erg/cm$^{2}$/s based on the ranges of observed flux values
\citep[$\sigma_{\rm out}$ was explored in logarithmic space following][]{Hogg2010}. With the adopted mixture
model, the likelihood function for photometric data is

\begin{equation}
\mathcal{L}_{\rm phot} = \prod_{i=1}^{N_{\rm phot}}\bigl(\mathcal{L}_{\rm SED,i} + \mathcal{L}_{\rm out,i}\bigr),
\end{equation}

where the index $i$ corresponds to each photometric measurement, and
$\mathcal{L}_{\rm SED,i}$ and $\mathcal{L}_{\rm out,i}$ are the
likelihoods of a given photometric point arising from the SED model
and from the outlier model, respectively. These can be written as

\begin{gather}
\mathcal{L}_{\rm SED,i}= \frac{1-P_{\rm out}}{\sqrt{2\pi [\sigma_i^2 + \sigma_{i, {\rm
        model}}^2]}} \exp{\biggl(-\frac{(y_i - y_{i,{\rm model}})^2}{2 [\sigma_i^2 +
    \sigma_{i, {\rm model}}^2]}\biggr)}\\
\mathcal{L}_{\rm out,i}= \frac{P_{\rm out}}{\sqrt{2\pi [\sigma_i^2 + \sigma_{\rm
        out}^2]}} \exp{\biggl(-\frac{(y_i - y_{\rm out})^2}{2 [\sigma_i^2 +
    \sigma_{\rm out}^2]}\biggr)},
\end{gather}

where $y_i$ and $\sigma_i$ are the observed fluxes and the corresponding uncertainties,
$y_{i,{\rm model}}$ is the flux predicted by the ANN at the same wavelength, and
$\sigma_{i,{\rm model}} = 0.1\times y_{i,{\rm model}}$ is the adopted uncertainty for the
ANN (10\,\%). 

{\bf Spectroscopic likelihood}: $\mathcal{L}_{\rm spect}$ evaluates the likelihood of the spectrocopic
data. \emph{Spitzer}/IRS spectra are available for all the objects in the sample, and
\emph{Herschel}/SPIRE spectra also exist for some of them. In this case, we chose the standard likelihood form
for Normal uncertainties, and did not account for possible outliers since all the data in the spectra are
taken simultaneously. Therefore, the adopted likelihood for the spectra is

\begin{equation}
\mathcal{L}_{\rm spect}= \prod_{i=1}^{N_{\rm spect}}\frac{1}{\sqrt{2\pi [\sigma_i^2 + \sigma_{i,{\rm
        model}}^2]}} \exp{\biggl(-\frac{(y_i - y_{i,{\rm model}})^2}{2 [\sigma_i^2 +
    \sigma_{\rm out}^2]}\biggr)},
\end{equation}

where $y_i$, $\sigma_i$, $y_{i,{\rm model}}$, and $\sigma_{i,{\rm model}} = 0.1\,
y_{i,{\rm model}}$ have the same meaning as in the photometric case. Here, $\sigma_i$ are
the weighted uncertainties of the spectra (see Sect.~\ref{sec:uncertainties}).

{\bf Stellar temperature likelihood}: while the fundamental parameters of stellar evolution are age and $M_*$, the
spectral type/effective temperature of stars is the easiest to measure. For our sample, $T_*$
estimates are available through spectroscopic observations (see Sect.~\ref{sec:sample}). Therefore, although
$T_*$ is not an input parameter of the models, we require that the $T_{*, \rm isochrone}$ value obtained from
the MIST isochrones (see step 1 in Sect.~\ref{sec:model}) is compatible with the observed $T_*$. For this
purpose, we include a term in the likelihood that accounts for the stellar temperature,
$\mathcal{L}_{\rm T_*}$. Assuming Normal uncertainties for $T_*$, this likelihood is

\begin{equation}
  \mathcal{L}_{\rm T_*}= \frac{1}{\sqrt{2\pi \sigma_{\rm T_*}^2}} \exp{\biggl(-\frac{(T_* - T_{*, \rm isochrone})^2}{2 \sigma_{\rm T_*}^2}\biggr)},
\end{equation}

where $T_*$ is the adopted stellar temperature from the SpTs in \citet[][see also Section
\ref{sec:sample}]{Luhman2017}, $\sigma_{\rm T_*}$ is the uncertainty of this value (set to
100\,K, see Sect.\ref{sec:sample}), and $T_{*, \rm isochrone}$ is the temperature predicted by the
MIST isochrones based on the input age and $M_*$ values for the model.

{\bf Parallax likelihood}: for most objects in our sample, a parallax estimate is available from the
\emph{Gaia} DR2 catalog. In those cases, we included a term in the likelihood to incorporate this
information: 

\begin{equation}
  \mathcal{L}_{\rm parallax}= \frac{1}{\sqrt{2\pi \sigma_{\varpi}^2}} \exp{\biggl(-\frac{(\varpi - \varpi_{model})^2}{2 \sigma_{\varpi}^2}\biggr)}.
\end{equation}

In this expression, $\varpi$ and $\sigma_{\varpi}$ are the observed parallax and its uncertainty, and
$\varpi_{model}$ is the parallax corresponding to the model distance. We note that the likelihood is written
in terms of parallax instead of distance to account for the fact that parallax uncertainties are not symmetric
in distance space. When no parallax measurement exists, this term is not included in the likelihood.

\section{Results for individual sources} \label{appendix:results}

Table~\ref{tab:modeling_params} lists modeling results for individual sources for the parameters of interest
in Sect.~\ref{sec:results_params}. We note that these should be considered with caution especially when
comparing some of them with observational results, since a direct comparison is not always straightforward:
For example, the accretion rate in the disk is not necessarily the same as the accretion rate onto the star
(which can be largely variable), and the disk radius is different for different grain sizes due to radial
migration whereas our models use one single value.

\begin{sidewaystable*}
\caption{Modeling results for the relevant (non-marginalized) parameters. The reported values correspond to
  the derived median of each parameter, while uncertainties are computed as 16\,\% and 84\,\% percentile levels.}
\label{tab:modeling_params}
\centering
\begin{tabular}{l c c c c c c c c}
\hline \hline
Source  &  $\alpha$  &  $\dot{M}$ [$\times 10^{-8} M_\odot$/yr]  &  $R_{\rm disk}$ [au]  &  $\epsilon$ [$\times
                                                                                       10^{-4}]$  &  $a_{\rm
                                                                                                    max,
                                                                                                    upper}$
                                                                                                    [$\mu$m]
  & $a_{\rm max, midplane}$ [mm]  &  $M_{\rm disk}$ [$\times 10^{-2} M_\odot]$  \\
\hline
BP Tau & $0.017_{-0.013}^{+0.006}$ & $2.0_{-1.6}^{+0.6}$ & $250_{-70}^{+40}$ & $1.27_{-0.2}^{+0.59}$ & $1.4_{-0.5}^{+3.0}$ & $2.4_{-1.5}^{+2.8}$ & $1.5_{-0.3}^{+0.4}$ \\
CIDA 7 & $0.003_{-0.002}^{+0.005}$ & $0.13_{-0.1}^{+0.17}$ & $160_{-110}^{+90}$ & $80_{-50}^{+170}$ & $0.42_{-0.11}^{+0.16}$ & $2.7_{-2.0}^{+5.2}$ & $0.27_{-0.13}^{+0.24}$ \\
CI Tau & $0.014_{-0.006}^{+0.007}$ & $4.0_{-0.9}^{+1.1}$ & $160_{-60}^{+90}$ & $11_{-5}^{+14}$ & $1.3_{-0.3}^{+1.0}$ & $0.6_{-0.5}^{+2.4}$ & $3.8_{-0.8}^{+1.3}$ \\
CW Tau & $0.06_{-0.02}^{+0.03}$ & $10.4_{-2.1}^{+1.8}$ & $53_{-15}^{+55}$ & $4.6_{-1.8}^{+3.4}$ & $5.0_{-1.7}^{+2.6}$ & $2_{-2}^{+4}$ & $1.1_{-0.3}^{+0.5}$ \\
CX Tau & $0.06_{-0.03}^{+0.02}$ & $2.4_{-0.6}^{+0.7}$ & $170_{-120}^{+90}$ & $2.2_{-0.9}^{+1.9}$ & $0.45_{-0.14}^{+0.21}$ & $3_{-3}^{+4}$ & $0.32_{-0.1}^{+0.12}$ \\
CY Tau & $0.00014_{-0.00003}^{+0.00005}$ & $0.042_{-0.009}^{+0.018}$ & $260_{-40}^{+30}$ & $1.11_{-0.08}^{+0.13}$ & $9.2_{-1.1}^{+0.6}$ & $0.8_{-0.3}^{+0.6}$ & $7.7_{-1.2}^{+1.1}$ \\
DE Tau & $0.0005_{-0.0003}^{+0.0159}$ & $0.04_{-0.03}^{+1.56}$ & $260_{-60}^{+30}$ & $10_{-5}^{+17}$ & $5_{-3}^{+2}$ & $7.3_{-3.1}^{+1.8}$ & $1.0_{-0.2}^{+0.4}$ \\
DG Tau & $0.025_{-0.008}^{+0.01}$ & $29.0_{-2.7}^{+1.9}$ & $43_{-5}^{+9}$ & $2000_{-2000}^{+5000}$ & $6.9_{-2.0}^{+1.9}$ & $1.0_{-0.8}^{+3.4}$ & $4.9_{-1.0}^{+1.7}$ \\
DN Tau & $0.0016_{-0.0012}^{+0.0025}$ & $0.4_{-0.3}^{+0.6}$ & $230_{-100}^{+50}$ & $1.6_{-0.5}^{+0.8}$ & $7.6_{-2.1}^{+1.7}$ & $3_{-2}^{+3}$ & $3.8_{-1.3}^{+1.5}$ \\
DS Tau & $0.019_{-0.009}^{+0.012}$ & $1.8_{-0.6}^{+0.8}$ & $230_{-80}^{+50}$ & $1.4_{-0.3}^{+0.7}$ & $1.0_{-0.3}^{+0.8}$ & $1.2_{-1.0}^{+2.4}$ & $1.2_{-0.3}^{+0.3}$ \\
FM Tau & $0.008_{-0.005}^{+0.006}$ & $6.5_{-1.8}^{+1.4}$ & $10.3_{-0.2}^{+0.5}$ & $1.5_{-0.4}^{+1.5}$ & $1.0_{-0.3}^{+0.6}$ & $0.5_{-0.3}^{+1.8}$ & $0.6_{-0.3}^{+1.0}$ \\
FP Tau & $0.00017_{-0.00006}^{+0.00013}$ & $0.08_{-0.04}^{+0.06}$ & $170_{-70}^{+80}$ & $3.7_{-1.8}^{+3.2}$ & $7.7_{-3.1}^{+1.6}$ & $0.9_{-0.7}^{+3.1}$ & $5.2_{-1.2}^{+2.0}$ \\
FY Tau & $0.0018_{-0.0013}^{+0.0082}$ & $0.07_{-0.05}^{+0.24}$ & $150_{-90}^{+110}$ & $21_{-12}^{+28}$ & $6_{-3}^{+2}$ & $1.6_{-1.3}^{+4.9}$ & $0.38_{-0.12}^{+0.2}$ \\
FZ Tau & $0.03_{-0.02}^{+0.02}$ & $11_{-3}^{+4}$ & $11.6_{-1.2}^{+3.0}$ & $3.0_{-1.5}^{+8.8}$ & $3.2_{-1.4}^{+2.7}$ & $1.1_{-0.8}^{+4.7}$ & $0.5_{-0.2}^{+1.6}$ \\
GI Tau & $0.07_{-0.03}^{+0.02}$ & $4.1_{-1.4}^{+1.4}$ & $34_{-20}^{+77}$ & $9_{-5}^{+11}$ & $0.34_{-0.07}^{+0.1}$ & $3_{-3}^{+5}$ & $0.17_{-0.06}^{+0.14}$ \\
GO Tau & $0.0004_{-0.0003}^{+0.0006}$ & $0.07_{-0.04}^{+0.09}$ & $190_{-70}^{+70}$ & $21_{-10}^{+18}$ & $0.64_{-0.16}^{+6.36}$ & $0.27_{-0.09}^{+0.22}$ & $2.2_{-0.3}^{+0.4}$ \\
Haro 6-13 & $0.02_{-0.006}^{+0.009}$ & $11.9_{-1.1}^{+1.1}$ & $37_{-5}^{+8}$ & $600_{-400}^{+800}$ & $3.5_{-1.0}^{+2.0}$ & $4_{-3}^{+4}$ & $2.9_{-0.7}^{+1.0}$ \\
HO Tau & $0.0018_{-0.0012}^{+0.0016}$ & $0.24_{-0.15}^{+0.17}$ & $220_{-80}^{+60}$ & $2.1_{-0.8}^{+1.8}$ & $1.5_{-0.5}^{+0.8}$ & $1.5_{-1.1}^{+3.2}$ & $1.6_{-0.5}^{+0.8}$ \\
HP Tau & $0.0009_{-0.0007}^{+0.0041}$ & $11.3_{-1.1}^{+1.4}$ & $17.3_{-2.1}^{+2.0}$ & $40_{-30}^{+240}$ & $1.6_{-0.5}^{+0.7}$ & $1.3_{-1.0}^{+3.0}$ & $30_{-20}^{+180}$ \\
IQ Tau & $0.0013_{-0.001}^{+0.0113}$ & $0.3_{-0.2}^{+2.4}$ & $190_{-100}^{+70}$ & $10_{-5}^{+9}$ & $7.6_{-2.4}^{+1.7}$ & $1.3_{-1.0}^{+3.0}$ & $2.1_{-0.5}^{+0.8}$ \\
IRAS 04385+2550 & $0.039_{-0.016}^{+0.018}$ & $12_{-3}^{+3}$ & $28_{-7}^{+9}$ & $1000_{-700}^{+1200}$ & $2.8_{-1.5}^{+2.7}$ & $3_{-3}^{+4}$ & $0.9_{-0.3}^{+0.5}$ \\
V710 Tau & $0.0008_{-0.0006}^{+0.0018}$ & $0.18_{-0.14}^{+0.36}$ & $150_{-60}^{+80}$ & $2.9_{-1.2}^{+2.4}$ & $4_{-3}^{+3}$ & $0.27_{-0.13}^{+0.54}$ & $2.2_{-0.5}^{+0.8}$ \\
V836 Tau & $0.00021_{-0.00008}^{+0.00026}$ & $0.036_{-0.017}^{+0.052}$ & $260_{-70}^{+30}$ & $1.6_{-0.4}^{+1.0}$ & $0.41_{-0.09}^{+0.53}$ & $3.3_{-1.9}^{+3.5}$ & $2.9_{-0.8}^{+0.8}$ \\
\hline 
\end{tabular} 
\end{sidewaystable*}

\section{Marginalized parameters} \label{appendix:marginalized_params}

Some parameters in our model are either nuisance ones (i.e., the three parameters required to model possible
outlier photometric points) or are included to account for our ignorance about their true values and their
impact on other parameters. Therefore, their posteriors are not really informative or relevant for our study,
and we do not analyze them in detail. These parameters are:

\begin{figure*}[h]
    \centering
  \includegraphics[height=5.3cm]{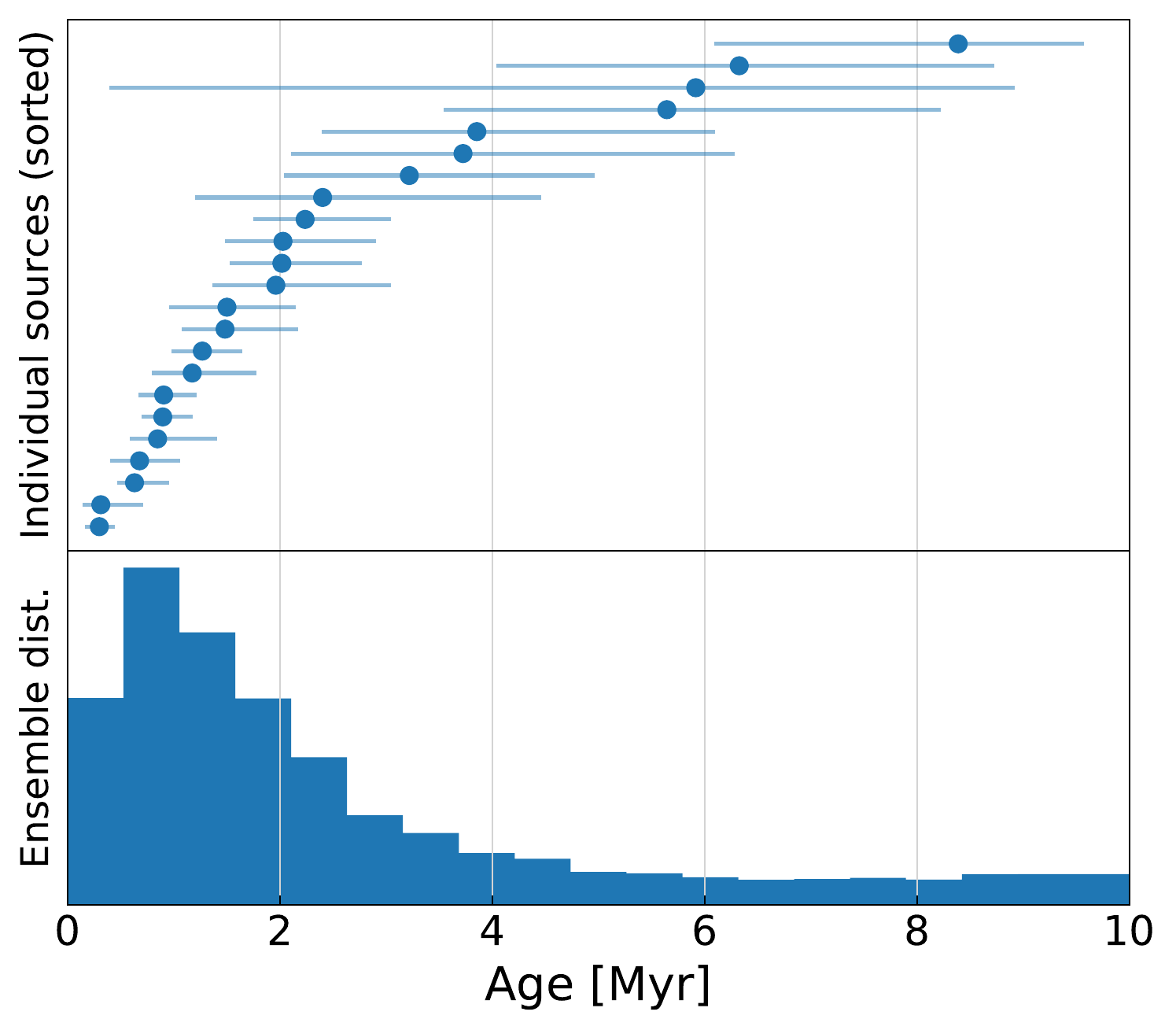}\hfill
  \includegraphics[height=5.3cm]{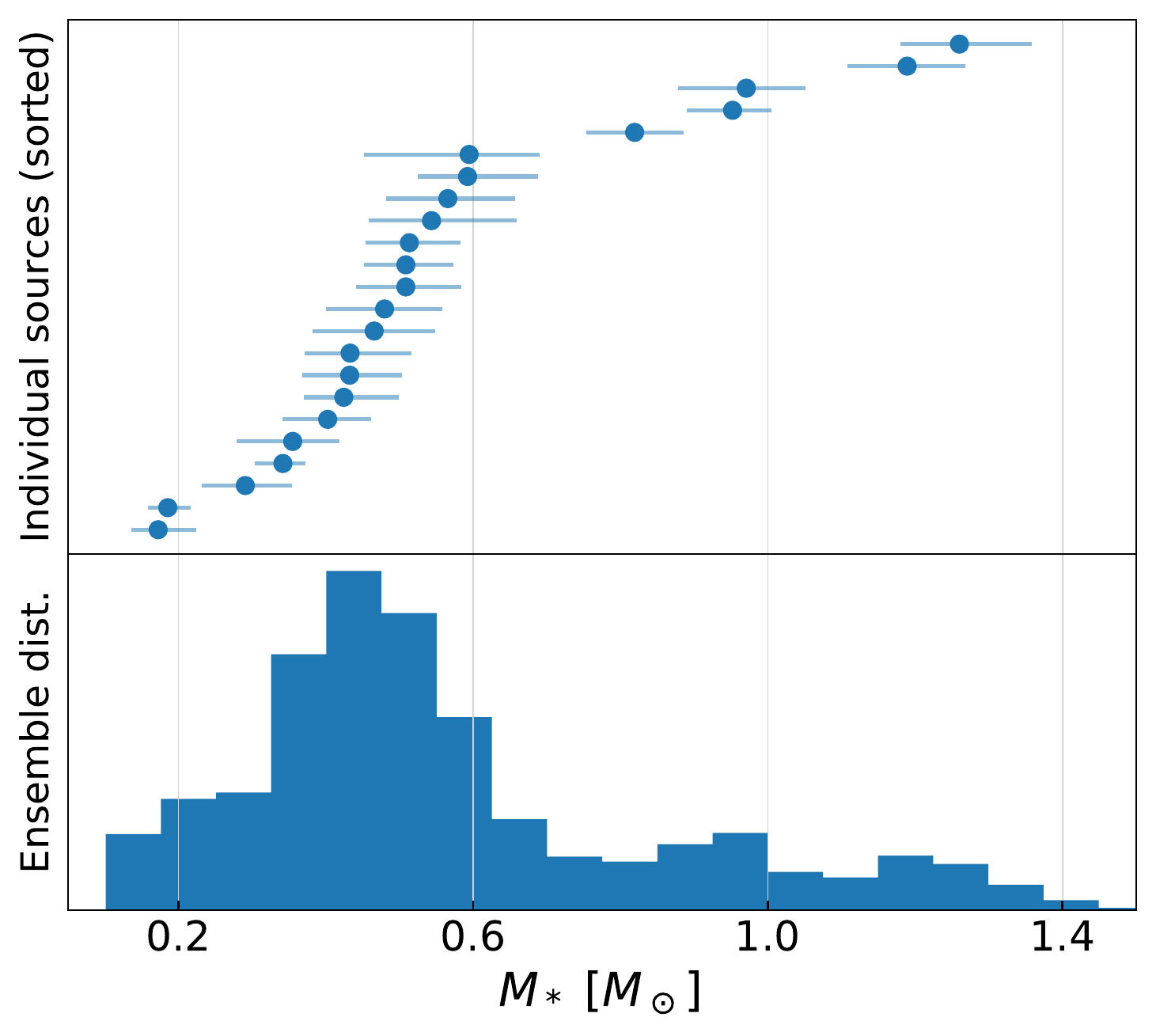}\hfill
  \includegraphics[height=5.3cm]{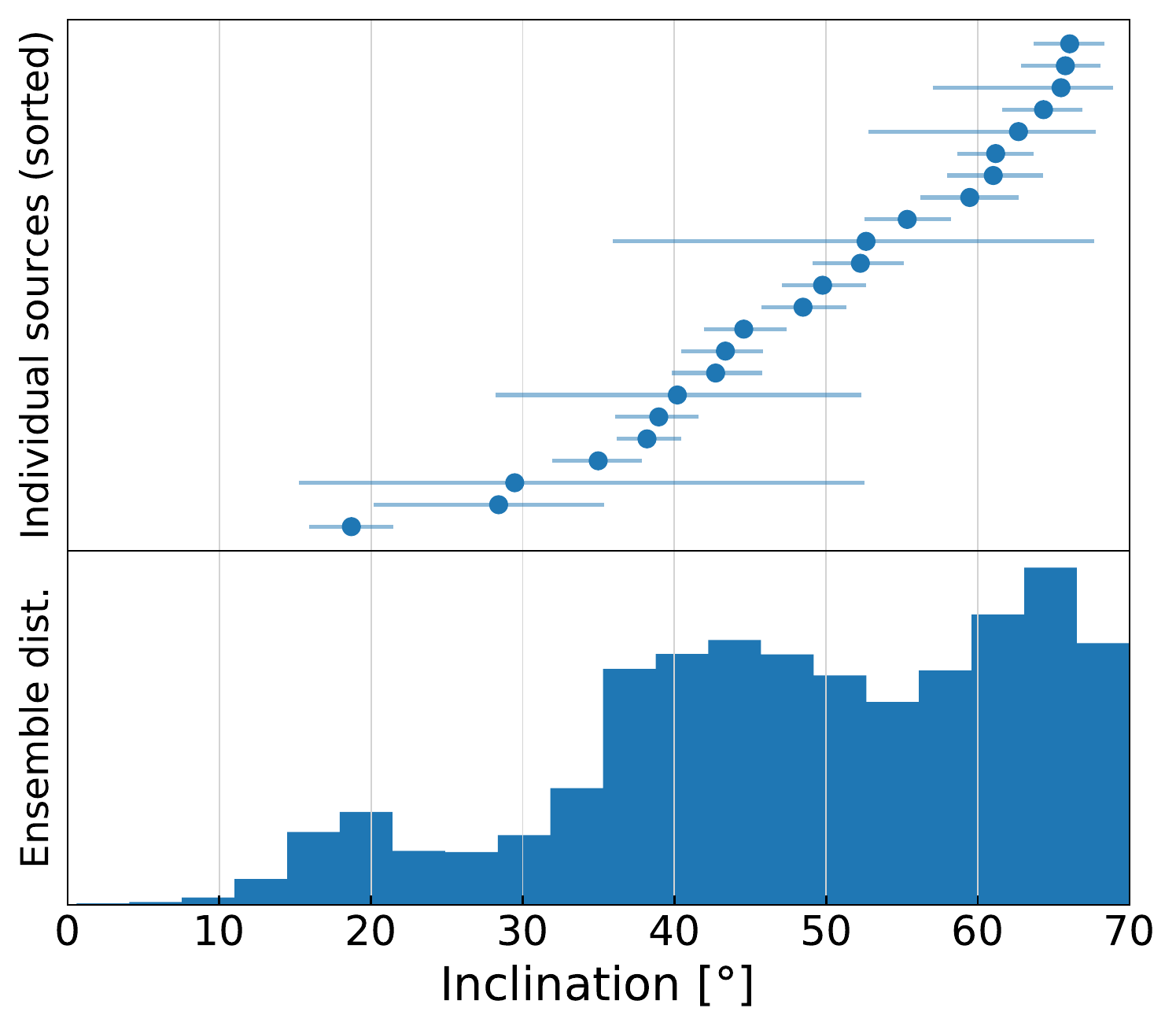}\\
  \includegraphics[height=5.3cm]{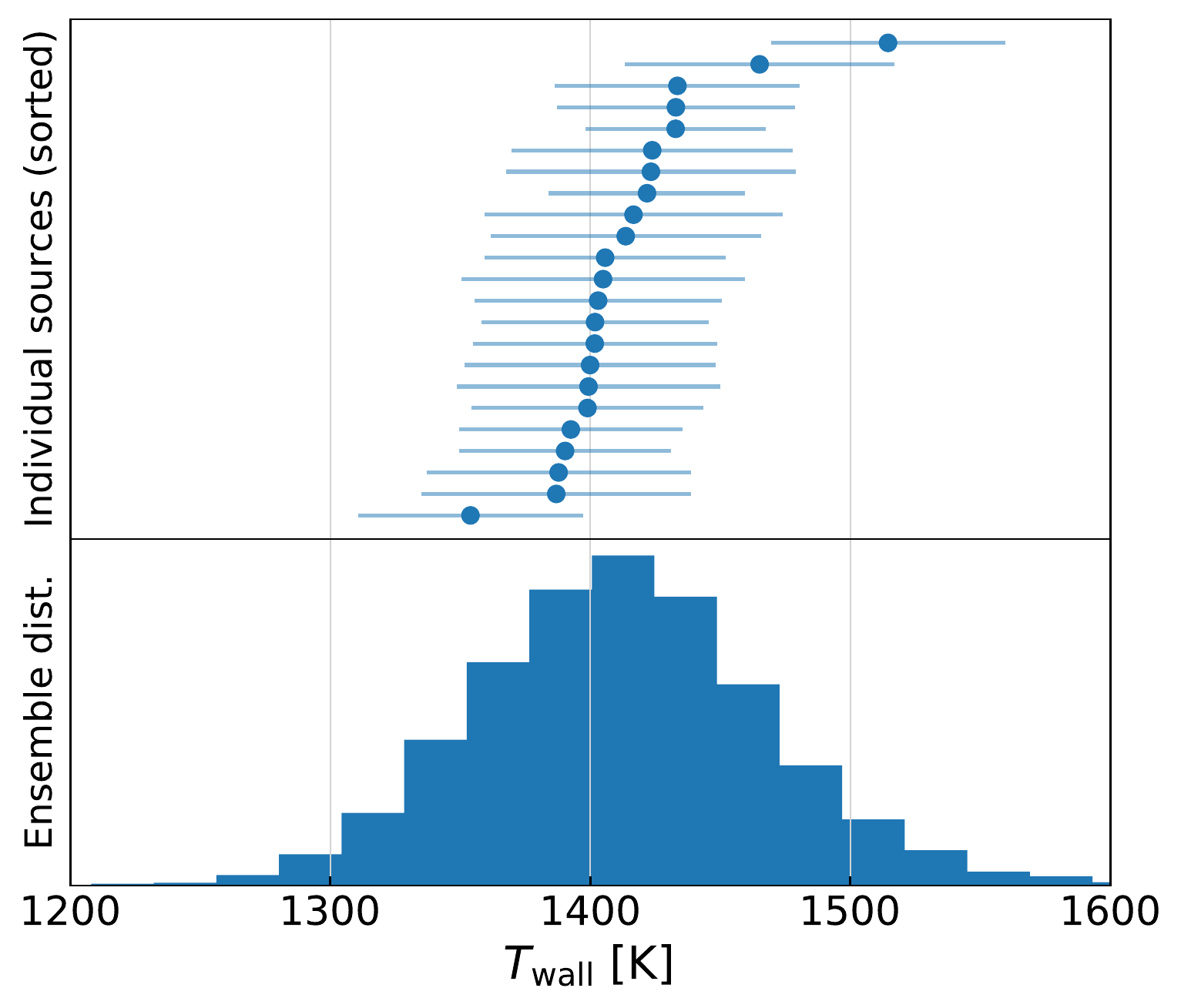}\hfill
  \includegraphics[height=5.3cm]{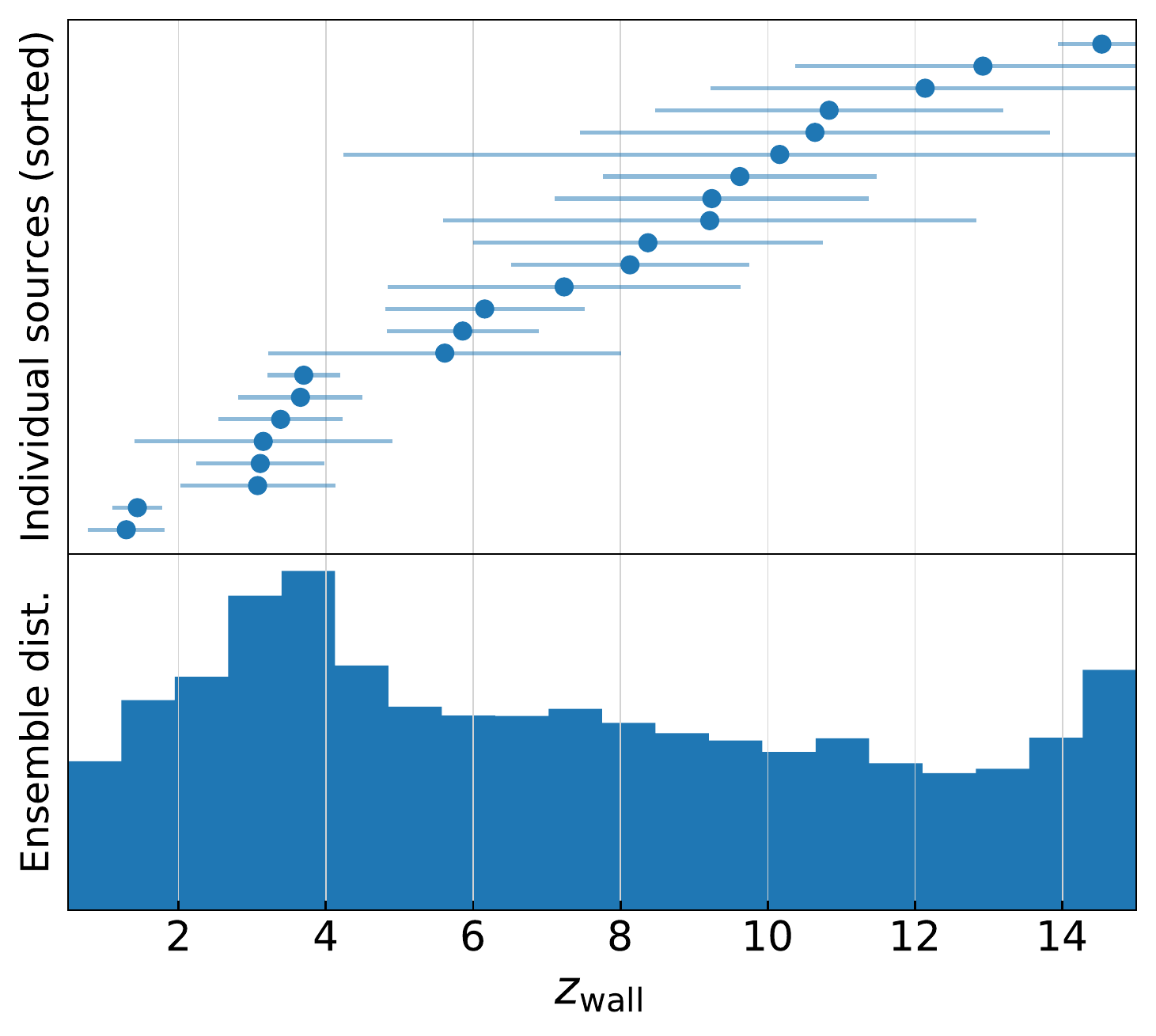}\hfill
  \includegraphics[height=5.3cm]{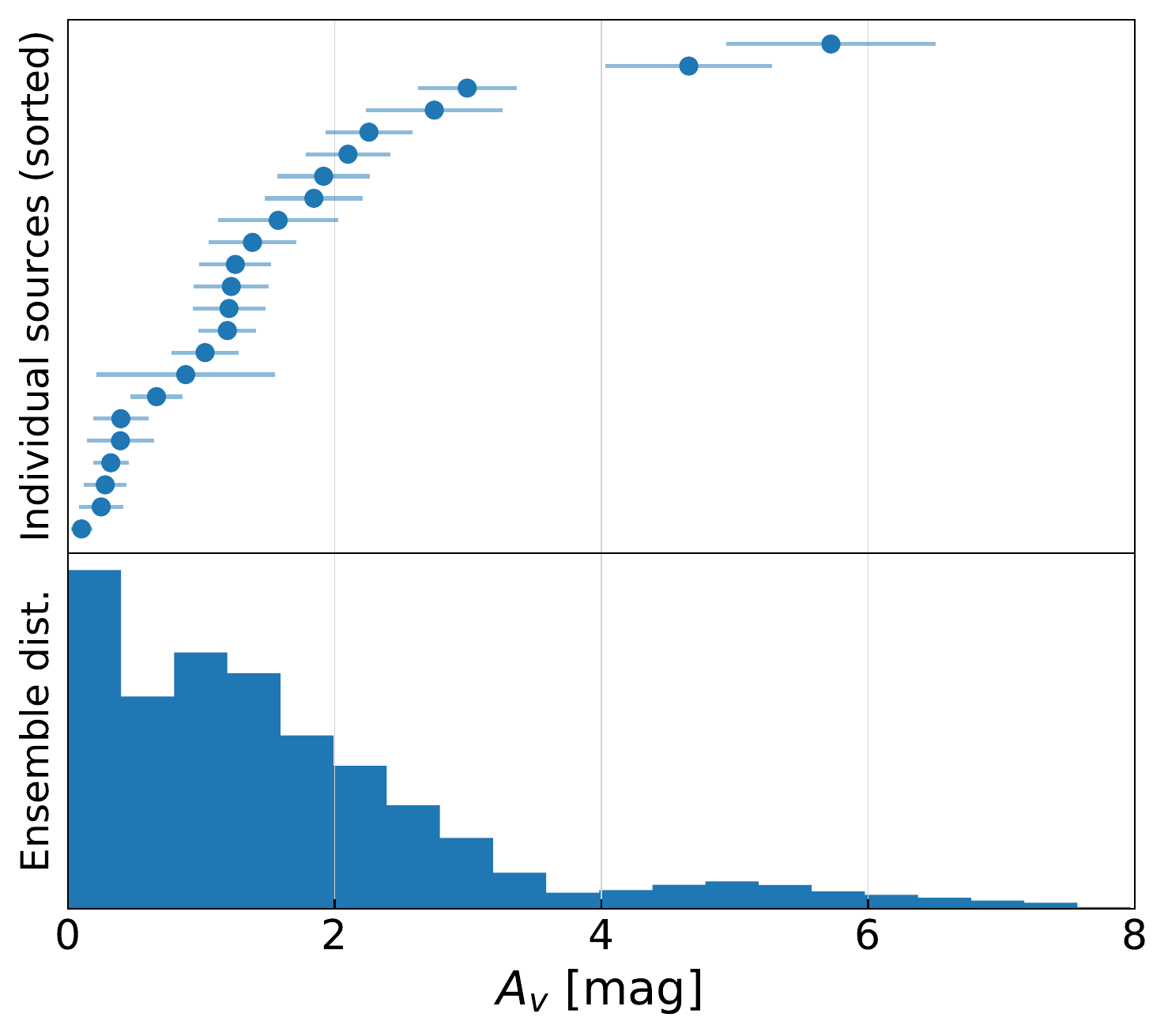}\\
  \includegraphics[height=5.3cm]{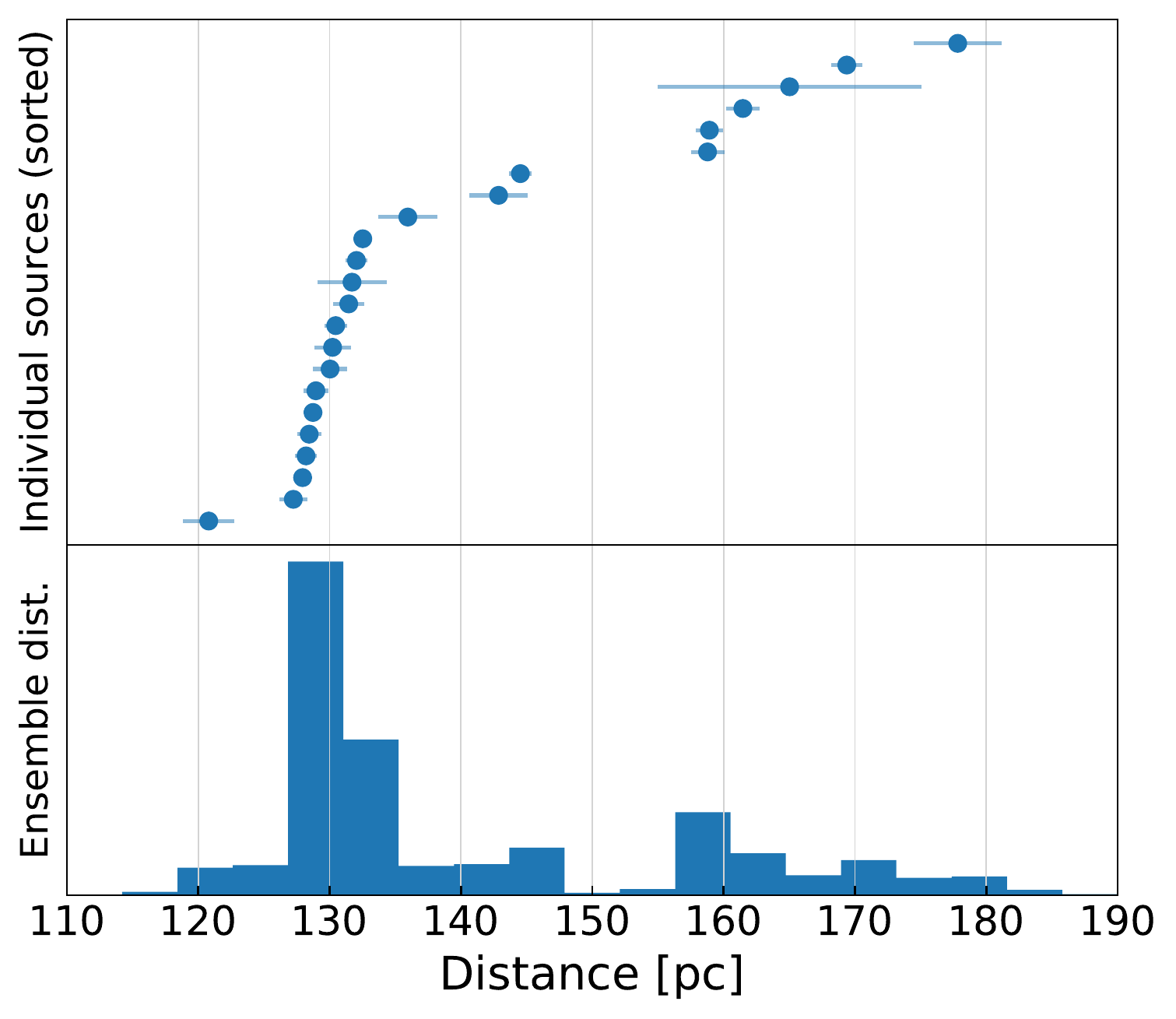}
  \caption{Results for individual sources (top panels) and ensemble distribution (bottom panels) of the
      marginalized parameters in our study for the sample of 23 modeled disks. The
      ensembles are produced similarly to the ones in
      Fig.~\ref{fig:ensemble_distributions}.}\label{fig:ensemble_posteriors_marginalized_params}
\end{figure*}

\begin{itemize}

\item Stellar parameters (age, $M_{*}$, $T_{*}$, and $R_{*}$): our model contains two star-related parameters
  (age and $M_{*}$), which are passed to the MIST isochrones to derive consistent $T_{*}$ and $R_{*}$
  values. While these are constrained in all cases, we do not expect their values to be truly accurate: this
  would require a more complete treatment of the photospheric fluxes during the fitting process, that is, the
  use of high-resolution synthetic spectra and convolutions with the corresponding photometric filters. Instead, the
  photospheric values in DIAD come from the empirical colors in \citet{Pecaut2013} and are interpolated at
  intermediate wavelengths when needed. Therefore, the derived stellar parameters are just approximate, and
  are used for internal consistency of the models and to account for the additional uncertainties they produce
  in other parameters. Nevertheless, they are compatible with more precise estimates from previous studies
  using spectroscopic measurements \citep[e.g.,][]{Luhman2017}. It is worth noting the resulting distribution
  for the age, with a clear preference for values$\sim$1-2\,Myr and steep but smooth decline toward older
  ages.
  
\item Inclination ($i$): there is little information about disk inclinations in unresolved photometry, except in
  the case of edge-on disks (which were excluded from our sample, Sect.~\ref{sec:sample}). In all cases, the
  posterior distribution of this parameter resembled the prior used, meaning that no useful information was
  grained during the modeling process.

\item The dust sublimation temperature ($T_{\rm wall}$) and scaling of the inner wall of the disk
  ($z_{\rm wall}$): these two parameters encapsulate our ignorance about the true shape and location of the
  inner wall. The posterior of $T_{\rm wall}$ resembles the prior used and is still centered around
  $\sim$1400\,K. On the other hand, the scaling factor $z_{\rm wall}$ is required to account for the
  additional solid angle of the wall because DIAD assumes a flat, vertical wall. This factor is correlated
  with the inclination, since disks close to face-on will require, in general, a larger $z_{\rm wall}$
  value. Overall, the posterior shows that some scaling is required in all cases, thus suggesting that the
  inner walls are indeed curved and/or puffed-up. However, given the ad hoc weighting of the \emph{Spitzer}/IRS
  spectra and the fact that we have not explored different dust compositions (which play a crucial role in the
  location and temperature of the inner wall), we do not consider the results for $z_{\rm wall}$ to be
  informative of its true value.

\item Distance ($d$): the posterior distributions for the distances are completely dominated by the \emph{Gaia} DR2
  data, as expected: Distances derived from photometric data are much more uncertain and do not improve upon
  \emph{Gaia} estimates, so our modeling process does not provide any additional information about this parameter.

\item Extinction ($A_V$): extinction values are constrained in all cases but, as in the case of the stellar
  parameters, a precise estimate of this quantity requires a proper convolution of each photometric filter
  with photospheric models. Therefore, the derived $A_V$ values only provide general estimates of extinction
  values. All sources objects have $A_V< 6$\,mag, in good agreement with other studies
  \citep[e.g.,][]{Andrews2013}.

\item Parameters modeling outlier photometric data ($y_{\rm out}$, $\sigma_{\rm out}$, and $P_{\rm out}$):
  these are the three parameters involved in the outlier rejection model. While the mixture model makes the
  fitting more robust to possible outliers, these parameters have no physical meaning.

\end{itemize}

The ensemble posterior distributions for the marginalized parameters are show in
Fig.~\ref{fig:ensemble_posteriors_marginalized_params}.

\newpage
\section{Correlations between $\dot{M}$, $\alpha$, and $M_{\rm disk}$}\label{appendix:Mdot_alpha_Mdisk_correlation}

\begin{figure}[h]
  \centering
  \includegraphics[width=\hsize]{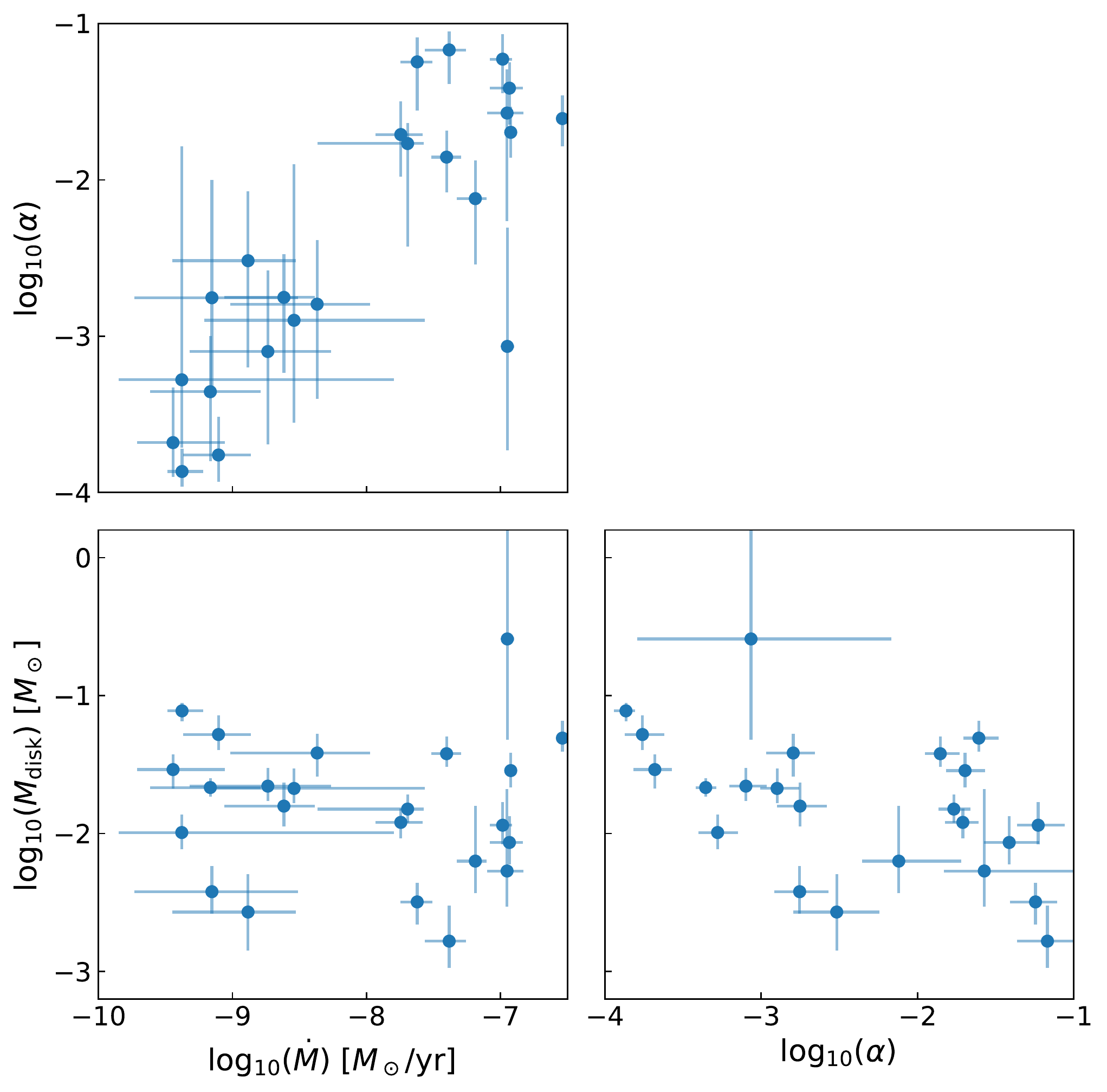}
  \caption{Comparison of the derived $\dot{M}$, $\alpha$, and $M_{\rm disk}$ values. $\dot{M}$ and
      $\alpha$ appear clearly correlated, as expected for the $\alpha$-disk prescription. However, $M_{\rm disk}$
      does not show strong correlations with these parameters.}\label{fig:Mdot_alpha_Mdisk_correlations}
\end{figure}

\end{appendix}

\end{document}